\documentclass[fleqn,usenatbib]{mnras}

\usepackage{newtxtext,newtxmath}

\usepackage[T1]{fontenc}
\usepackage{amsmath}
\DeclareRobustCommand{\VAN}[3]{#2}
\let\VANthebibliography\thebibliography
\def\thebibliography{\DeclareRobustCommand{\VAN}[3]{##3}\VANthebibliography}

\usepackage{graphicx}	
\usepackage{amsmath}	
\usepackage{floatrow,sidecap,multicol,booktabs,longtable}
\usepackage{caption}
\usepackage{makecell}
\usepackage{pifont}
\newcommand{\angstrom}{\text{\normalfont\AA}}
\usepackage[usenames,dvipsnames,table]{xcolor}
\usepackage{xltabular}
\usepackage{deluxetable}
\usepackage{multirow}
\usepackage{hhline}

\title[Luminosity function completeness]{A quantitative assessment of completeness correction methods and public release of a versatile simulation code}

\author[N. Leethochawalit et al.]
{
Nicha Leethochawalit,$^{1,2,5}$\thanks{E-mail:nicha.leethochawalit@unimelb.edu.au}
Michele Trenti,$^{1,2}$
Takahiro Morishita,$^{3}$
Guido Roberts-Borsani,$^{4}$
\newauthor
Tommaso Treu,$^{4}$
\\
$^{1}$School of Physics, Tin Alley, University of Melbourne, VIC 3010, Australia\\
$^{2}$ ARC Centre of Excellence for All Sky Astrophysics in 3 Dimensions (ASTRO 3D), Australia \\
$^{3}$Space Telescope Science Institute, 3700 San Martin Drive, Baltimore, MD 21218, USA\\
$^{4}$Department of Physics and Astronomy, UCLA, 430 Portola Plaza, Los Angeles, CA 90095-1547, USA\\
$^{5}$National Astronomical Research Institute of Thailand (NARIT), MaeRim, Chiang Mai, 50180, Thailand
}

\date{Accepted 2021 November 05. Received 2021 November 04; in original form 2021 May 11}

\pubyear{2015}

\begin{document}
\label{firstpage}
\pagerange{\pageref{firstpage}--\pageref{lastpage}}
\maketitle

\begin{abstract}
Having accurate completeness functions is crucial to the determination of the rest-frame ultraviolet luminosity functions (UVLFs) all the way back to the epoch of reionization. Most studies use injection-recovery simulations to determine completeness functions. Although conceptually similar, published approaches have subtle but important differences in their definition of the completeness function. As a result, they implement different methods to determine the UVLFs. We discuss the advantages and limitations of existing methods using a set of mock observations, and then compare the methods when applied to the same set of Hubble Legacy Field (HLF) images. We find that the most robust method under all our mock observations is the one that defines completeness as a function of both input and output magnitude. Other methods considering completeness only as a function of either input or output magnitude may suffer limitations in a presence of photometric scatter and/or steep luminosity functions. In particular, when the flux scatter is $\gtrsim0.2$ mag, the bias in the bright end of the UVLFs is on par with other systematic effects such as the lensing magnification bias. When tested on HLF images, all methods yield UVLFs that are consistent within $2\sigma$ confidence, suggesting that UVLF uncertainties in the literature are still dominated by small number statistics and cosmic variance. The completeness simulation code used in this study (\texttt{GLACiaR2}) is publicly released with this paper as a tool to analyse future higher precision datasets such as those expected from the James Webb Space Telescope. 

\end{abstract}




\section{Introduction}

Owing to several large extragalactic programs on the Hubble Space Telescope (HST), such as the Hubble Legacy Fields (HLF), the Hubble Frontier Fields (HFF), the Reionization Lensing Cluster Survey (RELICS) and the Brightest of
Reionizing Galaxies (BoRG) survey, we have discovered thousands of galaxy candidates at redshift $z\gtrsim6$ \citep[e.g.,][]{Bouwens2015,Ishigaki2018,Salmon2020,Schmidt2014,Morishita2018,Bowler2020,Roberts-Borsani2021}, leading to a better understanding of cosmic reionization in terms of its timing and the sources that governed it. 
Although star-forming galaxies are now considered the primary agents to reionize the universe \citep{FaucherGiguere2008, HaardtMadau2012, Robertson2015, Ma2020}, their actual contribution of ionizing photons remains to be observationally confirmed. The difficulty is due to the intergalactic medium absorption from neutral hydrogen in the reionizing epoch that makes it impossible to directly measure the ionizing UV radiation from star-forming galaxies. The ionizing UV radiation from star-forming galaxies, therefore, has to be calculated through integrating three quantities: the rest-frame UV luminosity density, an efficiency factor that converts the UV luminosity to Lyman-continuum photons, and the escape fraction. Each has its respective challenges.

The ultraviolet luminosity function (UVLF or $\phi(M_\textrm{UV})$) is defined as the number of galaxies, whose absolute magnitude --measured at $\sim1400-1600$\angstrom\ -- is between $M_\textrm{UV}$ and $M_\textrm{UV}+dM_\textrm{UV}$, at a specific redshift per unit comoving volume. It is a fundamental parameter that not only determines the rest-frame UV luminosity density but is also an indicator of many underlying physical properties of these galaxies such as star formation rate and stellar properties.

Much of our understanding in the high-redshift universe depends on the determination of the UVLF at the bright and the faint end. Both measurements are still observationally challenging. The state-of-the-art constraint for the faint-end come from the Hubble Frontier fields \citep{Ishigaki2018}, whose limiting UV absolute magnitudes reach $\sim-14$ mag. The current understanding is that the UVLFs at $z\gtrsim6$ have faint-end slopes that are steep and likely steepen with increasing redshifts \citep{McLure2013, Schenker2013, Ishigaki2018, Bouwens2021}, consistent with cosmological simulations \citep[e.g.,][]{Wilkins2017,Ma2018}. These results support the hypothesis that faint galaxies are the main contributor to ionizing photons. Nonetheless, to account for the ionizing photons needed for the reionization, we still need to extrapolate the UVLFs beyond the current observational limits by perhaps two orders of magnitude fainter in luminosity, i.e. $M=-12$ \citep{Mason2019}. Hence, there is an observational need to confirm this behaviour at the faint end and, besides, to observe the predicted flattening of UVLF at $M<-12$ due to increased ionizing background that suppresses star formation in dwarf galaxies \citep[e.g.,][]{Gnedin2016,Yue2016}. 

While the difficulty in constraining the UVLF at the faint end originates from limited exposure depth of the surveys, the problem at the bright end comes from a combination of limited survey volumes, and impact of cosmic variance on a single large-area field (e.g. see \citealt{Trenti2008} for a discussion on cosmic variance). Several surveys have endeavoured to address this issue by a combination of increased search area to find these rare bright objects, and observations among multiple lines of sight to limit cosmic variance. These are done by either taking advantage of pure-parallel opportunities with Hubble \citep[e.g.,][]{Yan2012,Morishita2021} or having dedicated surveys with long exposures on 10m-class ground telescopes \citep[e.g.,][]{Bowler2015,Ono2018,Stefanon2019}. These observations are providing tentative but increasing evidence that there are more luminous sources at the bright-end of UVLFs at high redshift $z>7$ than what the exponential decline component of the Schechter function predicts \citep{Bowler2014,Ono2018,Morishita2018, Bowler2020}. Many empirical models for the redshift evolution of UVLFs have successfully predicted UVLFs at high redshifts by starting from the dark-matter halo mass function, and assuming that star formation tracks the halo accretion rate, with a calibration step at a reference redshift to encode all the complex physics of conversion of gas into stars which is well beyond the scope of these models to consider  \citep[e.g.,][]{Trenti2010,Mason2015b, Mashian2016, Tacchella2018}. The assumption of a strong link between star formation rate (SFR) and halo accretion rate ($\dot{M_h}$) is well supported by the tight relationship observed between SFR$/\dot{M_h}$ and $M_h$, which does not evolve significantly with redshift, at least up to $z\sim7$ \citep[e.g.,][]{Behroozi2013,Harikane2018}. Under this assumption, the aforementioned empirical models predicted UVLFs that are well described by the Schechter function up to redshift as high as $z=20$. If there are indeed more luminous galaxies relative to what Schechter function describes, the models may need to be modified.  Many factors may cause the bright-end of the UVLFs to deviate from the Schechter functional form, such as reduced or different modes of AGN feedback \citep{Finkelstein2015}, the presence of substantial scatter in the star-formation rate at fixed mass in individual halos \citep{Ren2019}, starburst time-scale \citep{Lacey2011}, or change in shape of the underlying halo mass function with redshift \citep{Bowler2020}. Several studies are also identifying dust attenuation (temperature, extinction curve, dust-to-mass ratio etc.) as the primary factor influencing the precise shape of the bright-end UVLF \citep{Lacey2011,Somerville2012,Cai2014,Kimm2013, Cullen2017,Ma2018,Bowler2020}. But the dust contribution is still under debate. For example, \citet{Ma2019} used post-process dust radiative transfer calculations on high-resolution cosmological simulation and found that the dust-to-metal ratio $(>0.8)$ required to match the observed bright-end UVLF at $z>8$ is much higher than what is needed for lower redshift galaxies. Yet, such high dust-to-metal ratio would contradict the current understanding that high-redshift galaxies are relatively dust-poor \citep{Khusanova2020,Hou2019}.

Therefore, to accurately determine the underlying physical models that describe high-reshift UVLFs, one should maximize the accuracy of the observed UVLFs. Observationally, multiple factors can contribute to bias in UVLF: cosmic variance \citep{Trenti2008,Moutard2016, Bowler2020}, magnification bias \citep{Wyithe2011,Mason2015}, and contamination from objects at different redshifts \citep{Morishita2018}. If sources of bias are not accounted and corrected for, they can lead to incorrect interpretations of fundamental physical processes. 

One aspect that deserves further scrutiny in light of the imminent improvement in data quality expected from JWST is whether methods that derive the UVLFs at high redshift are robust and unbiased. Existing works in literature define completeness functions differently. Therefore, they use different methods to derive UVLFs. Concurrently, their measured UVLFs show occasional tension, albeit it is unclear whether the differences arise from the datasets used, the data reduction/source selection, and/or the completeness step. \citet{Bouwens2015} and \citet{Bouwens2021} found that the UVLF at $4<z<8$ is consistent with a Schechter functional form, $N(M)dM=\phi^* 0.4\ln{10} [10^{0.4(M^*-M)(\alpha+1)}\exp{-10^{0.4(M^*-M)}}]$, where the faint-end slope $\alpha$ and the characteristic number density $\phi^*$ significantly evolve with redshift. However, there was no significant evolution in the $M^*$ parameter prior to $z\sim3$. \citet{Finkelstein2015} found a similar conclusion regarding the evolution of the Schechter parameters with redshift but their UVLFs at $z>6$ show some mild tension with those of \citeauthor{Bouwens2015} at 1-2 $\sigma$ level. On the other hand, \citet{Bowler2015}, \citet{Calvi2016}, \citet{Rojas-Ruiz2020} and \citet{Bowler2020} found that the UVLFs at high redshifts either comparatively show an excess at the bright end (although they are all still consistent with each other within a few sigmas and therefore the differences are not conclusively significant) or are better fit with a double power-law function with little evolution at the bright end. They also detected a significant change in $M^*$ with redshift. Although these works use injection-recovery simulations that are conceptually similar to calculate effective comoving volumes in their observational surveys, they define completeness functions differently and hence use different calculations to derive the UVLFs. 

This paper aims to systematically and quantitatively assess the methods proposed to determine the UV luminosity functions using the same comparison sample, to characterize systematic differences in recovered UVLFs that are due to completeness modeling. We consider literature studies that calculate completeness functions via injection-recovery simulations, and determine whether the differences involving completeness functions can lead to different UVLFs, given the same observational situation.

This paper is organized as follows. In section 2, we describe the definitions of the completeness functions and the UVLF derivation methods that we consider in this paper. In section 3, we test these methods on a mock observation that covers a large area of the sky, using mock completeness simulations that mimic the observation perfectly. In section 4, we test the methods on $z\sim5$ galaxies in the Hubble Legacy Fields. In this section, we also introduce GLACiAR2, the publicly available completeness simulation used in this work\footnote{\url{https://github.com/nleethochawalit/GLACiAR2-master}\label{url_github}}.

\section{Details of previously defined completeness functions and methods to determine UVLF}
\label{sec:completeness_definition}

Determining a UVLF -- the number of galaxies per absolute UV magnitude per comoving volume within a redshift range of interest-- is conceptually straightforward. In basic terms, two components must be known: 1) the number of galaxies detected at the redshift of interest in an observational survey and 2) a mapping that translates the number of observed galaxies to the intrinsic number density of galaxies in the universe. A completeness simulation (or an injection-recovery simulation) is often used to calculate the latter. The simulation usually starts with the creation of galaxy stamps that are as realistic as possible (in terms of shapes, sizes, point-spread functions, and spectral energy distributions). It then randomly places these galaxy stamps onto the observed images and recovers them, using the same analysis process followed to obtain the observational sample. 
 
There are two broad approaches to apply the completeness simulation to determine the UVLFs or the best-fit functions that describe the UVLFs (such as the Schechter function). One is to iteratively simulate mock galaxies while searching for the best-fit UVLF parameters \citep[e.g.,][]{Ishigaki2015, Ishigaki2018}. In each iteration, this method samples a UVLF model, creates a set of mock galaxies accordingly, projects them on the images, and compares the recovered number counts to the number of galaxies in the observation. Although this method is straightforward, it can be computationally expensive
if the observational sample contains multiple images. Thus, most of the studies in the literature utilize a second different approach, where the completeness simulation is only performed once and is independent of the UVLF parameters derived from the observed sample. This approach stores the simulation's result in terms of completeness functions or effective volumes specific to each image in the survey, and then uses the information to recover a luminosity function from observed galaxy number counts. This paper focuses on the latter approach, i.e., on how these completeness functions are defined and subsequently used to calculate the UVLFs and their best-fit parameters. 

 \begin{figure*}
     \centering
     \includegraphics[width=0.8\textwidth]{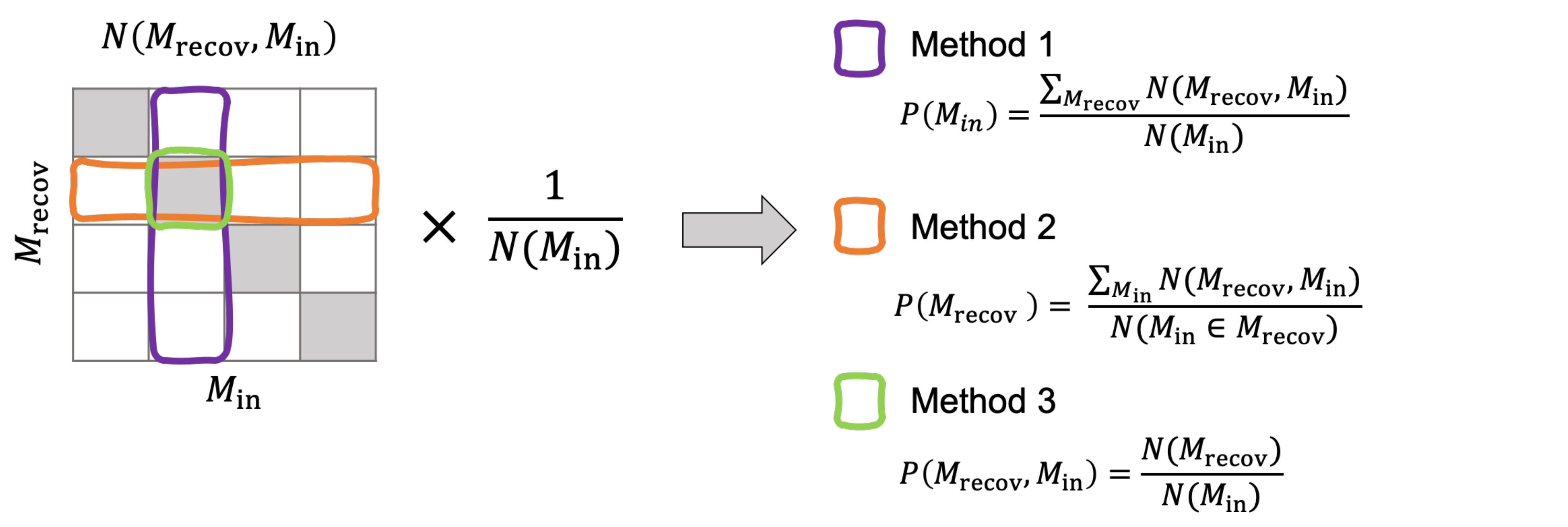}
     \caption{Schematic diagrams showing the completeness definition for the three methods considered in this paper. The matrix on the left shows the result from the injection-recovery simulation where each cell is the number of simulated galaxies with injected $M_\textrm{in}$ and recovered $M_\textrm{recov}$. $N(M_\textrm{in})$ is the total number of injected galaxies in each column $M_\textrm{in}$. The shaded grey cells show the 1-1 line, where $M_\textrm{recov} = M_\textrm{recov}$. The coloured rectangles show the direction of the summation for each definition.}
     \label{fig:scheme}
 \end{figure*}
 
Completeness functions map the number of observed galaxies to the intrinsic UVLF. The injection-recovery simulations introduce artificial galaxies with some specified properties (size, shape, surface brightness profile) into observed images and record their recovery status as well as their recovered properties. Based on these results, one can define a completeness function that maps the number of recovered galaxies with a given set of recovered properties to the number of injected galaxies with a given set of injected properties. They can be a function of multiple parameters, other than UV magnitude, such as redshift \citep[e.g.,][]{Bowler2015}, galaxy radius and UV spectral slope \citep[e.g.,][]{Finkelstein2015}. For simplicity, we only demonstrate the completeness as a function of magnitude in this work. I.e., we assume no cross-contamination between redshifts and that the detection is independent of galaxy size and UV spectral slope. With these assumptions, a 2-dimension matrix is sufficient to fully represent the results of the injection-recovery simulation (shown on the left side of Figure \ref{fig:scheme}). Its columns refer to the input magnitude, while its rows represent the output magnitude of the simulation. The value of each cell is the number of galaxies whose intrinsic luminosity matches the column's magnitude but are recovered to have the row's output magnitude. 

The differences among previous works come in under three main aspects. First is the type of input and recovered magnitude. Most works inject galaxies in bins of intrinsic UV absolute magnitude, while some inject galaxies in bins of theoretical apparent magnitudes -- the magnitudes of the artificial galaxies after applying the distance modulus and the K correction calculated with the input redshift and at the detection band \citep[e.g.,][]{Oesch2007,Oesch2018,Carrasco2018}. For the types of output magnitude, \citet{Oesch2007} and \citet{Bouwens2015} use apparent magnitudes in a detection band, while \citet{Bowler2014,Bowler2015,Bowler2020} and \citet{Finkelstein2015} use UV absolute magnitudes measured from the recovered spectral energy distributions (SEDs) and inferred redshifts. Second is the distribution function of the injected galaxies. \citet{Oesch2007} and \citet{Bouwens2015} inject artificial sources drawn from an underlying flat magnitude distribution. \citet{Finkelstein2015} and \citet{Rojas-Ruiz2020} inject galaxies with a faint-heavy distribution. \citet{Bowler2015} inject galaxies that either follow a Schechter distribution or a double-power law distribution. The third and final source of difference lies in the specific definition used for the completeness function itself, which directly affects the method used to calculate the best-fit UVLF parameters. In principle, the first difference (defining completeness functions as a function of apparent magnitude or as a function of UV absolute magnitude) is inconsequential because a given completeness function can be translated into another by combining it with a one-to-one function that maps between the two magnitudes. Therefore, in this work we will investigate only impact of the underlying distribution in the completeness simulations and of the definition of the completeness function used to derive UVLFs.

We first describe the three types of definitions and methods considered in this work in Section \ref{sec:finkelstein_method} to \ref{sec:bouwens_method}. A schematic summary is given in Figure \ref{fig:scheme}.

\subsection{Method 1: completeness as a function of input magnitude}\label{sec:finkelstein_method}
This method defines completeness as a function of input magnitude. For example, e.g. \citet{Finkelstein2015,Rojas-Ruiz2020} define the completeness $P(M_\textrm{in})$ as the fraction of recovered galaxies, regardless of their output magnitudes, in an input magnitude $M_\textrm{in}$ bin:
\begin{equation}
    P(M_\textrm{in}) = \frac{\sum_{M_\textrm{recov}} N(M_\textrm{recov}, M_\textrm{in})}{N(M_\textrm{in})}
\end{equation}
where $M_\textrm{in}$ is the intrinsic absolute UV magnitude, and $M_\textrm{recov}$ is the recovered magnitude of the simulated galaxies. As shown in Figure \ref{fig:scheme}, each cell in the matrix represents the number of galaxies with injected $M_\textrm{in}$ magnitude that are recovered to have $M_\textrm{recov}$ magnitude. The completeness defined by method 1 is the summation along the column (highlighted in purple) divided by the total number of injected galaxies in the injected magnitude $M_\textrm{in}$. We note that the completeness by \citet{Finkelstein2015} is also a function of two additional parameters: half-light radius and UV spectral slope, but these are not considered by \citet{Rojas-Ruiz2020}. 

With this completeness definition, the effective comoving volume in an observed image is also a function of $M_\textrm{in}$:
\begin{equation}
    V_\textrm{eff}(M_\textrm{in}) = \int \frac{dV}{dz}P(M_\textrm{in},z) dz,
\end{equation}
where $V$ is the comoving volume associated to the observed image. If we let a set of parameters $\theta$ describe a luminosity function, $\phi(\theta)$, the expected number of galaxies with an intrinsic $M_\textrm{in}$ is equal to 
\begin{equation}
    N^\textrm{exp}(M_\textrm{in},\theta) = V_\textrm{eff}(M_\textrm{in}) \phi(\theta,M_\textrm{in})dM.
    \label{eq:nexp_method1}
\end{equation}
Ultimately, one can use the maximum likelihood estimation to determine the best-fit parameters ($\theta$) by comparing the modelled number of galaxies $N^\textrm{exp}$ to the number of observed galaxies $N^\textrm{obs}(M_\textrm{in})$ in the image. 
One aspect of this method is that the number of observed galaxies with intrinsic $M_\textrm{in}$ magnitudes ($N^\textrm{obs}(M_\textrm{in})$) may be challenging to quantify. \citet{Finkelstein2015} and \citet{Rojas-Ruiz2020} found that the recovered fluxes of the galaxies in the injection-recovery simulation were typically fainter than the input fluxes. Therefore, they apply the typical offset found in the simulations to the fluxes measured for the observed galaxies to derive the best estimate of their intrinsic fluxes:
\begin{equation}
    N^\textrm{obs}(M_\textrm{in}) = N^\textrm{obs}(M_\textrm{recov}+\textrm{offset} \in M_\textrm{in}).
\label{eq:offset_method1}
\end{equation}
\subsection{Method 2: completeness as a function of recovered magnitude}\label{sec:bowler_method}
This method is arguably the more commonly used approach to determine high-redshift UVLFs \citep[e.g.,][]{Oesch2007, Oesch2009, Oesch2012, McLure2013,  Bowler2014, Bowler2015, Morishita2018, Bridge2019, Bowler2020}. It defines completeness as a function of recovered magnitudes, either in terms of recovered absolute UV magnitude \citep[e.g.,][]{Bowler2014,Bowler2015,Bowler2020} or in terms of recovered apparent magnitude \citep[e.g.,][]{McLure2009,McLure2013,Oesch2007,Oesch2009,Oesch2012}. It is the ratio of the number of recovered galaxies with recovered magnitudes in the bin $M_\textrm{recov}$ to the total number of injected galaxies with intrinsic magnitudes in the same bin:
\begin{equation}
    P(M_\textrm{recov}) = \frac{\sum_{M_\textrm{in,i}}N(M_\textrm{recov},M_\textrm{in,i})}{N(M_\textrm{in}\in M_\textrm{recov})}.
\end{equation}
The numerator sums all galaxies with the same recovered magnitude regardless of their intrinsic magnitudes. As represented in Figure \ref{fig:scheme}, this is the summation along the row of $M_\textrm{recov}$ divided by $N(M_\textrm{in})$. We note that depending on the number density distribution of the injected galaxies, the completeness function thus defined can exceed unity. For example, if simulated galaxies are injected following a steep underlying luminosity distribution function (such as the bright-end of the Schechter function), then the completeness can exceed unity due to an up-scattering of more numerous faint galaxies \citep{Bowler2020}. In contrast, if a simulation injects galaxies with an underlying flat distribution \citep[e.g.,][]{Oesch2007}, the completeness is  smaller than 100\% for realistic functional forms of photometric scatter (e.g. normal). We note that \citet{McLure2009,McLure2013,Bowler2014,Bowler2015} and \citet{Bowler2020} also allow their completeness to be a function of redshift i.e. the completeness is the ratio of the output to the input $M_\textrm{UV}$-$z$ grids. However, for a simple demonstration purpose here, we drop the redshift component, i.e. there is no cross-contamination between redshifts. 

Based on this definition of completeness, previous studies use two procedures to infer binned UV luminosity functions (also see \citet{Schmidt2014} for an unbinned method). First is via the calculation of the maximum volume occupied by each galaxy, i.e. the $V_\textrm{max}$ method \citep[e.g.,][]{Bowler2014, Bowler2015, Bowler2020}. Each galaxy $i$ occupies a maximum volume $V_\textrm{max,i}$, a shell with the largest distance at which a galaxy with absolute magnitude $M_\textrm{in}$ can be selected into the sample. It is usually the distance at which its signal-to-noise ratio would drop below a selected signal-to-noise threshold. The luminosity function is then defined as the sum of these maximum volume densities, corrected by the completeness:
\begin{equation}
     \phi(M_\textrm{in})dM_\textrm{in}  =  \sum\limits_{i=1}^N \frac{C(M_{\textrm{recov},i})}{V_{\textrm{max,i}}},
\label{eq:phinominal_method2}
\end{equation}
where $M_\textrm{in}=M_\textrm{recov}$ and $C(M_\textrm{recov}) =  1/P(M_\textrm{recov})$. The summation is over all galaxies with recovered absolute magnitude in bin $M_\textrm{recov}$.

Alternatively, the UVLFs can be inferred via the calculation of an effective comoving volume \citep{Oesch2007, Calvi2016, Morishita2018}:
\begin{equation}
    V_\textrm{eff}(M_\textrm{recov}) = \int\frac{dV}{dz}P(M_\textrm{recov}) dz.
\end{equation}\footnote{These works generally separate their completeness functions into the ``completeness" term ($C(m_\textrm{obs})$) and the selection function term ($S(m_\textrm{obs},z)$). Here we simply call the combination as completeness function ($P(M_\textrm{recov}$).}
For a given modeled luminosity function $\phi(\theta)$, the expected number of galaxies with recovered magnitude $M_\textrm{recov}$ is
\begin{equation}
    N^\textrm{exp}(M_\textrm{recov},\theta) =  V_\textrm{eff}(M_\textrm{recov})\phi(\theta,M_\textrm{in}=M_\textrm{recov})dM.
    \label{eq:nexp_method2}
\end{equation}
One can then compare $N^\textrm{exp}(M_\textrm{recov},\theta)$ to the number of observed galaxies in the survey $N^\textrm{obs}(M_\textrm{recov})$ to find the best-fit UVLF. 

\subsection{Method 3: completeness as a function of input and recovered magnitude}\label{sec:bouwens_method}
This method relies on the full matrix produced by the injection-recovery simulation in Figure \ref{fig:scheme} (see Table 19 in \citet{Bouwens2006} for an example). The completeness as well as the effective volume are a function of both injected magnitude $M_\textrm{in}$ and recovered magnitude $M_\textrm{recov}$:
\begin{equation}
     P(M_\textrm{recov},M_\textrm{in})  = \frac{N(M_\textrm{recov},M_\textrm{in})}{N(M_\textrm{in})}; 
\end{equation}
\begin{equation}
       V_\textrm{eff}(M_\textrm{recov},M_\textrm{in})  = \int \frac{dV}{dz}P(M_\textrm{recov},M_\textrm{in}) dz, 
\end{equation}
where $M_\textrm{in}$ is typically an intrinsic UV magnitude, while $M_\textrm{recov}$ is typically an observed apparent magnitude \citep{Bouwens2011,Bouwens2015, Bouwens2019, Bouwens2021}. Here, we use the recovered $M_\textrm{UV}$ magnitude instead, so that we can present a consistent comparison with the other two methods. For a given model $\phi(M_\textrm{in},\theta)$, one can calculate the expected number of observed galaxies in a bin centered on $M_\textrm{recov}$ through a dot product between the effective comoving volume and $\phi(M_\textrm{in},\theta)$:
\begin{equation}
     N^\textrm{exp}(M_\textrm{recov},\theta) =  V_\textrm{eff}(M_\textrm{recov},M_\textrm{in}) \cdot \phi(M_\textrm{in},\theta)dM.
     \label{eq:nexp_method3}
\end{equation}
Then, one can directly use a maximum likelihood estimation to find the best-fit parameters by comparing this modeled $ N^\textrm{exp}(M_\textrm{recov},\theta) $ to the observed number of galaxies $N^\textrm{recov}(M_\textrm{obs})$.

 \section{Mock observation assessment of completeness corrections to UVLF determination}
 \label{sec:section3}
In this section, we investigate whether the difference in the definition of completeness, and the underlying distribution of the injected galaxies in the source recovery simulation, as presented in Section~\ref{sec:completeness_definition}, lead to any systematic bias across methods. For this purpose, we use a set of mock observations where we vary the amount of photometric (flux) scatter, hence this section is based on mock data only. 

\subsection{Mock observation setup}
\label{sec:mock_setup}

\begin{figure}
    \centering
    \includegraphics[width=\textwidth]{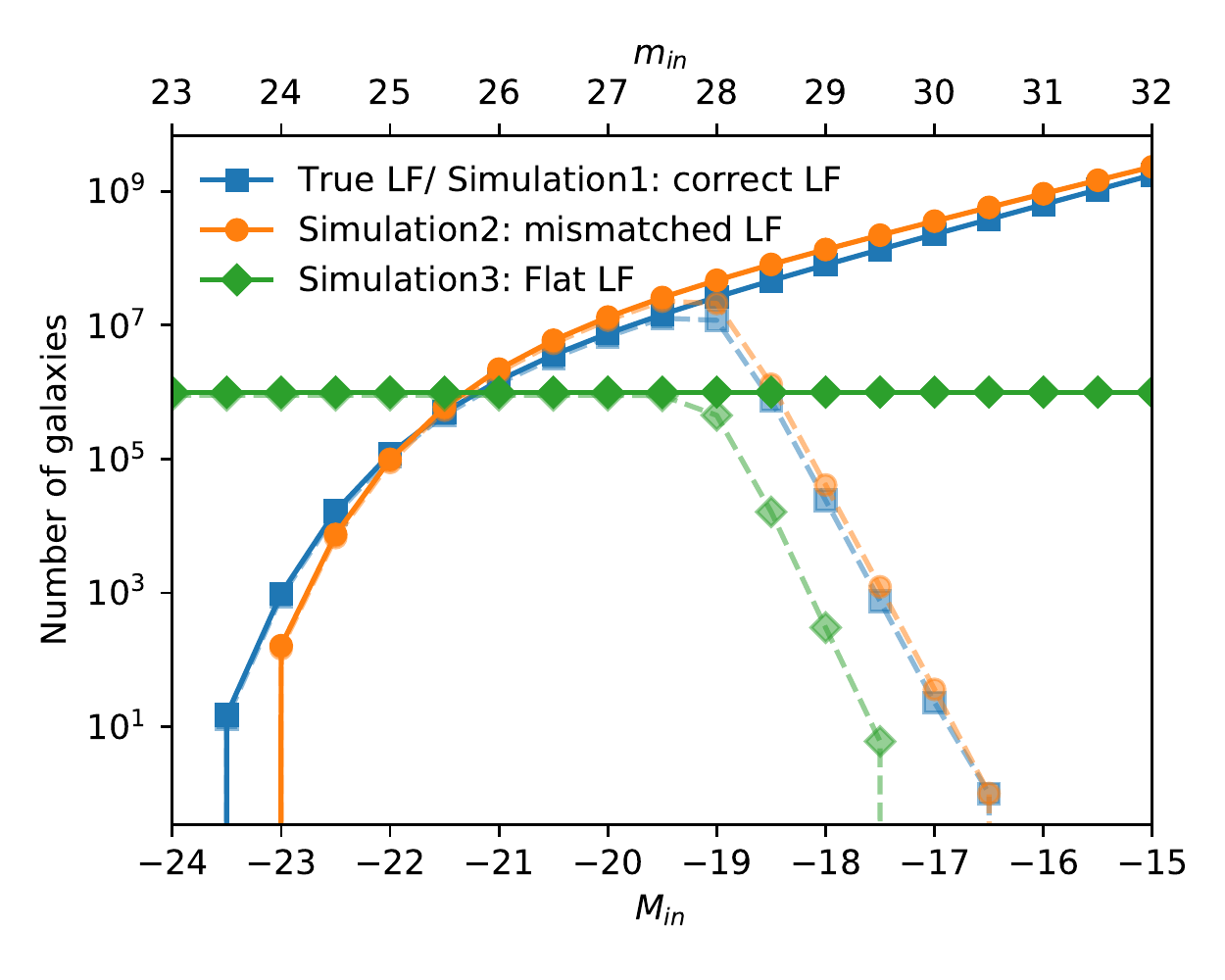}
    \caption{Luminosity function in mock data. Dark blue squares with solid blue lines show the true numbers of galaxies in the mock universe, while the light blue squares with dashed lines show the numbers of galaxies observed in the mock survey. The dark and light blue data squares also represent the number of injected and recovered galaxies in the completeness simulation that simulates galaxies with an input distribution equal to the correct one for the mock universe. Dark orange circles (dark green diamonds) show the number of the injected galaxies in the simulation whose injected sources were populated with a distribution (flat) which substantially differs from the correct one, while light orange circles (light green diamonds) show the number of the recovered galaxies in the simulation.}
    \label{fig:n_uni}
\end{figure}

We begin by assuming a mock universe and an observational survey for $z=9$ galaxies. The mock universe contains $z=9$ galaxies whose intrinsic luminosity function follows the Schechter luminosity function from \citet{Morishita2018}: $\alpha=-2.1,\ M^*_\textrm{UV}=-21.0$, and $\log\phi^* = -4.2$. We set the combination of distance modulus to $z=9$ and the k-correction factor to a round-number of $m=47$ mag (AB). We populate 0.5-magnitude-wide bins with mock sources such that the galaxies' intrinsic magnitudes are exactly equal to the bin center values.  

We assume that the mock observation is a half-sky survey with a finite exposure time. The comoving volume of the survey is $1.4\times10^{11}$ Mpc$^3$. We ignore contamination by galaxies at different redshifts. We further set the completeness of the survey to be a Sigmoid (S-shaped) function of theoretical apparent magnitude, 
\begin{equation}
C(m_\textrm{in}) = \frac{ 0.9}{1+\exp[8\times(m-28)]}.
\label{eq:sigmoid}
\end{equation} 
The completeness is $\sim90\%$ at the brightest magnitude bins, since in typical telescope images (e.g. from HST) foreground sources or bad pixels block $\sim10\%$ of the bright $z\sim 9$ galaxies that are well above the survey detection limit. We set the completeness to quickly drop off at $m\sim28$ AB, which is approximately equal to the limiting magnitude of a typical deep HST survey such as the Cosmic Assembly Near-infrared Deep Extragalactic Legacy Survey \citep[][]{Koekemoer2011}. With this configuration, there are 15 galaxies with $m_\textrm{in} = 23.5$ (or $M=-23.5$) and $\sim 2.3\times10^8$ galaxies with $m_\textrm{in} = 30$ (or $M=-17$), in the patrol area. However, after applying the completeness function (equation \ref{eq:sigmoid}), the numbers of observed galaxies are 13 at the $m_\textrm{in} = 23.5$ bin and zero at $m_\textrm{in} > 30.5$ bins. The true numbers of galaxies in the patrol area are shown as dark-blue squares in Figure \ref{fig:n_uni}, while the numbers of observed galaxies are shown in light-blue squares.

\begin{figure*}
    \centering
    \includegraphics[width=0.245\textwidth]{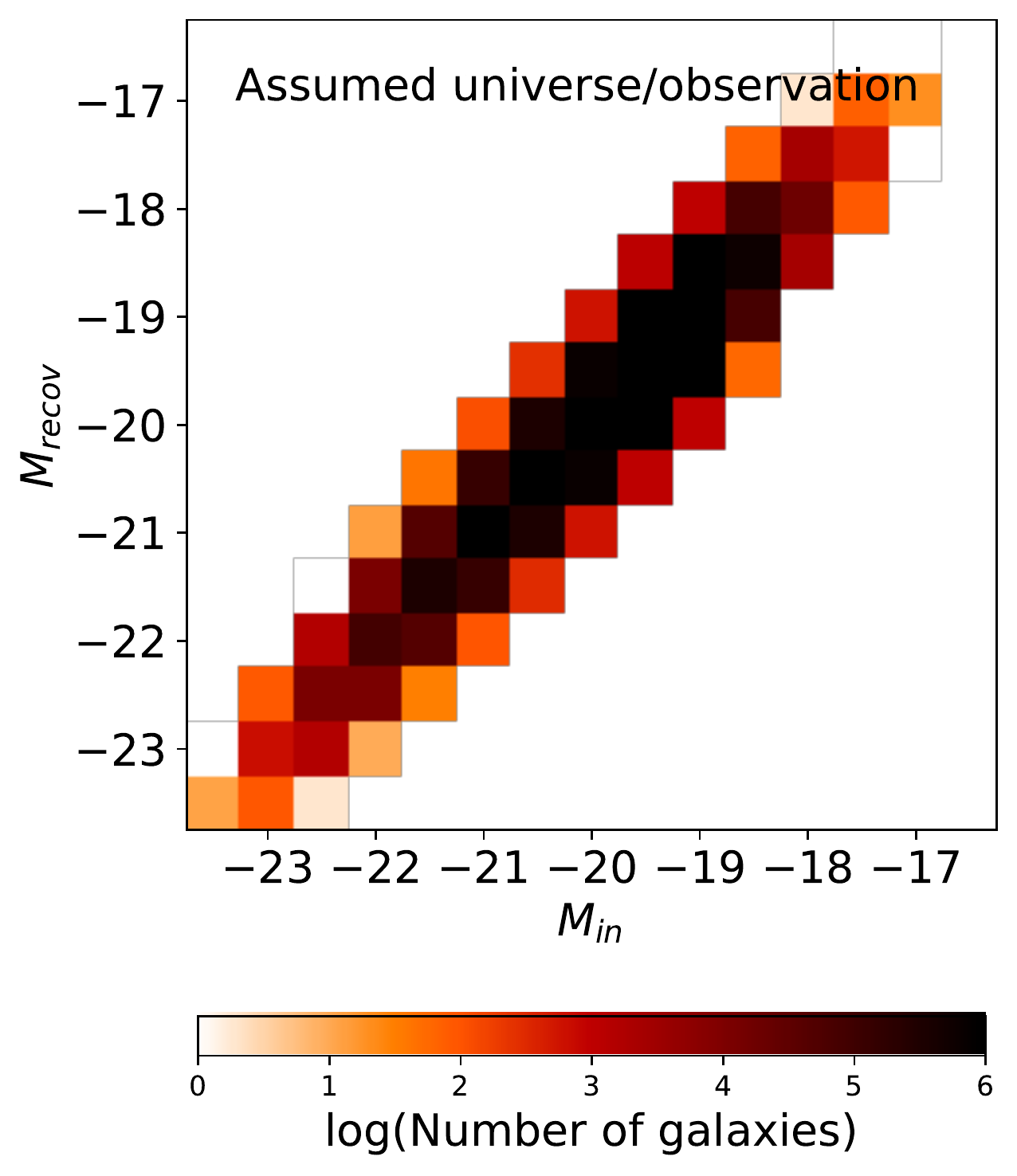}
    \includegraphics[width=0.245\textwidth]{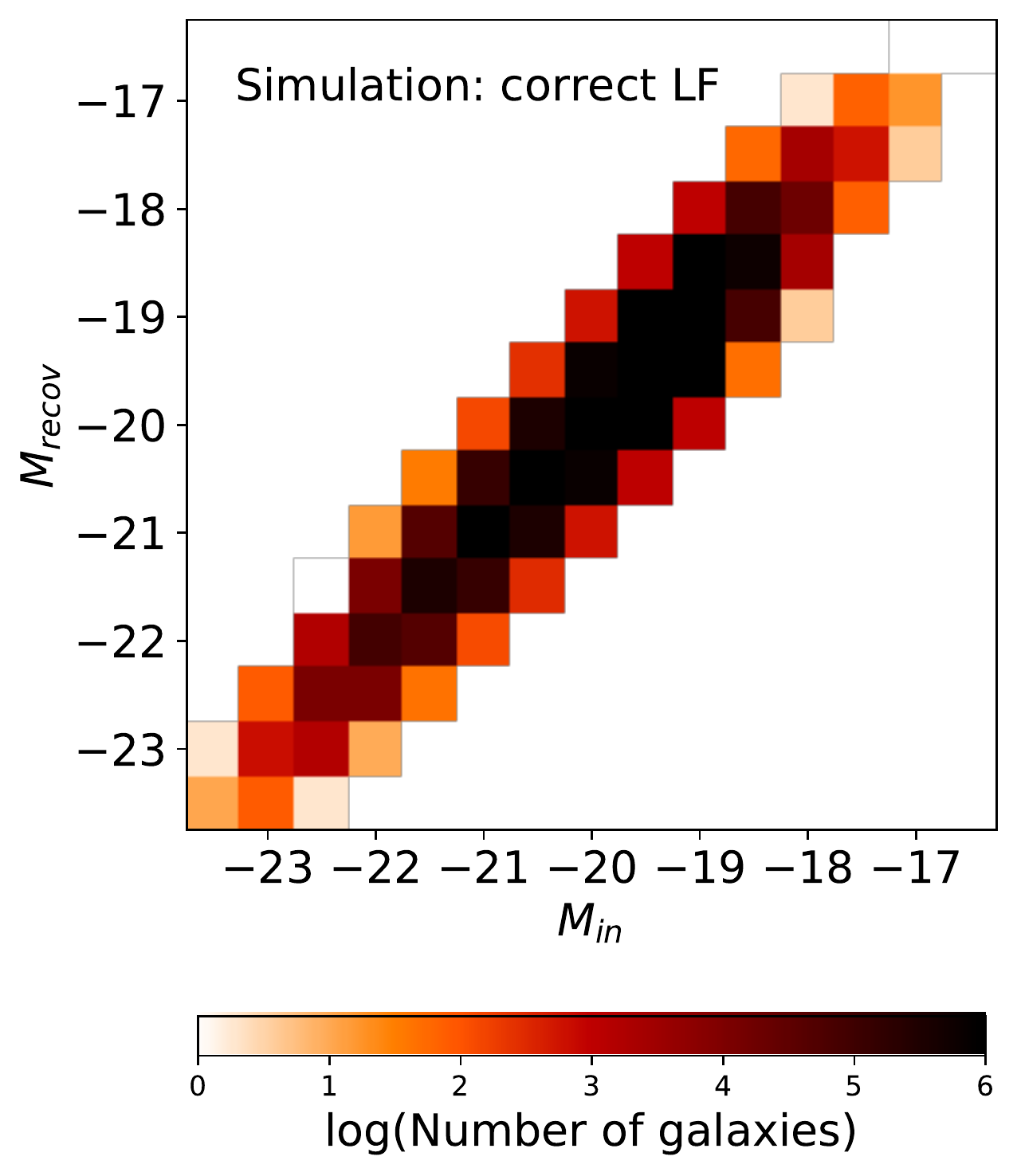}
    \includegraphics[width=0.245\textwidth]{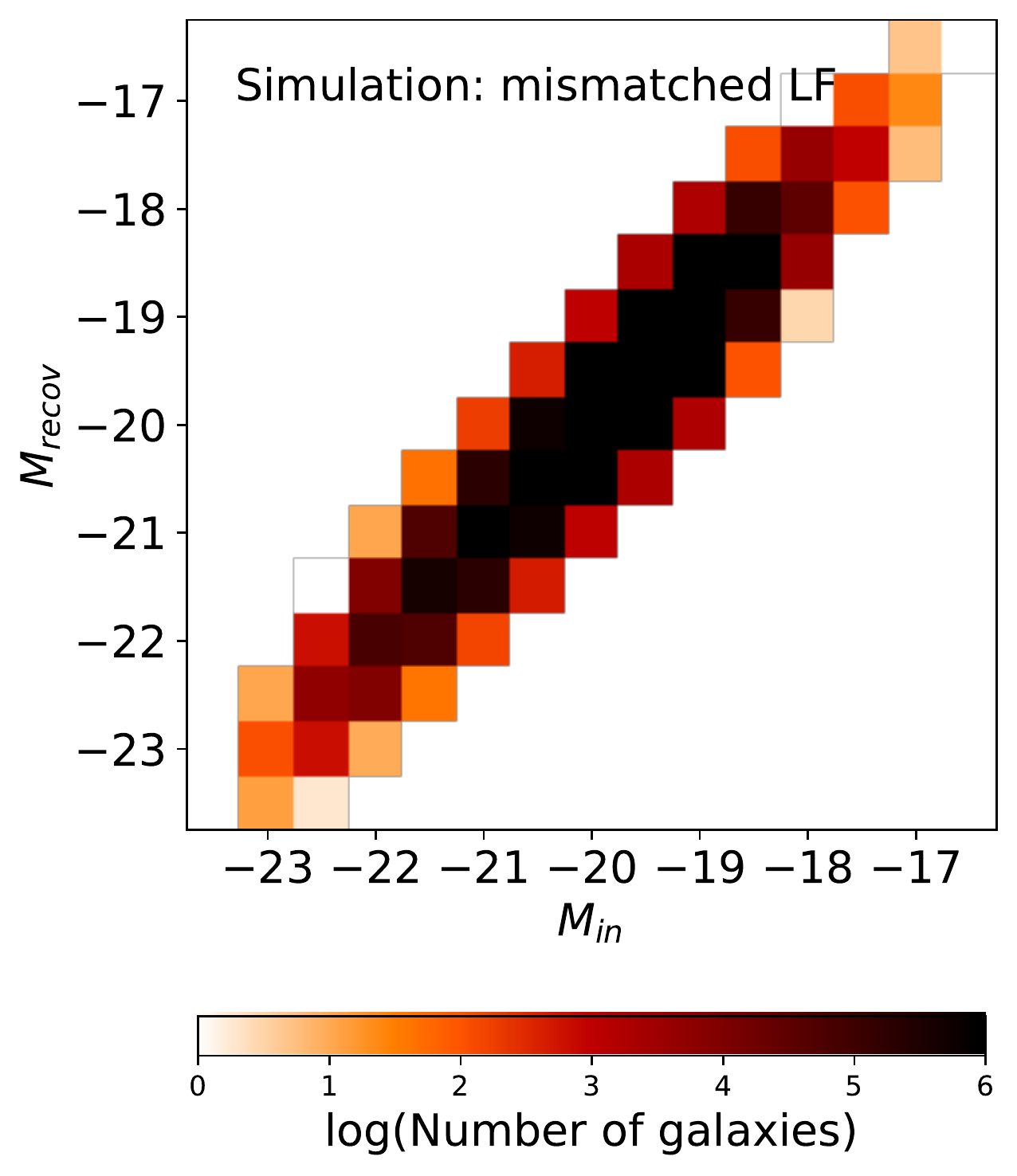}
    \includegraphics[width=0.245\textwidth]{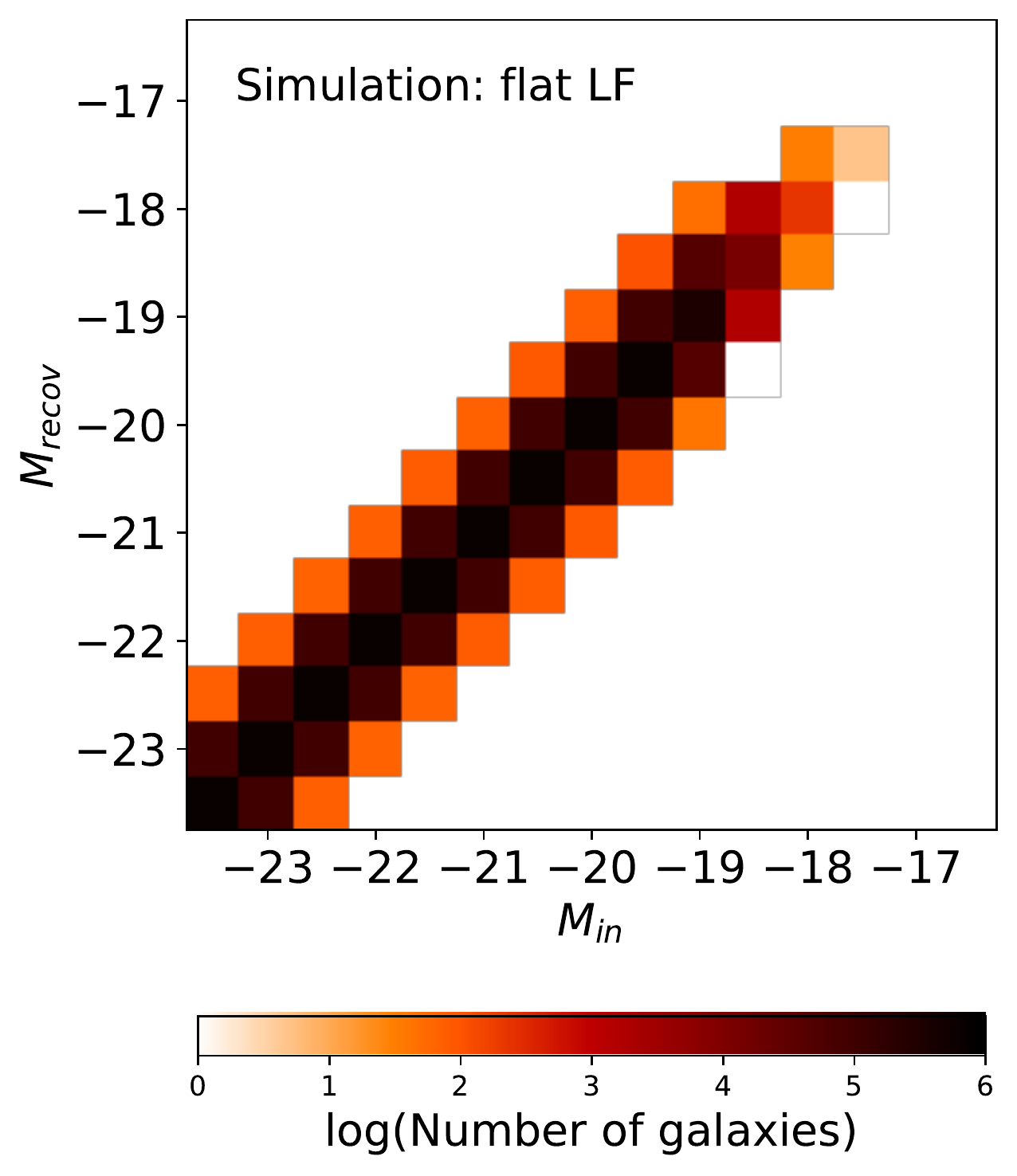}
    \caption{Number of observed galaxies as a function of intrinsic magnitude ($M_\textrm{in}$) and recovered magnitudes ($M_\textrm{recov}$) when there is a $\sigma=0.2$ magnitude scatter in $M_\textrm{recov}$. The left panel shows the actual mock universe and the correct simulation when the correct LF is used. The middle and right panels show the results of the simulations when a mis-matched Schechter luminosity function and a flat luminosity function is used, respectively.}
    \label{fig:situation2}
\end{figure*}

We further assume that the recovered fluxes do not perfectly match the input one due to some measurement errors. In real observation, this might arise from a combination of image processing, background noise, finite aperture effects, and/or uncertainties from SED fitting that determine $M_\textrm{UV}$. We set the flux scatter to be Gaussian with standard deviations ranging from 0 to 0.5 mag. For each data set, the standard deviation is constant cross the magnitude range for simplicity. There is no flux bias between the average recovered magnitudes and the intrinsic magnitudes. When there is no flux scatter ($\sigma=0$), the recovered magnitudes are therefore equal to the intrinsic magnitudes. An example for $\sigma =0.2$ mag is shown in the left-most panel of Figure \ref{fig:situation2}. A choice of flux scatter of $\sim =0.2$ mag is on a par with, if not conservative for, typical magnitude uncertainties of $z>8$ galaxies observed with the Hubble Space Telescope \citep[e.g.,][]{ Morishita2018,Stefanon2019, Rojas-Ruiz2020}. 

Lastly, for each observational situation (or for each $\sigma$ flux scatter), we create three mock injection-recovery simulations that inject galaxies with different underlying distributions (Figure \ref{fig:n_uni}). The first simulation injects galaxies with the same luminosity function as that of the mock universe. The result from this simulation, shown in the second panel of Figure \ref{fig:situation2}, is therefore statistically the same as the real observations. The second simulation injects galaxies with an underlying Schechter function that is slightly different from that of the mock universe.  This `mismatched'  distribution is described by the parameters for $z=8$ galaxies from \citet{Bouwens2015}: $\alpha=-2.0,\ M^*_\textrm{UV}=-20.6$, and $\log\phi^* = -3.7$. For the last simulation, we inject galaxies with an underlying flat distribution, where $10^6$ galaxies are injected to each magnitude bin. In all cases, we use the same $C(M_\textrm{in})$ as in the mock universe. We show examples of these three simulations, when the amount of flux scatter is $\sigma = 0.2$mag, in the three right panels of Figure \ref{fig:situation2}.

\subsection{Results of the mock observations}
\label{sec:mock_result}
For each amount of flux scatter, we apply different definitions of completeness functions to the `simulation' results, e.g. the matrices in Figure \ref{fig:situation2}, and recover the luminosity functions accordingly. If possible, we calculate nominal luminosity functions directly and avoid fitting a functional form. 

\smallskip

\begin{figure}
    \centering
    \includegraphics[width=\textwidth]{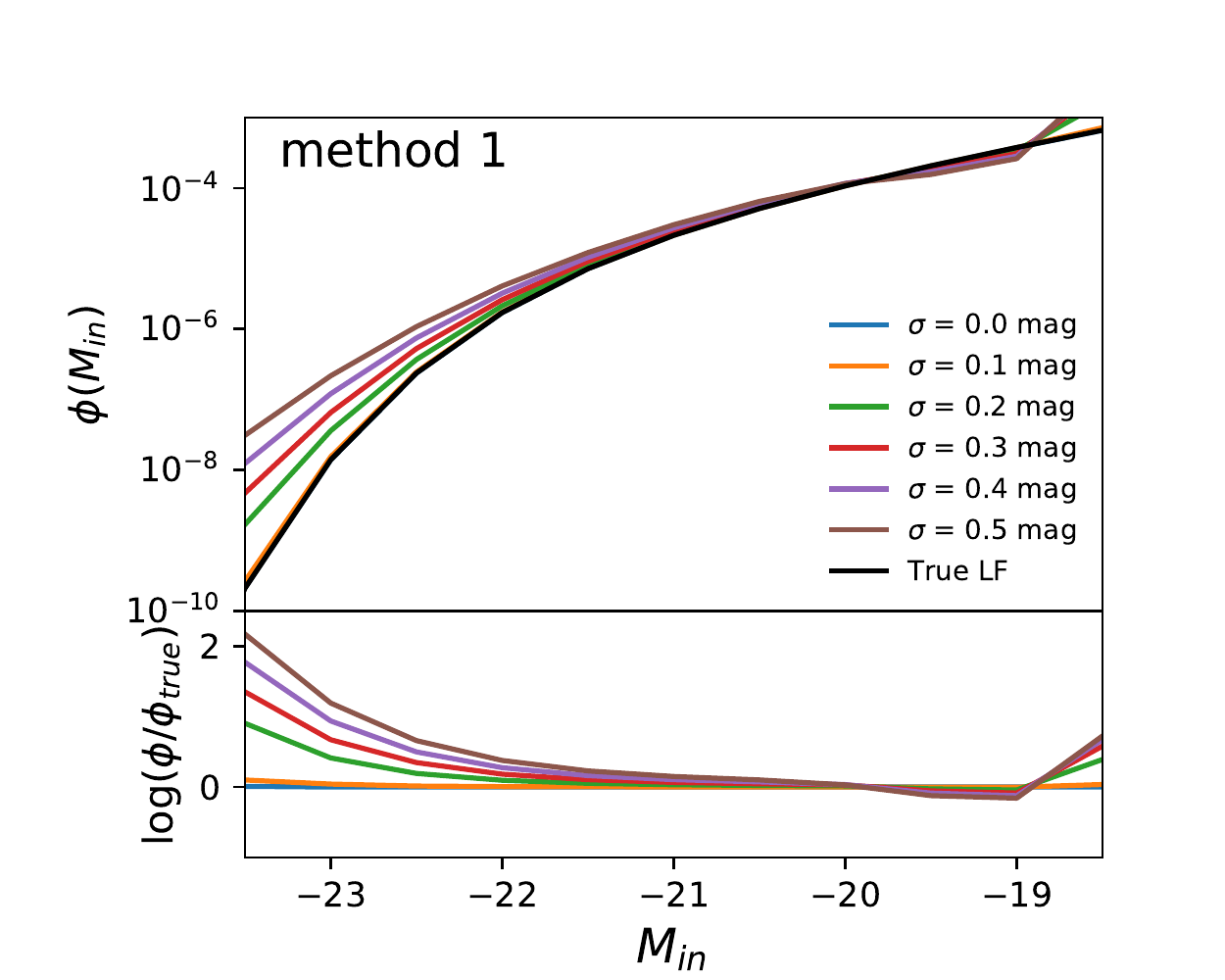}
    \caption{The recovered luminosity functions derived by method 1, in which the completeness function is defined as a function of $M_\text{in}$. When flux scatter is present, the derived luminosity function at the bright end biases high due to the more numerous fainter objects. A similar bias at the faint end is due to the detection limit that causes a sharp drop in the number of fainter sources. The bottom panels show the offset from the intrinsic luminosity function in log scale.}
    \label{fig:nominal_phi_method1}
\end{figure}

\textbf{Method1:} For this method, the nominal luminosity function is the inverse of Equation \ref{eq:nexp_method1} when the observed number of galaxies is equal to the expected number:
\begin{equation}
    \phi(M_\textrm{in}) = N^\textrm{obs}(M_\textrm{in})/V_\textrm{eff}(M_\textrm{in})/dM.
\end{equation}
$N^\textrm{obs}(M_\textrm{in})$ is calculated from Equation \ref{eq:offset_method1}, setting the magnitude offset to zero. Results are shown in Figure~\ref{fig:nominal_phi_method1}. 

We find that the recovered UVLFs do not depend on the underlying distribution used in the injection-recovery simulations. This is because the completeness definition is a function of input magnitude only. The completeness is the number of recovered galaxies in each input magnitude bin normalized by the number of injected galaxies in that bin. Therefore, it does not matter how many galaxies were injected in the input magnitude bin. Instead, the derived UVLFs depend on the amount of flux scatter, especially at the bright end where the luminosity function is steep and at the faint end where the completeness sharply drops off. In our test cases, the number density at both ends are overestimated. The larger the scatter, the larger the overestimation. At the fiducial amount of flux scatter $\sigma=0.2$ mag, the overestimation at the brightest magnitude bins $M_\textrm{in}=-23.0, -22.5, -22.0$ mag are $0.4, 0.2, 0.1$ dex, respectively.  When the flux scatter is $\sigma=0.3$ mag, the overestimation at $M_\textrm{in}=-22$ mag is 0.2 dex. These amounts of the overestimation are comparable to the bias induced by gravitational lensing. For example, \citet{Mason2015} found that the magnification bias can induce an $0.15$ mag overestimation at $M_\textrm{UV}\sim-22$ mag. We note that the typical photometric scatter in the measured $M_\textrm{UV}$ of $z\gtrsim8$ galaxies from the HST/WFC3 observations are in the range of 0.1-0.3 mag \citep[e.g.,][]{Calvi2016,Morishita2018,Bouwens2019}.

The observed overestimate is the well-known Eddington bias \citep{Eddington1913}. Flux scattering causes galaxies in a magnitude bin with a larger number of galaxies to fill up its adjacent magnitude bins with fewer number of sources. In general, method 1 may not fully correct this. As seen in Figure \ref{fig:nominal_phi_method1}, the bias is larger when the luminosity function is steep, which is the case for the Schechter luminosity function at the brightest magnitude bins. The overestimation at the faint end is caused by the steep drop in the completeness that we used in the mock analysis (see the light-colour data points in Figure \ref{fig:n_uni}). At $M\sim-19$, the completeness is set at roughly $50\%$. In real observation, this problem is generally not serious as long as faint magnitude bins that are highly incomplete are discarded, although Eddington bias should be considered carefully in situations where completeness is very low such as faint-end slope determination from gravitational lenses \citep[][]{Livermore2017,Atek2018}. 

The overestimation at the bright magnitude end may explain some of the excess found in the literature. \citet{Rojas-Ruiz2020}, which used method 1 to calculate effective volumes, found that their nominal UV luminosity function at $M_\textrm{in}\approx-22$ bin is larger than the Schechter form in \citet{Finkelstein2016} by $\sim0.5$ dex at $z\sim8$ and $\sim1$ dex at $z\sim9-10$. As discussed earlier, the amount of the possible overestimation depends on the slope of the intrinsic luminosity function at the bright ends, and the intrinsic flux scatter ($\sigma$). Both effects are difficult to quantify precisely and beyond the scope of this paper. Nonetheless, if we assume that intrinsic observational flux scatter is 0.3 dex, the overestimation found in Figure \ref{fig:nominal_phi_method1} is 0.2--0.3 dex at $M_\textrm{in}\sim-22$ bins. This amount cannot account for all the excess found in \citet{Rojas-Ruiz2020} but is still a significant fraction. The bias can be significantly higher if the intrinsic flux scatter is larger. 

We note that \citet{Finkelstein2015} and \citet{Rojas-Ruiz2020} took flux uncertainties into account via a Monte Carlo sampling. Fluxes of the observed galaxies were sampled based on the observational flux uncertainties and the magnification uncertainties. The fluxes were  also corrected with the average offset between the input and output fluxes found in their completeness simulations. However, the Monte Carlo sampling and the average offset may not totally correct the bias if the input distribution in the completeness simulation is different from the intrinsic distribution. In other words, if we let $F$ be the intrinsic flux of a galaxy and $F_o$ be the observed/recovered flux of the galaxy, according to the Bayesian rule, the probability of the intrinsic flux $F$ is $P(F|F_o) \propto P(F_o|F)P(F)$. The average offset and the Monte Carlo sampling only take care of the $ P(F_o|F)$ part but leave the $P(F)$ term out.\footnote{\citet{Stefanon2017a} use this Bayesian correction to correct their rest-frame optical fluxes of high-z galaxies.}

Although we can apply a correction function to the recovered fluxes to correct for the Eddington bias \citep[e.g.,][] {Hogg1998}, we technically already know how the recovered fluxes behave relative to the intrinsic fluxes based on the injection-recovery simulation. If we trust that the injection-recovery simulation can replicate the observation, we should be able to recover the intrinsic fluxes statistically. This means that \textit{the completeness needs to be a function of both intrinsic and recovered magnitudes}. In fact, method 2 and method 3 already take this into account.

\begin{figure}
    \centering
    \includegraphics[width=\textwidth]{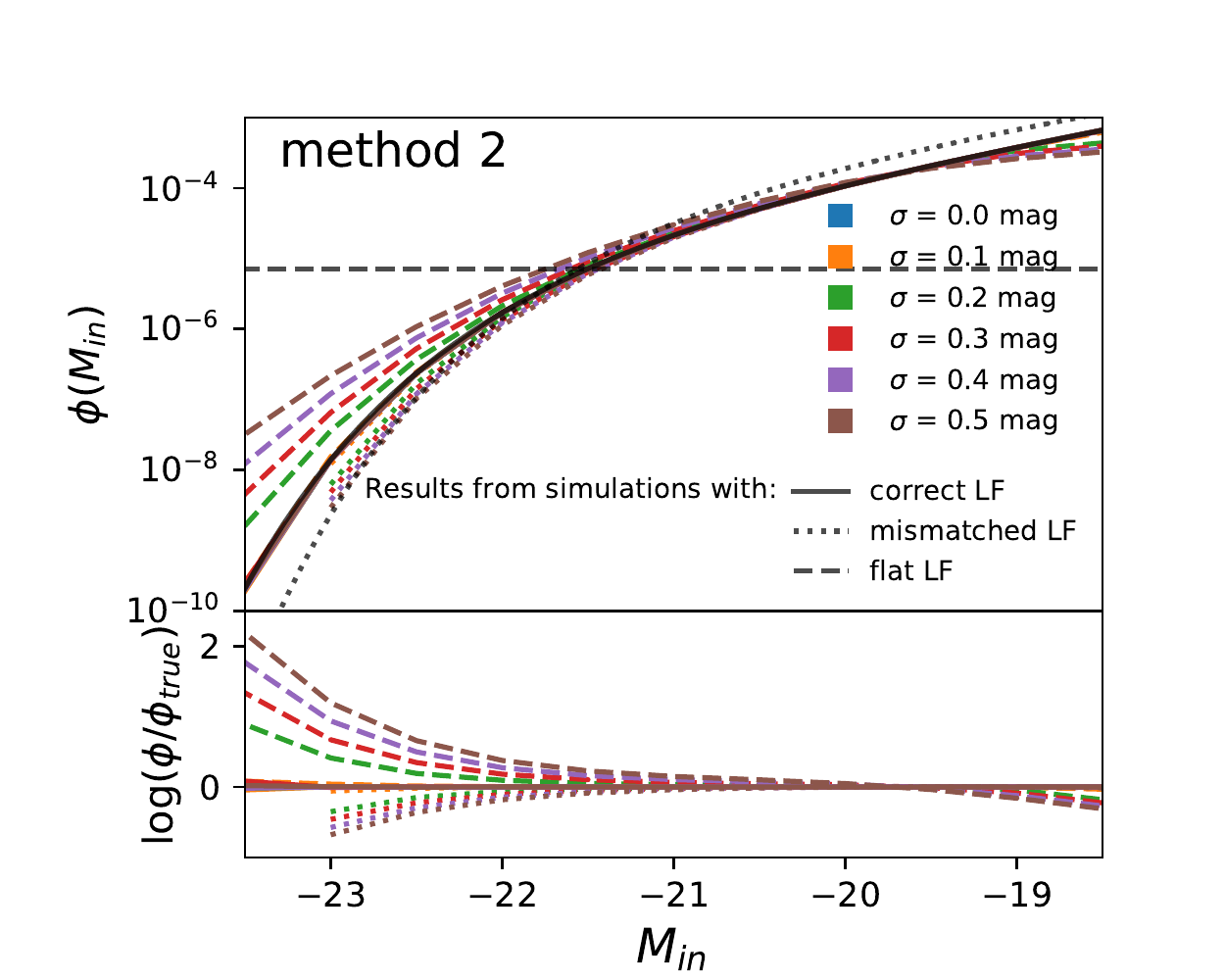}
    \caption{Similar to Figure \ref{fig:nominal_phi_method1} but shows the recovered UVLFs derived by method 2, in which the completeness function is defined as a function of $M_\text{recov}$. This method can recover the intrinsic luminosity function if the underlying distribution used in the completeness simulation is the same as the intrinsic luminosity function. Otherwise, the recovered luminosity functions will be biased toward the underlying distribution used in the simulation when the flux scatter is present. The gray dotted and dash lines show the flat and the mismatched distributions used in the simulations, respectively.}
    \label{fig:nominal_phi_method2}
\end{figure}

\smallskip

\textbf{Method 2:} For this method, we directly use Equation \ref{eq:phinominal_method2} to recover the luminosity functions. The results are shown in Figure \ref{fig:nominal_phi_method2}. In the presence of flux scatter, this method can only recover the intrinsic luminosity function if the underlying distribution used in the completeness simulation is the same as the intrinsic luminosity function.  However, when the underlying distribution function used in the simulation is different from the intrinsic function, the derived UVLFs differ from the intrinsic UVLF. The bias is in the direction toward the underlying distribution function. As seen in the figure, the coloured dashed lines show the recovered UVLFs using the completeness functions derived from the simulations with an underlying flat distribution. When the flux scatter $\sigma$ is greater than zero, both the faint end and the bright end of the LF show a bias toward the flat distribution (grey dashed line).  The derived UVLFs based on the simulations with mismatched LF distribution show a similar behavior. The derived UVLFs (the coloured dotted lines) always have a bias toward the underlying distribution used in the injection simulations (the grey dotted line). The amount of the offset depends on the amount of the flux scatter and on the difference between the shapes of the intrinsic and underlying luminosity function for the magnitude bin considered. The recovered UVLFs in our test cases are resilient to flux scatter up to $\sigma=0.1$ (the differences are less than 0.1 dex at all magnitude bins). Based on Figure \ref{fig:nominal_phi_method2}, when the flux scatter of $\sigma=0.2$ mag, the offset is $\sim-0.3$ dex at magnitude bin $M_\textrm{in}=-23$ for the simulation with an underlying mismatched distribution (green dotted line) and $\sim+0.4$ dex for the simulation with an underlying flat distribution (green dashed line). 

Method 2 relates the number of observed galaxies with recovered magnitudes in bin $M$ to the number of galaxies with intrinsic magnitudes in bin $M$ based on the number of pairs found in the simulation. In the presence of flux scattering, if the simulation uses a different underlying distribution from the intrinsic distribution, such mapping will not work properly. In our test cases, the  UVLFs derived from simulations with an underlying flat distribution are similar to those derived with method 1. The excess at the bright ends of the dashed lines in Figure \ref{fig:nominal_phi_method2} is quantitatively the same as the excess found in Figure \ref{fig:nominal_phi_method1}. Although this similarity is due to the assumption that the flux scatter is Gaussian, it can lead to a false consistency between the UVLFs derived by the two methods.

\citet{Calvi2016} used method 2 to calculate completeness as a function of observed apparent magnitude $m_\textrm{obs}$ with a flat distribution as an input distribution in the simulation. They found that the number densities of $z\sim9$ galaxies at the bright end of the luminosity function, $M\lesssim-22$, show some excess compared to the predicted Schechter fit \citep[e.g.,][]{Bouwens2016} by $\sim0.4$--$0.5$ dex. This excess is consistent with what \citet{Rojas-Ruiz2020} found using method 1. According to Figure \ref{fig:nominal_phi_method2}, when we apply a simulation with an underlying flat luminosity distribution to the observation with a flux scatter of 0.3 mag (red dashed line), the induced bias is in the order of $0.2$--$0.3$ dex for magnitude bins with brightness comparable to the \citet{Calvi2016} sources. Thus, a significant fraction (approximately half) of the excess found in \citet{Calvi2016} may come from completeness-simulation bias induced by flux scattering.

\begin{figure*}
    \centering
    \includegraphics[width=0.45\textwidth]{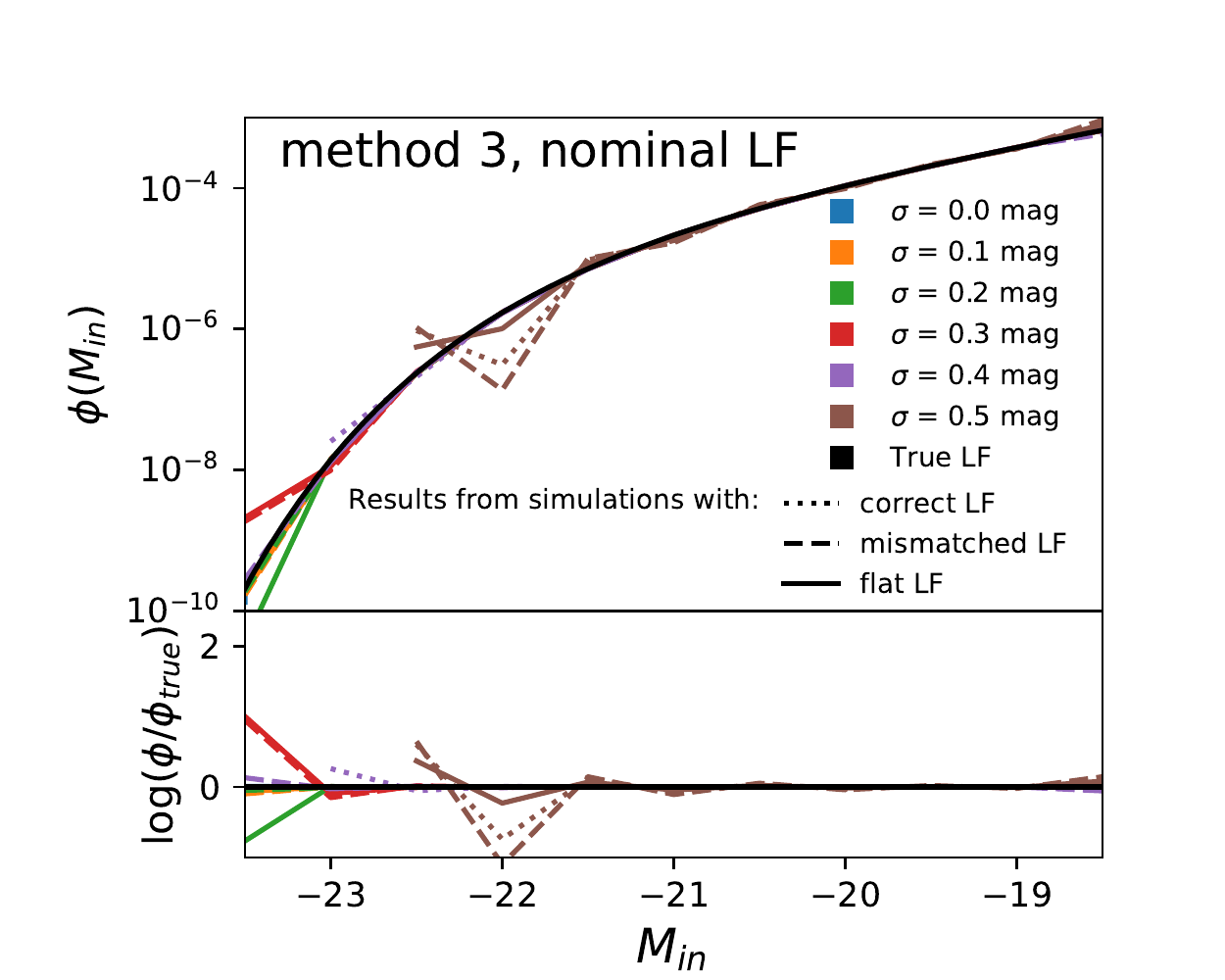}
    \includegraphics[width=0.45\textwidth]{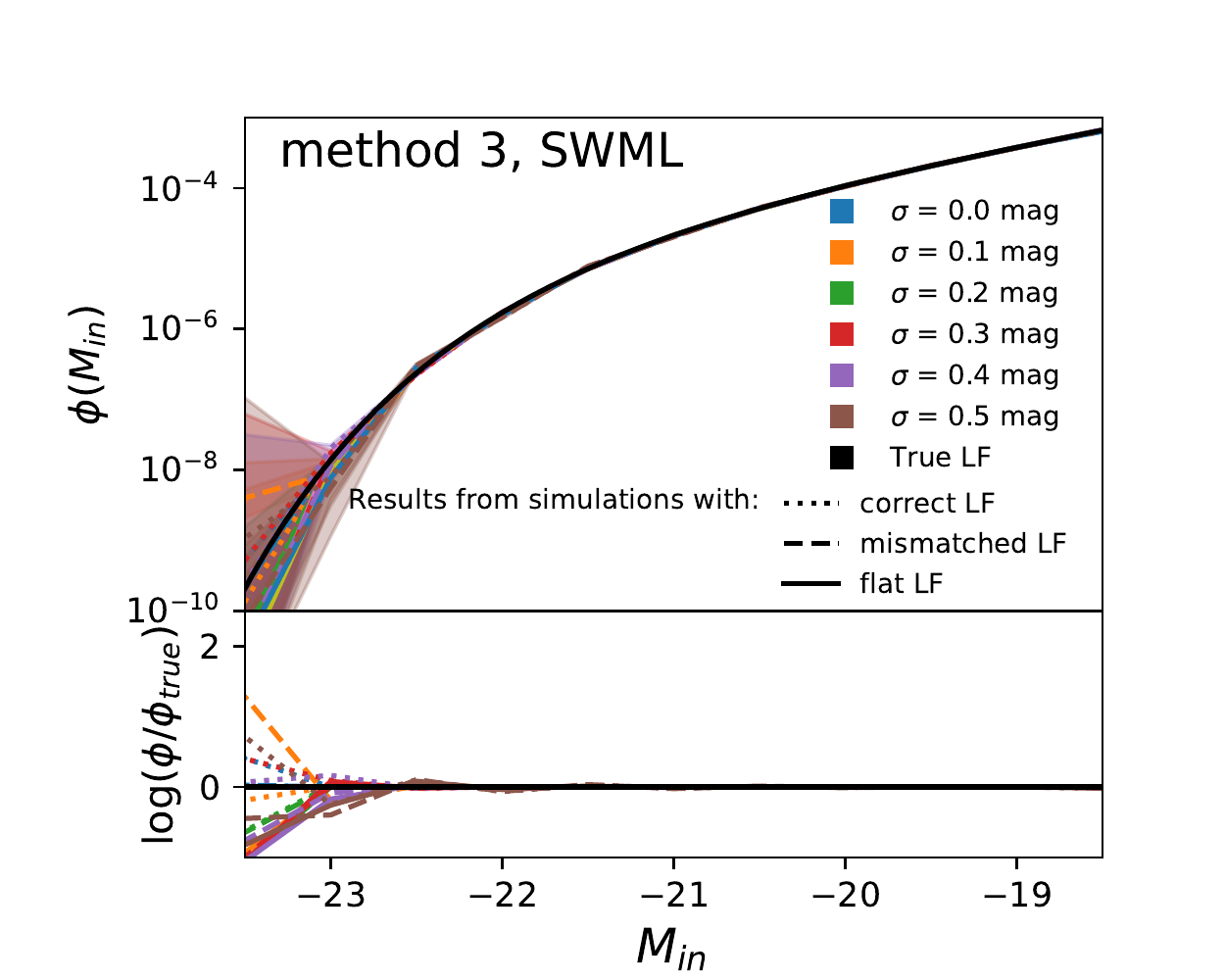}
    \caption{Similar to Figure \ref{fig:nominal_phi_method1} and \ref{fig:nominal_phi_method2} but for the luminosity functions derived with method 3. The left panel shows the nominal luminosity functions derived by solving equation \ref{eq:nexp_method3}. The right panel shows luminosity functions derived with the stepwise maximum likelihood approach, similar to that in \citet{Bouwens2015}.}
    \label{fig:nominal_phi_method3}
\end{figure*}

Indeed, \citet{Bowler2020} recently noticed that their derived UVLFs may depend on the underlying distribution of the galaxies in the completeness simulation. They found that the bright ends of their derived $z\sim$8--10 UVLFs based on the simulation with underlying double power-law distribution differ from those based on the simulation with underlying Schechter distribution. Nonetheless, the derived UVLFs based on both simulations still show an excess of sources at the bright end compared to a Schechter form. This suggests that at least some of the excess detected at the bright end of the LF is likely intrinsic. 

\smallskip

\textbf{Method 3:} For this method, we calculate nominal luminosity functions by directly taking an inverse of equation \ref{eq:nexp_method3}:
\begin{equation}
    \phi(M_\textrm{in})dM =  V^{-1}_\textrm{eff}(M_\textrm{in},M_\textrm{recov}) \cdot N^\textrm{obs}(M_\textrm{recov}),
    \label{eq:method3_nominalcal}
\end{equation}
where $V^{-1}_\textrm{eff}$ is the inverse of the $V_\textrm{eff}$ matrix. Generally, a square matrix is invertible if the determinant is non-zero. This suggests that we need to remove the $M_\textrm{in}$ columns that are zero vectors. However, because of the flux scatter, the matrix $V_\textrm{eff}$ after the removal of zero columns is often not a square matrix. Therefore, we can only take a left inverse $V_\textrm{eff}^{-1}$ of the $V_\textrm{eff}$ matrix. The recovered nominal UVLFs are shown in the left panel of 
Figure \ref{fig:nominal_phi_method3}.

  While this method appears formally robust, the direct calculation of luminosity functions using Equation \ref{eq:method3_nominalcal} has not been commonly used in the literature. Hence, we additionally calculate the best-fit luminosity function by using a more commonly used method: the stepwise maximum likelihood (SWML) approach \citep{Bouwens2015}. The SWML approach finds the best-fit luminosity function by comparing the expected number of galaxies at each recovered magnitude bin (as predicted by Equation \ref{eq:nexp_method3}) to the number of observed galaxies at the same recovered magnitude bin. It does not assume any functional form for the modelled luminosity function. Instead, $\phi(M_\textrm{in})$ at all magnitude bins are free parameters. Here, we use the Markov chain Monte Carlo (MCMC) method to find the best-fit UVLF, and we opt to calculate the likelihood with a binomial likelihood function, instead of the usual choice of the Poisson likelihood function, because our mock survey covers a large area of the sky rendering the Poisson estimation inappropriate: 
\begin{equation}
    \mathcal{L} =  \Pi \binom{n}{k} p^k(1-p)^{(1-k)},
\end{equation}
where the multiplication sum is over all $M_\textrm{recov}$ bins. $n$ and $k$ are a function of $M_\textrm{recov}$ where n is the expected number of galaxies in the universe. Since we assumed that our observation is half-sky, $n$ is therefore equal to $2\times N^\textrm{exp}(M_\textrm{recov})$, which is calculated in equation \ref{eq:nexp_method3}. $k$ is the observed number of galaxies in that $M_\textrm{recov}$ bin. $p$ is the probability of the galaxy to be in the patrol area, which is 1/2 in our case. The best-fit functions, as well as there $1\sigma$ uncertainties, are shown in the right panel of Figure \ref{fig:nominal_phi_method3}.

Based on Figure \ref{fig:nominal_phi_method3}, method 3 can recover the intrinsic luminosity function well, regardless of the flux scatter and the distribution function used in the injection-recovery simulation. An exception is at the brightest end. One contributing factor is stochasticity. At the bright end, there simply are not enough observable galaxies in the survey. For example, at the bright end $M_\textrm{in}=-23.5$, there are 13 observable galaxies with $M_\textrm{in}\sim 23.5$. In real observations, the survey area is usually much smaller, which suggests that stochasticity starts to dominate at much fainter magnitudes. Another contributing factor to the large uncertainties at the bright end is intrinsic to the deconvolution of Equation \ref{eq:nexp_method3}. Unlike methods 1 and 2, method 3 does not sum up the matrix in Figure \ref{fig:scheme} in either direction to create completeness functions. Therefore, it suffers the most from correlation between adjacent bins (i.e. the binning problem). \citet{Bouwens2015} also identified the correlation between magnitude bins to introduce substantial uncertainty at the bright ends of their derived UVLFs. To limit the correlation between adjacent bins, studies generally use a bin size that is larger than the typical flux scatter in the data. For example, \citet{Bouwens2015} used a wider binning scheme at the faint-end to determine UVLFs.


In general, method 3 works well irrespective of the amount of flux scatter and is independent of the distribution functions used in the completeness simulation. It is robust because it uses the most information produced by the completeness simulation to map between the number of observed galaxies and the intrinsic distribution of galaxies. 

\subsection{Applicability to actual observations}
\label{sec:mock_real_application_suggestion}
One limitation on the mock observation analysis from this Section is the assumption that within each magnitude bin, galaxies all have the same luminosity (i.e. we neglect the intra-bin luminosity function). In fact, for each input magnitude bin, the simulation injects galaxies with the same magnitude -- the value of the bin's center. As a result, the output matrix (e.g. those in Figure \ref{fig:situation2}) from simulations with underlying Schechter distribution is identical to the output matrix from the simulation with underlying flat distribution whose columns are multiplied by the value of the Schechter function at each bin center. If the simulation samples the flux of each injected galaxy instead, the scatter in the observed magnitude bins would likely increase. Especially when the underlying distribution is non-uniform, there will be more galaxies whose fluxes are close to edges of that bin. In turn, there will be more galaxies that scatter into adjacent recovered magnitude bins. 

Practically, it is often computationally difficult to simulate galaxies according to an underlying distribution with a steep slope, such as the Schechter distribution. For example, if one simulates 100 galaxies in the $M_\textrm{UV}=-23$ bin according to the UV luminosity function of $z=5$ galaxies \citep{Bouwens2015}, one has to simulate $\sim1.6$ million galaxies at $M_\textrm{UV}=-20$ bin. A leeway is to simulate equal numbers of galaxies in all magnitude bin but sample the magnitudes of simulated galaxies according to the desired underlying probability distribution. We implement this technique in our injection-simulation analysis on observational data in the next section.

\section{Test on observational data}
\label{sec:section4}
In this section, we compare $z\sim5$ UVLFs derived from Hubble Space Telescope observations using the three methods described in Section \ref{sec:completeness_definition}. For this, we resort to the publicly-released Hubble Legacy Fields (HLF) images from \citet{Whitaker2019}. Also, we artificially add noise to these data to create additional data sets that mimic observational surveys with larger flux scatter (i.e. mimic observations at shallower depths)
. We select the HLF as the testbed due to its exceptional depth and relatively wide area coverage (relative to Hubble observations) so that we can cover high redshift galaxies a wide range of magnitudes within one homogeneous data set. We choose to measure UVLFs at $z\sim5$ simply because there are more $z\sim5$ galaxies in the HLF fields than sources at yet higher redshifts, which minimizes the Poisson uncertainties in the derived UVLFs. Furthermore, \citet{Barone-Nugent2015} determined that, unlike galaxies at higher redshifts, $z\sim5$ galaxies in the HLF field are not significantly affected by magnification bias at $M_\textrm{UV}\gtrsim -23$, removing another source of uncertainty and systematic bias in the LF determination.

We describe the observational data, source detection, flux measurements, and completeness simulation in Section \ref{sec:observational_data} -- \ref{sec:recovered_MUV}. In Section \ref{sec:lowsn_data}, we generate additional sets of data with larger flux scatters by adding noise to the measure $M_\textrm{UV}$ magnitudes. We then derive the UVLFs and discuss the results in Section \ref{sec:observational_results}.
We provide their summary in Section \ref{sec:observation_summary}.

\subsection{Hubble Legacy Fields application}
\label{sec:observational_data}
 We use the V2.1 version data release of the Hubble Legacy Fields (HLF) project for the GOODS-S region \citep{Whitaker2019}. The data release homogeneously combines all imaging of the GOODS-S extragalactic field from over 30 different HST programs such as the Hubble UltraDeep Field \citep[e.g.,][]{Beckwith2006,Bouwens2011}, the Cosmic Assembly Near-infrared Deep Extragalactic Legacy Survey \citep{Koekemoer2011}, and the Early Release Science observations \citep{Windhorst2011}. The final product includes ACS, WFC3/IR and WFC3/UVIS imaging that are taken over 2635 orbits. The images are sky-subtracted, and point-spread function (PSF) matched to the F160W band. We use the images with 60-mas pixel scale, and trim them to the common overlap area in the F850LP, F125W, F140W, and F160W bands. 
 
The full GOODS-S field contains several patches of relatively deeper regions. We separate the XDF/HUDF region out of the GOODS-S field to separately run the injection-recovery simulation. Beside the XDF region, there are four parallel ultra-deep fields: HUDFP1 to HUDFP4. We disregard the parallel HUDFP1 and HUDFP2 region because the HUDFP1 region does not have $B_{435}$ images required for the $z\sim5$ dropout selection. The HUDFP2 region contains multiple bright stars that might affect the measurement of fluxes and colours. We do not separate the HUDFP3 and HUDFP4 out because they have similar depths to the rest of the GOODS-S field in the near-infrared bands. Hereafter, we identify the two regions as the XDF and the GOODS region. Finally, we create a detection image for each region by combining noise-equalized images in four HST bands (F850LP, F125W, F140W, F160W), following the procedure of \citet{Whitaker2019}. 
 The final area is equal to 5.1 arcmin$^2$ for the XDF region, and 135 arcmin$^2$ for the GOODS region.
 
\subsection{Source detection and flux measurements}
\label{sec:flux_measurement}
For each region, we run \texttt{SExtractor} \citep{BertinArnouts1996} on the PSF-matched HST images with a dual-image mode, using the detection image described above and the root mean square (rms) map specific to each filter. For the GOODS region, we use the same \texttt{SExtractor} parameters as that in \citet{Whitaker2019}. For the deep XDF/HUDF region, we lower the detection threshold (\texttt{DETECT\_THRESH}) from $1.8 \sigma$ to $1 \sigma$ and lower minimum number of pixels for detection (\texttt{DETECT\_MINAREA}) from 14 to 9 pixels in order to detect fainter galaxies (see also \citealt{Trenti2011} for similar parameter choices in the context of detection of high-z sources with low/modest signal to noise). From \texttt{SExtractor} catalogues, we obtain the \texttt{AUTO}, and \texttt{APER} photometry with aperture of diameter $0\arcsec.35$, and $0\arcsec.7$. We then correct the fluxes for the Galactic dust extinction, using the NASA/IPAC infrared archive's Galactic dust extinction calculator \citep{Schlegel1998, SchlaflyFinkbeiner2011}. Lastly, we set negative aperture fluxes to a nominal zero value ($m=50)$.

We use a $0\arcsec.35$-diameter circular aperture magnitudes to calculate colours of the detected galaxies and use the following procedure to calculate their total fluxes. Following \citet{Whitaker2019}, each galaxy's total flux is its $0\arcsec.7$-diameter circular aperture flux corrected by a correction factor, which is a product of two numbers. The first number approximately corrects the measured \texttt{FLUX\_AUTO} for the flux that may fall outside of the Kron (\texttt{AUTO}) aperture. It is the ratio between the total PSF flux to the PSF flux enclosed in a circularized Kron radius of the galaxy. 
To obtain this number, we directly interpolate the PSF curve of growth in \citet[][their Figure 6]{Whitaker2019}, assuming that a 2 \arcsec aperture should encompass all PSF light. The second number is the ratio between \texttt{FLUX\_AUTO} to the flux in the $0\arcsec.7$ diameter circular aperture of each galaxy in the reference F125W band. While \citet{Whitaker2019} used F160W as a reference band, we use F125W instead because the images in F125W band are $\sim0.3-0.4$ mag deeper than the images in F160W band. If we instead used F160W as a reference band, we would have lost many dropouts as they do not have measurable \texttt{FLUX\_AUTO} in F160W.  F125W therefore better suits the selection of $z\sim5$ dropout galaxies. 

We estimate flux uncertainties and signal-to-noise ratios using ``empty apertures" analysis instead of directly using the errors returned by \texttt{SExtractor}. This is to avoid the correlated noise between adjacent pixels that renders flux uncertainties returned by \texttt{SExtractor} underestimated \citep[e.g.,][]{Trenti2011, Whitaker2011}. Detail of this procedure are reported in Appendix \ref{appendix:flux_uncertainties}.

\subsection{Dropout selection}
\label{sec:selection}
To select $z\sim5$ Lyman-break galaxies (LBGs), we use the colour and signal-to-noise criteria from \citet{Bouwens2015} together with a photometric redshift selection. Starting from the total fluxes and colours derived in Section \ref{sec:flux_measurement}, we first select the initial set of candidates with the following criteria:
\begin{equation}
    \begin{array}{l}
        (i_{606}-z_{775}>1.2)\ \wedge\ (z_{850}-H_{160}<1.3) \ \wedge\\
        (V_{606}-i_{775}>0.8(z_{850}-H_{160})+1.2) \ \wedge\\
        (\chi^2_\textrm{NIR}>25) \ \wedge\
        (\textrm{SN}(B_{435})<2)  
        
    \end{array}
    \label{equation:color_cut5}
\end{equation}
And that the candidates must not satisfy the $z\sim6$ dropout criteria:
\begin{equation}
    \begin{array}{l}
        (i_{775}-z_{859}>1.0)\ \wedge\ (Y_{105}-H_{160}<1.0) \ \wedge\\
        (i_{775}-z_{850}>0.78(Y_{105}-H_{160})+1.0)\ \wedge\\
        (\chi^2_\textrm{NIR}>25) \ \wedge\ (\textrm{SN}(B_{435})<2) \ \wedge\\
        ((V_{606}-z_{850}>2.7) \vee (\textrm{SN}(V_{606})<2))
    \end{array}
    \label{equation:color_cut6}
\end{equation}
The $\chi^2_\textrm{NIR}$ is the signal-to-noise in the NIR bands, defined as $\sum_i\textrm{SGN}(f_i)(f_i/\sigma_i)^2$where the summation index $i$ is over all bands in [$Y_{105}, J_{125}, JH_{140}$, and $H_{160}$]. $f_i$ and $\sigma_i$ are the total flux and its uncertainty in band $i$. $\textrm{SGN}(f_i)$ is equal to 1 if $f_i$ is greater than zero, and equal to $-1$ otherwise. These criteria are the same as those in \citet{Bouwens2015} except that we only check for stellarity if the candidate has $\chi^2_\textrm{NIR} > 1000$. This is because we find that the stellarity measured from \texttt{SExtractor} may not work reliably with the multi-band PSF-matched detection images used in this study unless the detection signal-to-noise ratio is sufficiently high.

We further check whether the candidates blend with other objects. Suppose a candidate is adjacent to another brighter source which is not a dropout candidate. In that case, we count the number of pixels of the candidate source that are adjacent to the other objects, using the segmentation map. If the number of the adjacent pixels is larger than 25\% of the total number of pixels on its circumference, we consider our object to be blended and discard the dropout candidate. We set this criterion so that it is similar to the blending criterion used in our GLACiAR2 injection-recovery simulation (see Section~\ref{sec:injection-recovery_sim}). This step removes 15 objects in the XDF region and 67 objects in the GOODS region. In addition, we remove 56 objects in the GOODS region that we visually flag as artifacts from bright objects (e.g. outer regions of diffraction spikes). With these criteria, we obtain 170 initial candidates from the XDF region and 1176 initial candidates from the GOODS region.

We then pass these initial candidates to the EAZY photometric redshift code \citep{Brammer2008}, to remove low-redshift interlopers via spectral energy distribution (SED) fitting. The input of EAZY is the total fluxes and their corresponding uncertainties in F435W, F606W, F775W, F814W, F850LP, F105W, F125W, F140W, and F160W. We simultaneously fit with all templates in the v1.3 template set, which yields the least bias for high-redshift galaxies \citep{Brinchmann2017, Morishita2018}. The templates consist of the P\`{E}GASE stellar population synthesis model library (FiocRocca-Volmerange1997), emission lines \citep{Ilbert2009}, dusty SEDs \citep{Maraston2005,BC03}, and a SED with high equivalent width nebular emission lines \citep{Erb2010}. We use a flat prior and adopt the redshift at which the likelihood is maximized ($z_p$) as the best-fit redshift.

Based on the returned redshift probability distribution functions $P(z)$ from EAZY, we select the candidates whose probability $P(z>4)$ is greater than 84\% (i.e. those with 1-sigma lower limit redshift larger than 4). With these photometric redshift criteria, the final number of $z\sim5$ candidates are 128 galaxies in the XDF region, and 906 in the GOODS region. We calculate $M_\textrm{UV}$ for these galaxies by taking an average of their best-fit spectra over the 100\angstrom\ region centring at 1600\angstrom (rest-frame). We provide a comparison between our candidates with the candidates in \citet{Bouwens2015} in Appendix \ref{appendex:compare_candidates}. 

\begin{figure}
    \centering
    \includegraphics[width=\textwidth]{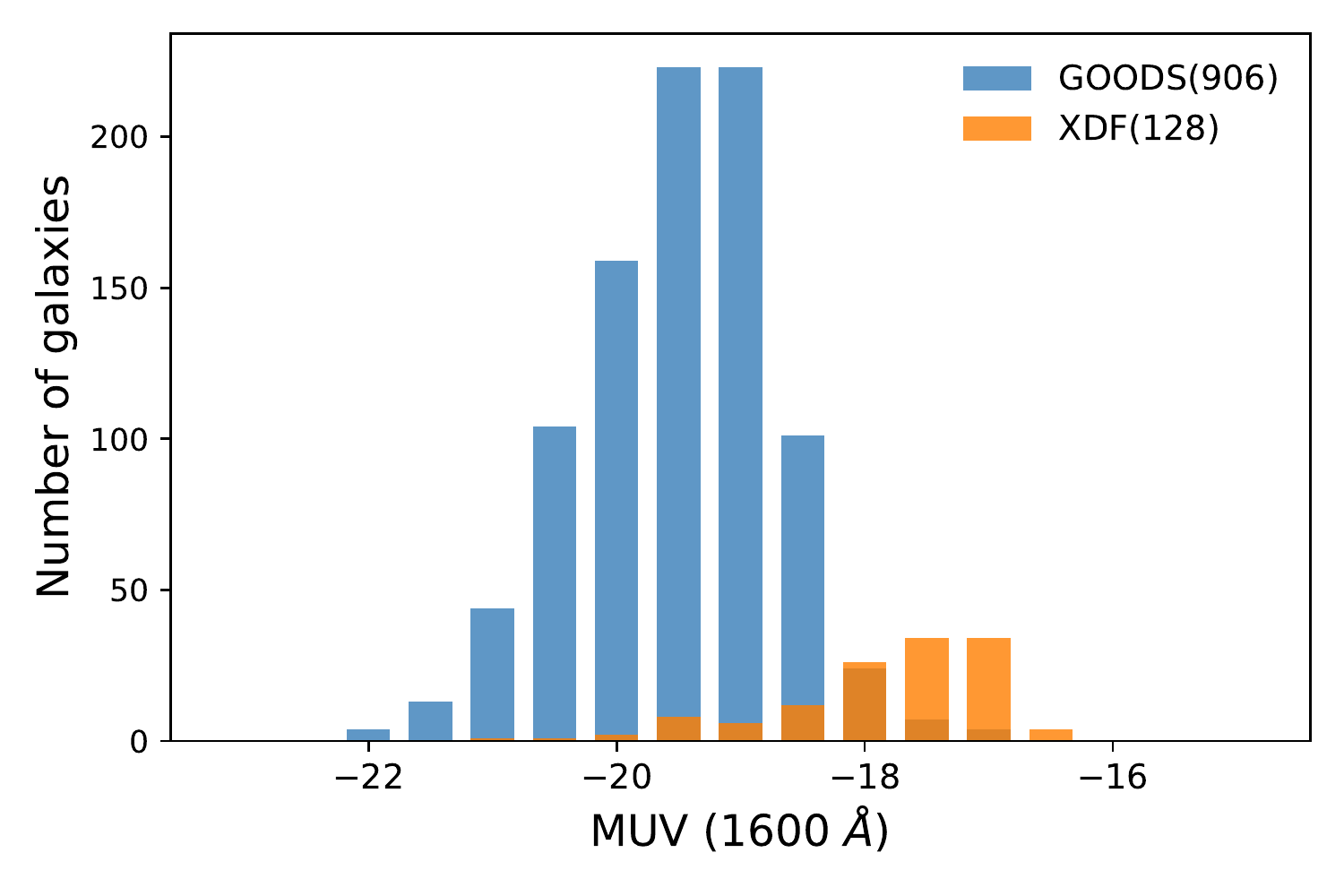}
    \caption{Number of $z\sim5$ candidates as a function of $M_\textrm{UV}$ for each field. Most of candidates overall come from the GOODS-S field, while the majority of the faintest sources are from the XDF region.}
    \label{fig:histogram}
\end{figure}

\subsection{Completeness simulation}
\label{sec:injection-recovery_sim}
We build our completeness simulation tool on a publicly available completeness simulation code written in python: the GaLAxy survey Completeness AlgoRithm \citep[][\texttt{GLACiAR}]{Carrasco2018}. The new version (\texttt{GLACiAR2}) is now available on \texttt{GITHUB} (see footnote \ref{url_github}). The modified code is capable of 1) injecting galaxies with different underlying probability distributions 2) accepting different detection image types, e.g. a coadd of multiple-band images or a single-band image as in the original code and 3) producing outputs suitable for various definitions of the completeness functions. The code is flexible and allows a wide range of input parameters.  More detailed description of \texttt{GLACiAR2} can be found in Appendix \ref{appendix:glaciar}. For the rest of this section, we describe the specific prescriptions used for the science images considered in this paper.

We perform two sets of simulations, injecting galaxies with different underlying distributions in each magnitude bin: (1) a flat distribution, and (2) a Schechter distribution described by the best-fit $z\sim5$ parameters derived by \citet{Bouwens2015}. The simulations inject galaxies in bins of redshift and intrinsic UV magnitude, with 13 redshift bins ranging from $z=4.1$ to $z=6.5$, and 23 $M_\textrm{UV}$ bins ranging from $-24.5$ mag to $-13.5$ mag. For each $M_\textrm{UV}$ and redshift bin, we sample 20 $M_\textrm{UV}$ values according to the selected underlying distribution. We synthesize a spectrum for each $M_\textrm{UV}$ value with a piecewise function: zero flux blueward of the Lyman break, and a power-law function with UV spectral slope $\beta$ redward of the Lyman break. The $\beta$ value is randomized from a Gaussian distribution with a mean $\beta = -2.2$ and a standard deviation of 0.4 \citep{Carrasco2018}. The spectrum is then multiplied with the HST transmission curves to generate the total fluxes for each HST filter.  For each SED, we create 100 artificial randomized Sérsic light profiles with a 50-50\% mix of $n=1$ and $n=4$ Sérsic indices. The effective radii of the injected galaxies scale with redshift, $R_e(z) = 1.075 \times 7/(1+z)~\mathrm{kpc}$, where we set the effective radius at $z=6$ as anchor point \citep{Bradley2012,Bernard2016}. The chosen scaling relation approximates the LBG size evolution determined by \citet{Bouwens2004} and \citet {Oesch2010}, which is based on $L\gtrsim 0.3 L_*$ LBGs in the range $z\sim3-7$. At any given redshift, we do not vary the effective radius of the injected sources. We assign random inclinations and ellipticities to the light profiles. Finally, we convolve the light profiles with the F160W PSF image, add the galaxy stamps to the science images, and set \texttt{GLACiAR2} to generate a detection image in the same manner as that for the science images. The code then runs SExtractor, using the same SExtractor parameters that we used for the science images. 

The output of \texttt{GLACiAR2} are the SExtracted properties of the injected sources, their intrinsic properties ($M_\textrm{UV}$ and $\beta$ slope), and their detection status (whether they were SExtracted or considered as blended with foreground galaxies). In total, we inject 598,000 galaxies into each field per simulation.

\begin{figure*}
    \centering
    \includegraphics[width=0.65\textwidth]{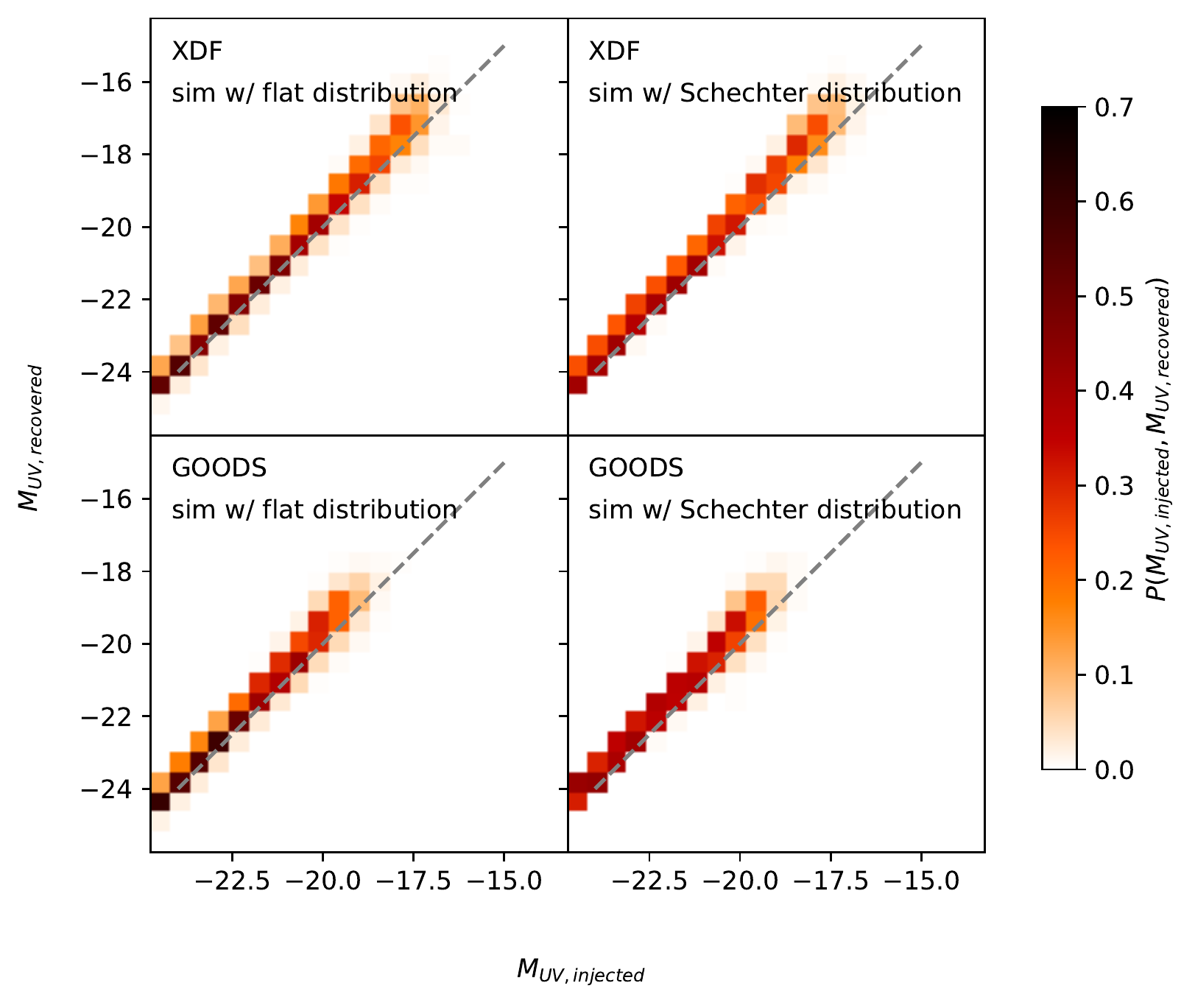}
    \caption{Examples of completeness functions as defined by method 3 for the XDF region (upper) and the GOODS region (lower) at $z=4.9$. The completeness is a function of both injected and recovered magnitudes. The left (right) panels show the results from simulations whose luminosities of the injected sources are sampled from an underlying flat (Schechter) distribution. We show the completeness function at $z=4.9$ here as it is the closest simulated bin to $z\sim5$.}
    \label{fig:completeness_method3}
\end{figure*}

\begin{figure}
    \includegraphics[width=0.9\textwidth]{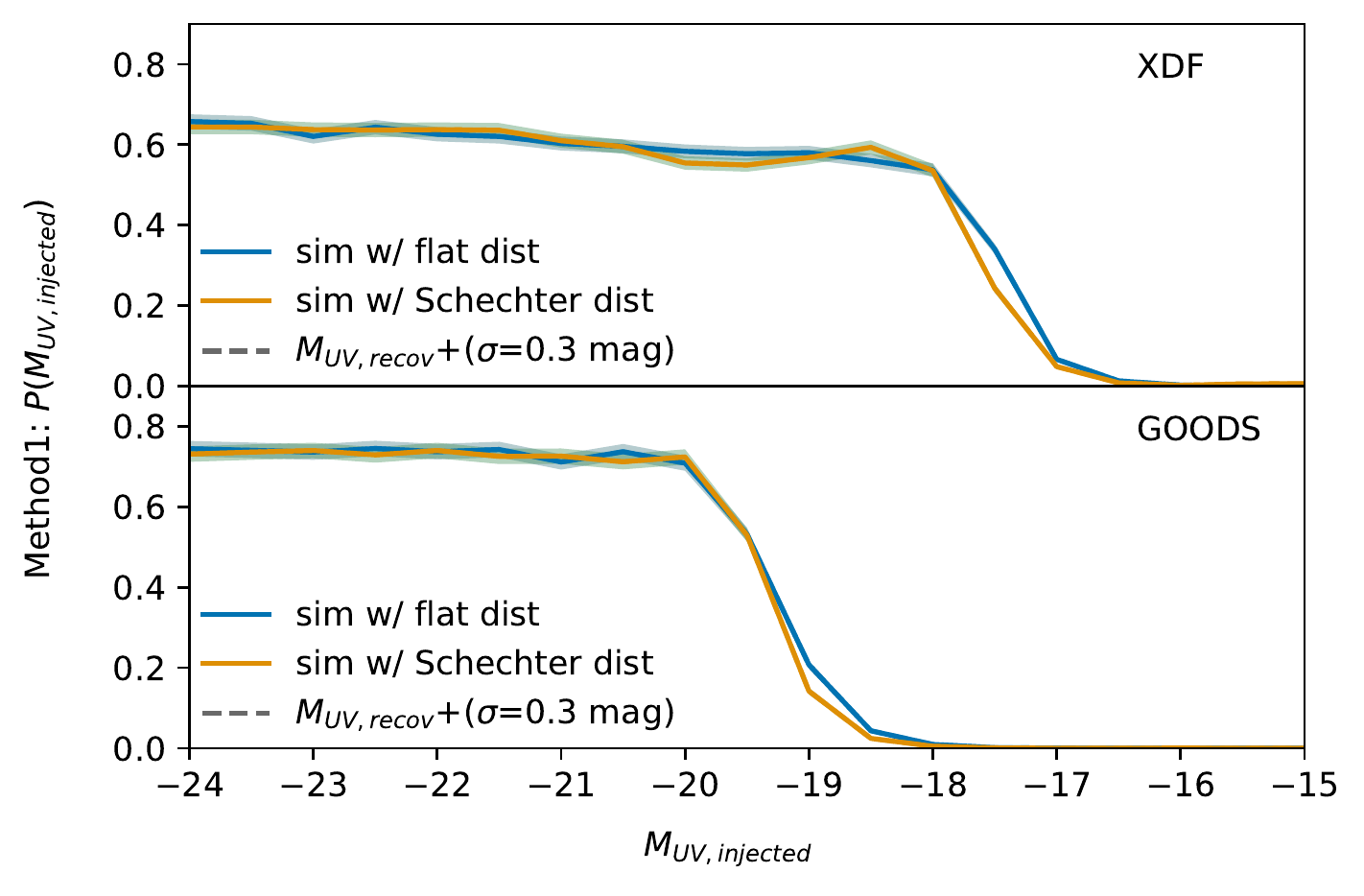}
    \caption{Examples of completeness as a function of intrinsic $M_\textrm{UV}$ magnitude, defined by method 1. In the upper panel, the solid lines show the completeness of the XDF region based on the simulation with different underlying distribution. The solid lines in the lower panel show the completeness for the GOODS region at $z=4.9$. It is clear that the completeness does not depend on the underlying distribution used in the completeness simulation. The dashed lines show the completeness for data with large flux scatters (see Section \ref{sec:lowsn_data}). }
    \label{fig:completeness_method1}
\end{figure}

\begin{figure}
    \includegraphics[width=0.9\textwidth]{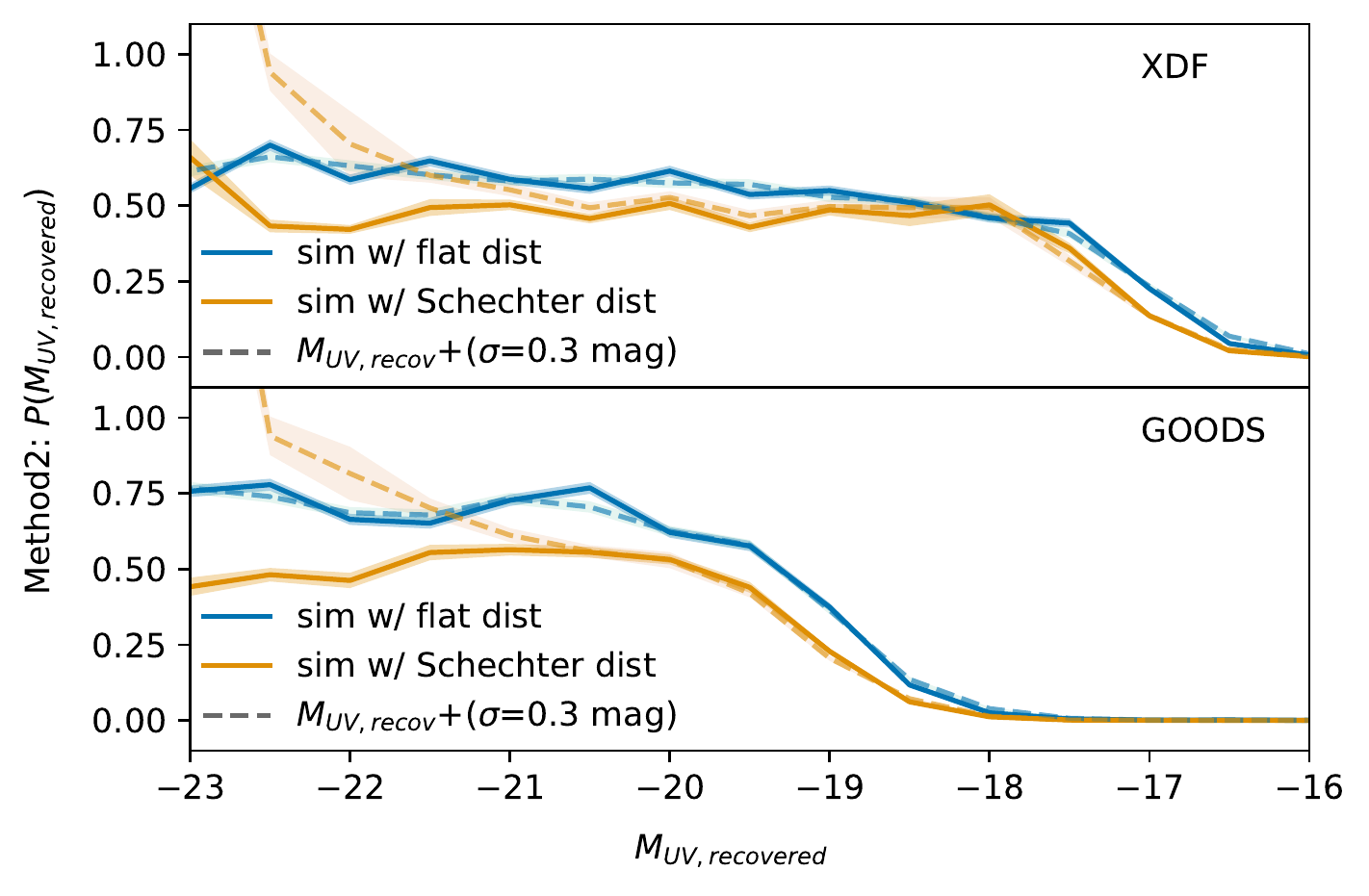}
    \caption{Examples of completeness as a function of recovered $M_\textrm{UV}$ magnitude at $z=4.9$, defined by method 2. The completeness depends on the underlying distribution used in the completeness simulation. The colours and line styles are the same as those in Figure \ref{fig:completeness_method1}.}
    \label{fig:completeness_method2}
\end{figure}

\subsection{Recovered UV magnitudes and completeness functions based on the simulations}
\label{sec:recovered_MUV}

We follow the same procedures applied to real sources to determine whether the simulated sources pass the dropout selection criteria and to recover their $M_\textrm{UV}$. First, we measure the colours and total fluxes of the simulated galaxies by repeating the method in Section \ref{sec:flux_measurement}. We then apply the same dropout and signal-to-noise criteria (Equation \ref{equation:color_cut5}) to determine whether those injected sources are classified as dropouts. We require that these injected galaxies must be detected by \texttt{GLACiAR2} as either isolated sources or minimally blended with existing objects that are fainter than themselves (detection status $\geq 0$). For each injected source that passes the dropout selection, we recover its $M_\textrm{UV}$ by fitting for the UV spectral slope $\beta$ and absolute magnitude at 1600 \angstrom ($M_\textrm{UV}$) to the recovered total fluxes. We fix the redshift to a random Gaussian distribution around the injected redshift with $\sigma_z=0.27$ to mimic the redshift uncertainties from EAZY photometric redshift \citep{Brinchmann2017}.\footnote{According to \citet{Brinchmann2017}, when using the \texttt{eazy\_1.3} template set, the best-fit photometric redshift $z_p$ has a median absolute deviation $(z_\textrm{spec}-z_p)/(1+z_\textrm{spec}) = 0.045$.}  We only use these redshifts to calculate the galaxies' distance measures.

Based on the simulation results, we calculate the completeness functions according to the methods described in Section \ref{sec:completeness_definition}. Examples of the three completeness functions at $z=4.9$ are in Figure \ref{fig:completeness_method3} -- \ref{fig:completeness_method2}. In particular, Figure \ref{fig:completeness_method3} shows the completeness as a function of both $M_\textrm{injected}$ and $M_\textrm{recovered}$ as defined by method 3. As expected, the XDF region is a few magnitudes deeper and has smaller flux scatter than the GOODS region. The recovered fluxes mostly scatter toward the direction of fainter magnitude bins compared to the injected values, i.e. upward of the grey dashed 1-1 line, indicating that the identification and recovery steps tends to underestimate systematically the flux of faint sources (this is expected to apply equally for real and simulated sources).  The left and the right panels show the results from the two simulations with different underlying distribution (flat and Schechter, respectively). More galaxies from the simulations with an underlying Schechter distribution (the right panels) scatter into fainter recovered magnitude bins than the galaxies from the simulations with underlying flat distribution (the left panels). This is expected because the intrinsic magnitudes  $M_\textrm{UV,injected}$ of galaxies that are sampled from a Schechter distribution are likely already close to the faint-edge values of the injected magnitude bins.  Therefore, their recovered magnitudes are more easily scattered into the fainter bins (Eddington bias). 

We show examples of the completeness function at $z=4.9$ as defined by method 1 in figure \ref{fig:completeness_method1}. Because the completeness definition from method 1 sums up all recovered galaxies regardless of the recovered magnitudes, there is no difference between two completeness functions derived from simulations with different underlying distributions. The completeness of the GOODS region is $\sim70\%$ at the brightest magnitude bins. The XDF region is deeper and appears more crowded. Thus, its completeness at the brightest magnitude bins is somewhat smaller ($\sim65\%$ ). The completeness drops by half at $M_\textrm{UV}\sim -19.5$ bin for the GOODS region and at $M_\textrm{UV}\sim -18$ bin for the XDF region. To avoid magnitude bins that are highly incomplete, we henceforth consider only the candidates in the magnitude bins that are brighter or equal to the earlier mentioned thresholds.

Figure \ref{fig:completeness_method2} shows examples of the completeness as a function of recovered magnitudes at $z=4.9$ as defined by method 2. The completeness from the simulations with underlying flat distribution (shown as blue lines) is essentially the summation of the matrix in the left panels of Figure \ref{fig:completeness_method3} along the row. In contrast, to derive the correction factor from the simulations with an underlying Schechter distribution similar to what done in \citet{Bowler2020}, we need to multiply our output matrices with a Schechter function. This is because although we sample the galaxy's $M_\textrm{UV,injected}$ from a Schechter distribution, we injected the same number of galaxies in each injected magnitude bin into the simulation. For this reason, we multiply each row of the matrices in the left panels of Figure \ref{fig:completeness_method3} with the Schechter function of $z\sim5$ galaxies from \citet{Bouwens2015}. We then calculate the completeness as usual. The resulting completeness functions are shown as yellow lines in Figure \ref{fig:completeness_method2}. Based on this figure, the completeness function is dependent on the underlying distribution functions used in the completeness simulation. The difference is slightly smaller for the XDF region whose flux scatter is smaller.

We contend that our assumption regarding the source sizes in the completeness simulation would at most affect the completeness functions at the faint end. In the completeness simulations, we assume that all injected galaxies at a given redshift have the same effective radius regardless of luminosity. However, studies have shown that LBGs follow a size-luminosity relation \citep[e.g.,][]{Huang2013,Shibuya2015,Liu2017}. Based on the size-luminosity relation of $z\sim5$ galaxies determined by \citet{Liu2017}, our chosen size of $R_e = 1.25$ kpc at $z=5$ is within their $1\sigma$ range at all magnitude bins brighter than $M_\textrm{UV}\sim-18$. This means that, at the faint end, the sizes of our injected galaxies are larger than the average of the actual values. Since it is easier to detect a point source than to detect an extended source of the same luminosity, our injected galaxies at the faint end are less likely to be recovered than most of the real galaxies in the same magnitude bin. Consequently, we likely underestimate our completeness functions at the faint end. Nonetheless, because we restrict our UVLF determinations to $M_\textrm{UV}<-18$, the UVLFs derived in this work should not be impacted by the choice of galaxy size.

\subsection{Flux scatters}\label{sec:flux_scatter}
\begin{figure*}
    \centering
    \includegraphics[width=0.45\textwidth]{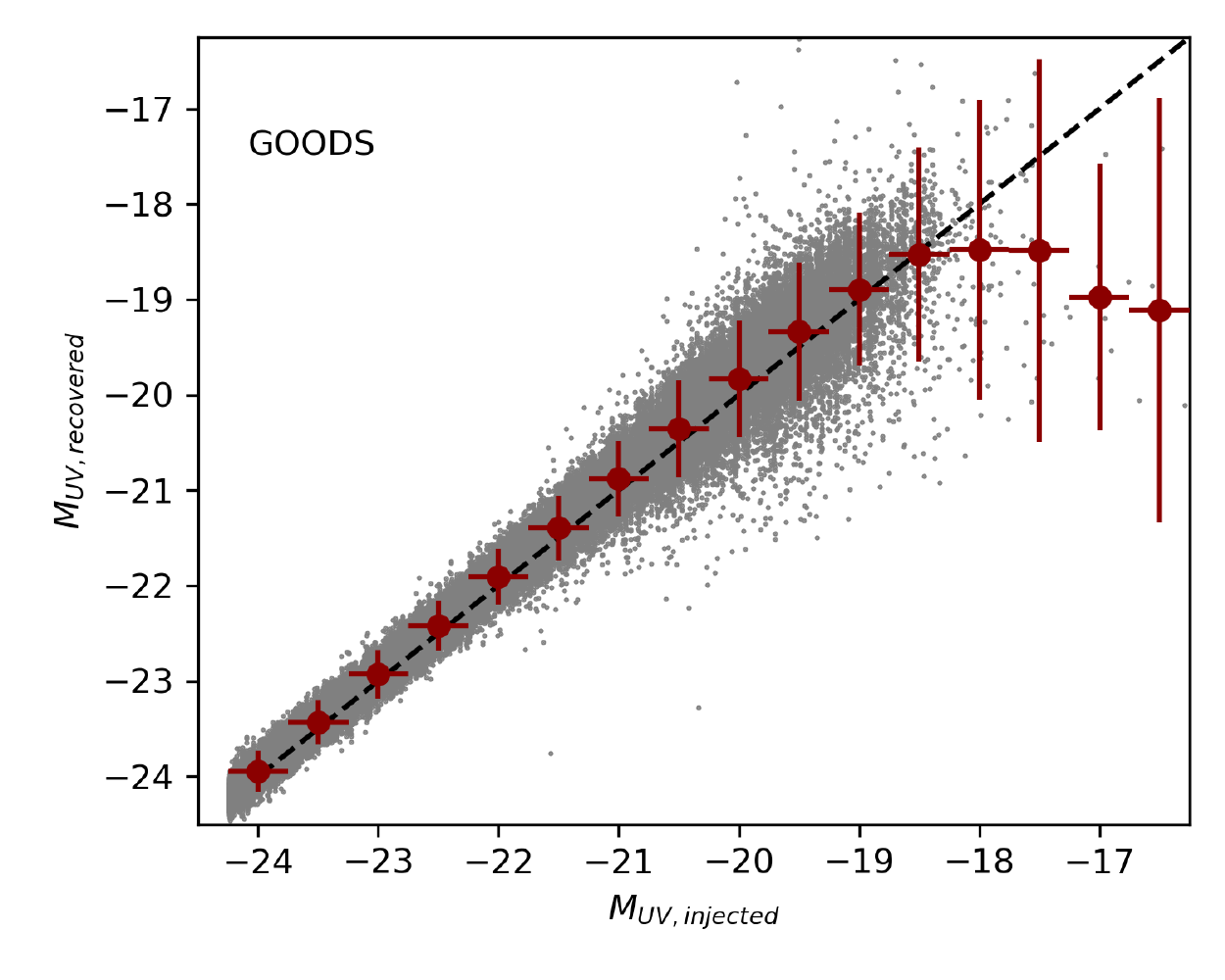}
    \includegraphics[width=0.45\textwidth]{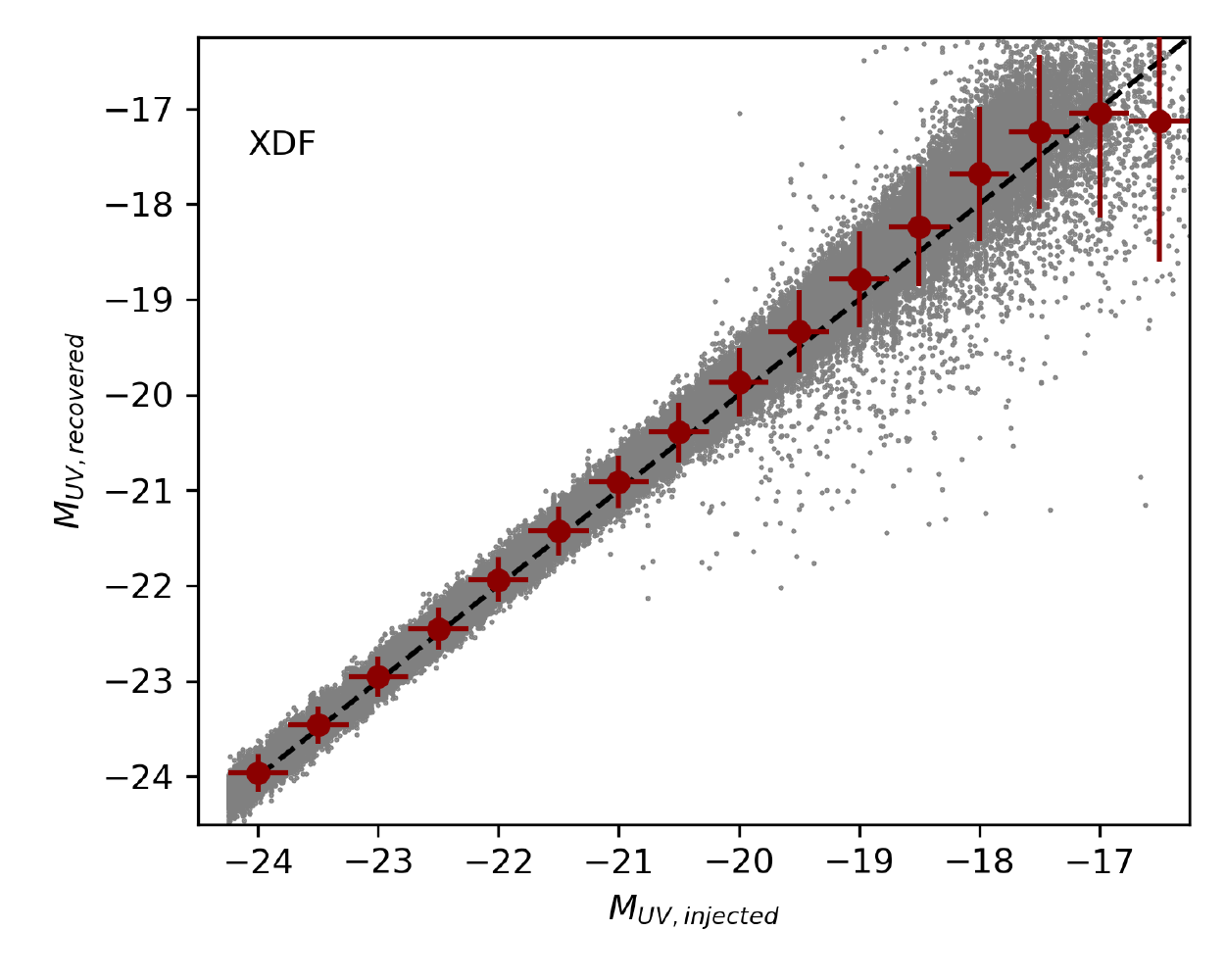}
    \caption{Injected and recovered $M_\textrm{UV}$ magnitudes of the simulated galaxies in simulations whose underlying distribution of the injected galaxy is flat. The gray points show all injected galaxies that are recovered as $z\sim5$ dropouts. The red errorbars represent the magnitude bin size (x-axis) and the 2$\sigma$ uncertainties (y-axis).}
    \label{fig:magnitude_scatter}
\end{figure*}

\begin{figure}
    \centering
    \includegraphics[width=0.95\textwidth]{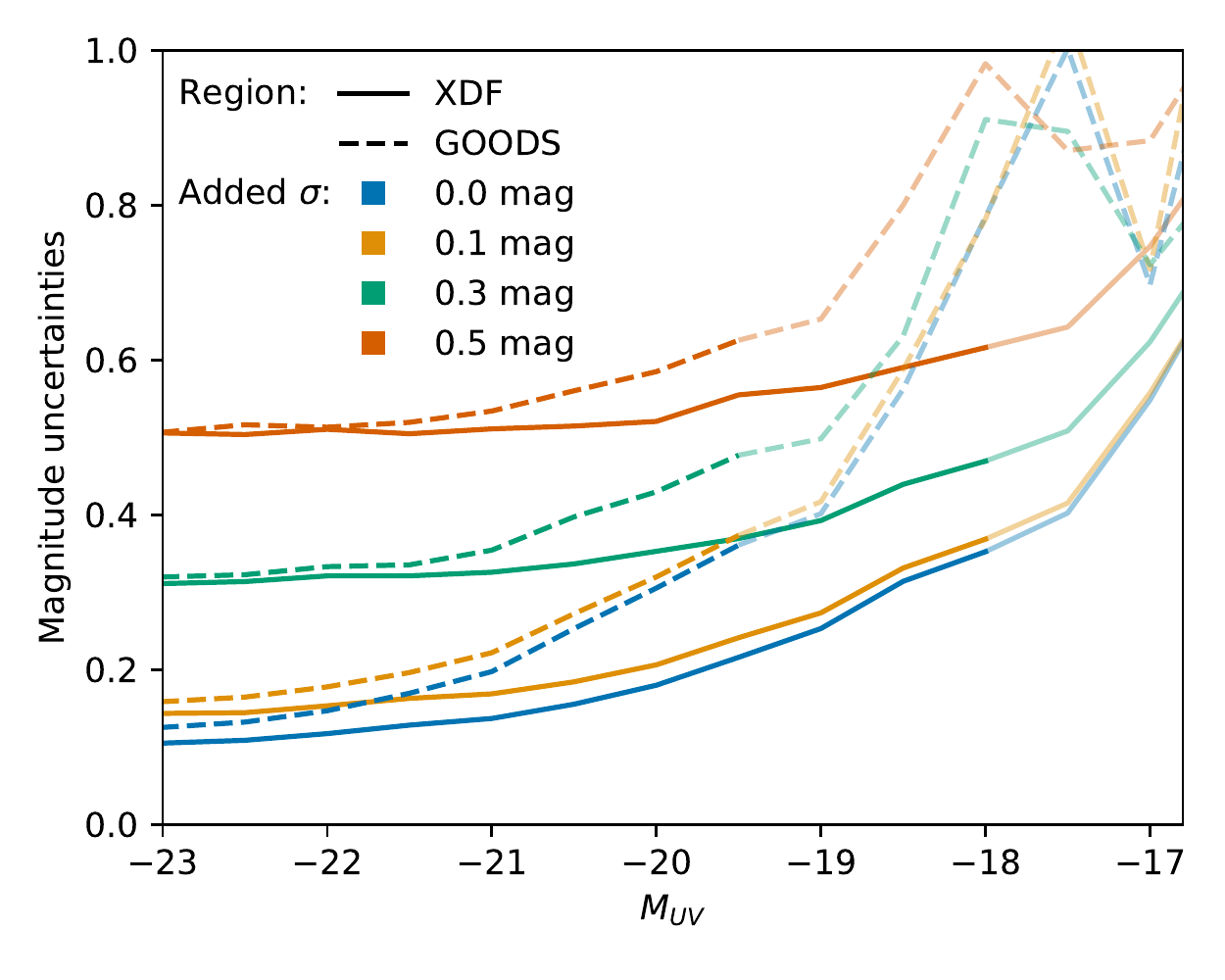}
    \caption{Estimated flux scatters as a function of $M_\textrm{UV}$. The blue solid and dashed lines show the intrinsic scatters estimated based on the completeness simulations as a function of intrinsic $M_\textrm{UV}$ (see Section \ref{sec:flux_scatter}). Lines of other colours represent the flux scatter estimated based on the completeness simulations, using the data sets with additional synthetic noise (see Section \ref{sec:lowsn_data}). The portions of the lines that are in lighter shade indicate the magnitude bins that are highly incomplete and not included in the UVLF calculations.
    }
    \label{fig:mag_uncertainties}
\end{figure}

As shown in Section \ref{sec:mock_result}, large flux scatters can cause bias in the measured UVLFs. It is therefore important to correctly estimate the flux scatter in a given data set. We estimate flux scatter from recovered galaxies in the completeness simulations. Figure \ref{fig:magnitude_scatter} shows the recovered UV magnitude as a function of the injected UV magnitude. We measure the differences between the injected and recovered magnitudes and calculate the mean and standard deviation in each magnitude bin. The results are plotted as red error bars in the figure. We re-plot the standard deviations (i.e., the flux scatters) as blue lines in Figure \ref{fig:mag_uncertainties}. We find that the flux scatter increases with magnitude. The $1\sigma$ flux scatters in the GOODS (XDF) region are approximately 0.12 (0.10) mag in the brightest magnitude bins and increase to 0.2 mag at $M_\textrm{UV}=-21$  ($M_\textrm{UV}=-19.5$) magnitude bin (see the blue solid and dashed lines).

\subsection{Mock data sets: data with different levels of flux scatter}
\label{sec:lowsn_data}

We have shown in Section \ref{sec:mock_result} that the bias in the derived luminosity functions increases with flux scatter in the recovered $M_\textrm{UV}$. When the flux scatter is less than or equal to 0.1 mag, the bias is minimal. In Section \ref{sec:flux_scatter}, we have shown that the flux scatter of $z\sim5$ galaxies in the HLF field is $\sim0.1$ mag in most of the magnitude bins. Consequently, we expect not to see differences in the UVLFs derived using different methods. Although this is assuring for $z\sim5$, it may not be indicative of the UVLF determination at higher redshifts. For example, the shot noise in the detected photo-electrons of $z\sim9$ galaxies would be $\sim 1.6$ times larger than the shot noise of $z\sim5$ galaxies of the same $M_\textrm{UV}$ that are in the same image (calculated based on the luminosity distances and the approximate ($1+z$) k-correction), even though in practice this is hardly relevant in current surveys, since the photometric noise in dominated by sky background and detector contributions (readout and dark current primarily) for the faint sources in question.
Moreover, we must observe galaxies at higher redshift with longer wavelength, where instrumental sensitivity might be lower, to obtain the same rest-frame wavelength. Galaxies at higher redshift might also be intrinsically fainter (i.e. $M_*$ evolves with redshift). Lastly, the characterization of high-redshift galaxies' spectra is still uncertain as there are limited numbers of observed spectra. All these factors may contributes to larger flux scatter in the measured $M_\textrm{UV}$ of high-redshift galaxies as they are usually determined based on SED fitting and/or photometric redshifts.

Therefore, we generate additional data sets with larger flux scatter starting from the original HLF data set. There are two options to do so. One is to add noise to the images. The other is to directly add noise to the measured $M_\textrm{UV}$. The first option mimics the observations with shallower depth more directly. However, by adding noise to the images, we will end up with a lower number of data points ($z\sim5$ candidates). In turn, the derived UVLFs will suffer from small number statistics and it will be difficult to disentangle the impact of increased flux scatter from the impact of having lower number of galaxies. Thus, in this study, we focus on adding noise to the measured $M_\textrm{UV}$ and keep the number of the candidates in the simulated samples the same. Nonetheless, we provide a brief discussion on the first approach (direct injection of noise in the images) in Appendix \ref{appendix:lowsn_data2}.

We add Gaussian noise with zero mean and $\sigma$ = 0.1,0.3, and 0.5 mag to the measured $M_\textrm{UV}$ of real sources. For each data set, we also add noise to the measured $M_\textrm{UV}$ of the simulated galaxies, to create completeness simulations that reflect the level of flux scatter of the data. These new data sets essentially have the same candidates and completeness simulations as the original data set. The only difference is flux scatter in the measured $M_\textrm{UV}$. This procedure can be viewed as mimicking studies that use different methods to convert observed colours to $M_\textrm{UV}$ (e.g. various assumptions of the spectral fitting procedures), or, to some extent, observations with lower depth but larger area (so that the number of the candidates are the same). The final flux scatters as a function of input magnitudes of these data sets are shown as orange, green and red lines in Figure \ref{fig:mag_uncertainties}.

\subsection{Comparing UVLFs derived with different methods}\label{sec:observational_results}
We are now equipped with the completeness functions output and the $z\sim5$ candidates. In this section, we use them to derive the UV luminosity functions for each data set, and underlying distribution in the completeness simulations according to the three methods in Section \ref{sec:completeness_definition}. For each setting, we fit for the UVLF with two functions: the Schechter function and the double power law function. We do not include candidates in the faint magnitude bins that are highly incomplete, i.e. the candidates that are in bins fainter than $M_\textrm{UV}=18$ and $M_\textrm{UV}=19.5$ for the XDF and the GOODS region, respectively. We provide more specific detail of the fitting procedure for each method in Appendix \ref{appendix:fitting_detail}. The best-fit parameters for all the cases considered are listed in Table \ref{table:schechterparameters}. 

For each set of data and method, we use the widely applicable information criterion \citep[][WAIC]{Watanabe2013} to determine the relative quality of the models for a given likelihood function. WAIC is a type of Bayesian information criteria that does not require the posterior probability distribution of the best fit parameters to be Gaussian, which is suitable for our models. We highlight the most preferred model in green colour. We find that the Schechter function is preferred over the DPL function in all of the tested cases, regardless of the flux scatter. 

\begin{figure}
    \centering
    \includegraphics[width=\textwidth]{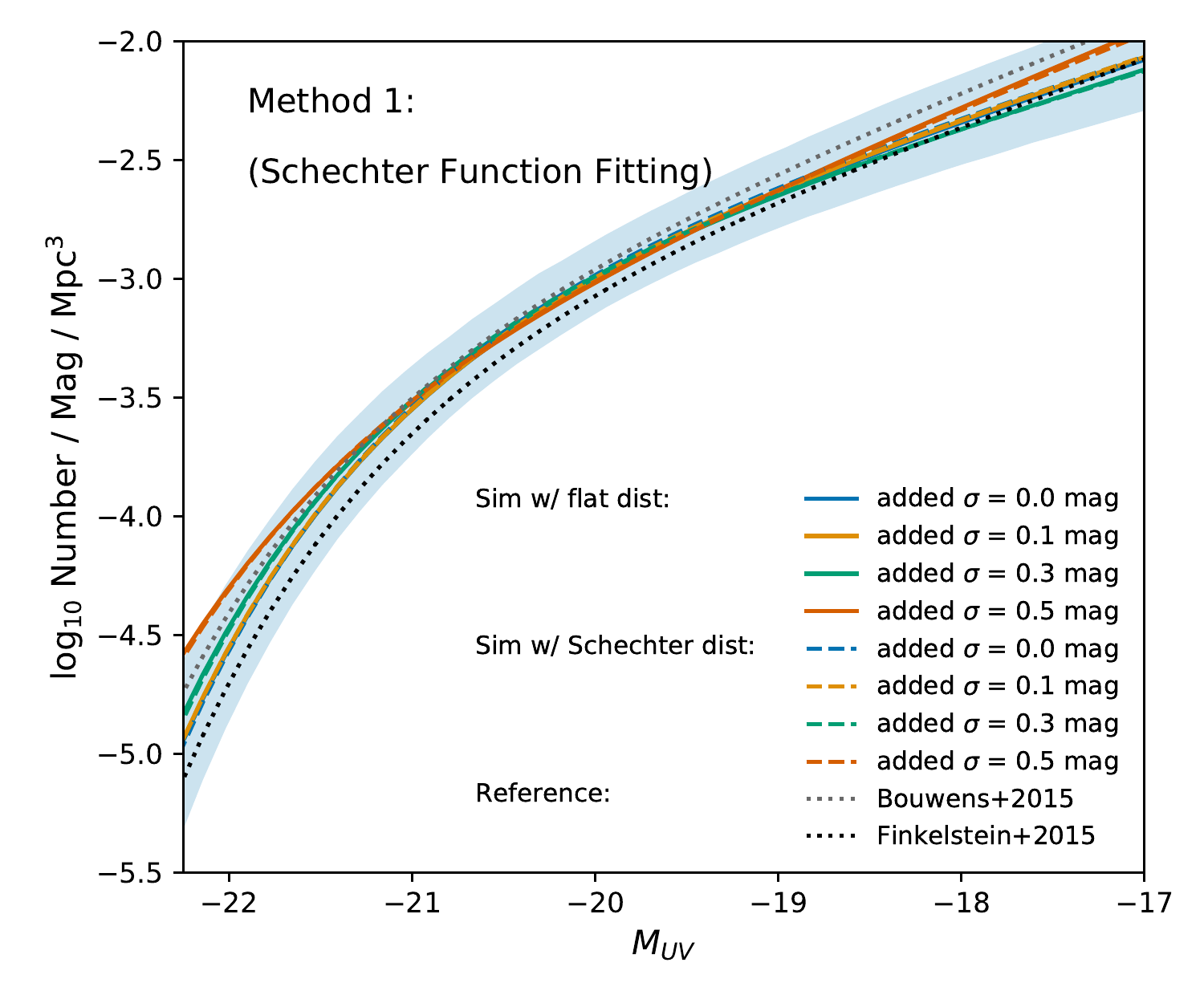}
    \caption{The best-fit UV luminosity functions for $z\sim5$ galaxies by method 1. The solid (dashed) lines show the best-fit functions derived from the completeness simulations with underlying flat (Schechter) distribution. The colours represent the best-fit functions derived from different data sets with different amount of flux scattering. We only shade $1\sigma$ uncertainty region for the line in the first entry of the legend. }
    \label{fig:schehter_method1}
\end{figure}

\begin{figure}
    \centering
    \includegraphics[width=\textwidth]{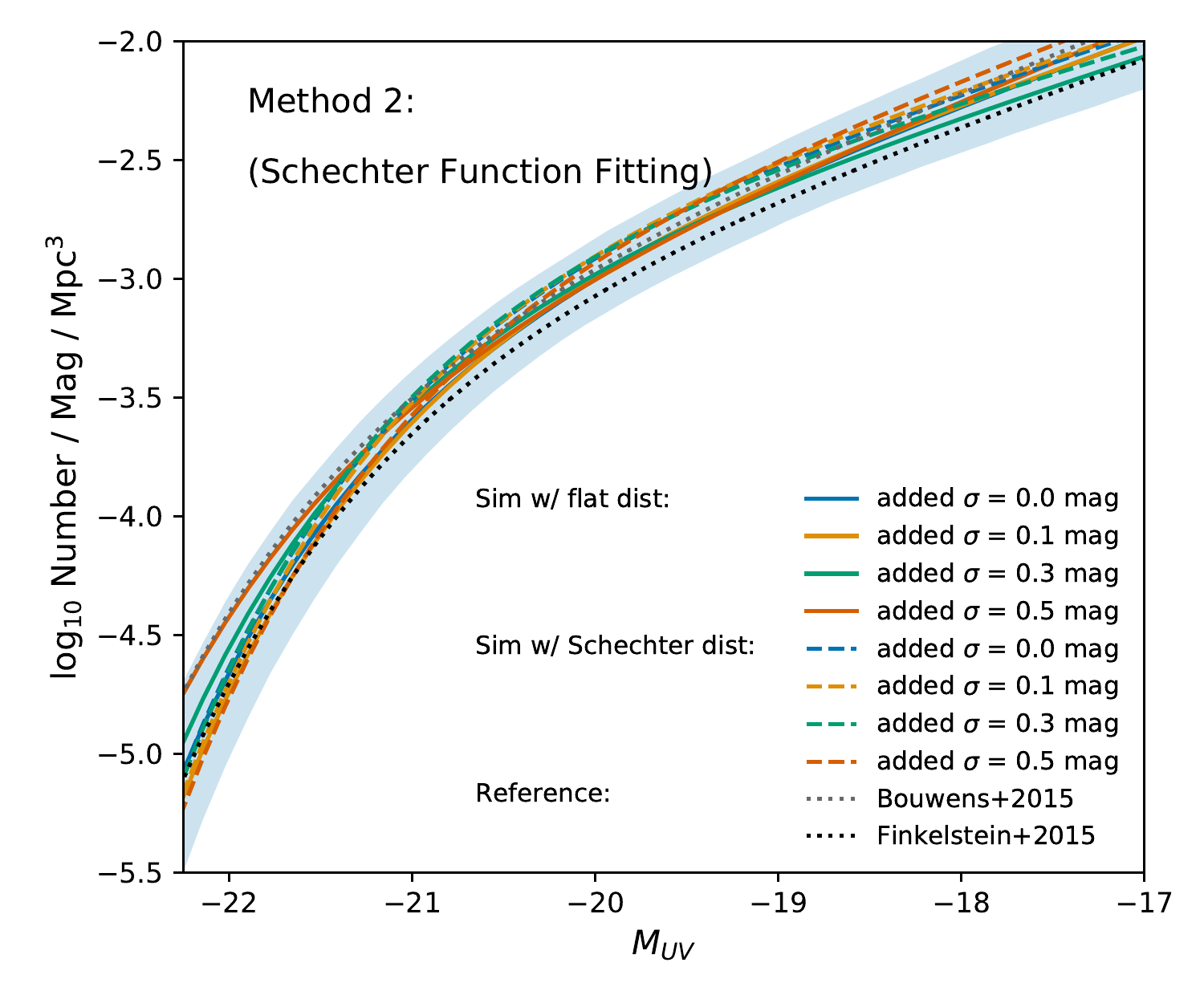}
    \caption{Same as Figure \ref{fig:schehter_method1} but for the best-fit functions derived with method 2. }
    \label{fig:schehter_method2}
\end{figure}

\begin{figure}
    \centering
    \includegraphics[width=\textwidth]{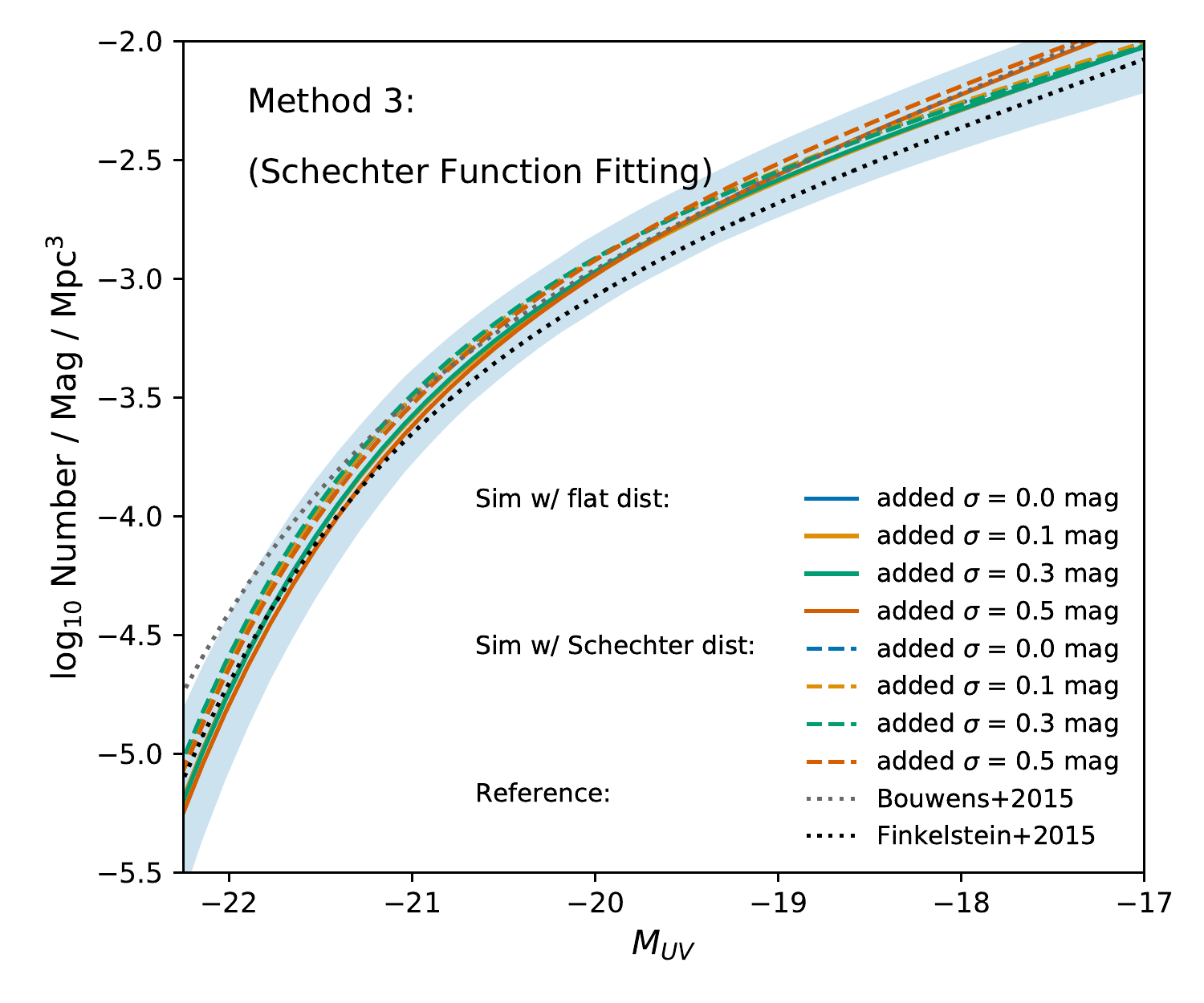}
    \caption{Same as Figure \ref{fig:schehter_method1} and \ref{fig:schehter_method2} but for the best-fit functions derived with method 3. }
    \label{fig:schehter_method3}
\end{figure}

We plot the best-fit Schechter functions of each method and simulation in Figure \ref{fig:schehter_method1} -- \ref{fig:schehter_method3}. Each colour represents a data set of certain flux scatter. The solid (dashed) lines show the best-fit functions derived from the completeness simulations with an underlying flat (Schechter) function. To avoid overfilling the figures, we only shade the $1\sigma$ uncertainty for one of the cases: best-fit function derived from the original data set with the completeness simulated with flat distribution (top entry in the legend). The uncertainties of other cases are similar in size. Generally, the best-fit Schechter functions are not significantly different from each other and are consistent with both best-fit functions derived in \citet{Finkelstein2015} and in \citet{Bouwens2015} within $2\sigma$. The  preferred UVLFs that are derived from the data without additional noise by each method (the blue dashed lines in Figure \ref{fig:schehter_method1} -- \ref{fig:schehter_method3}) are nearly identical to each other. However, at larger flux scatters, we observe the behaviors of the derived UVLFs that are predicted by our mock results in Section \ref{sec:mock_result} for all three methods. We emphasize that we do not aim to constrain the best UVLF at $z\sim5$ in this work but rather to compare the UVLFs derived with different methods given the same set of data (with an amount of flux scatter) and completeness simulations. 

The UVLFs derived from method 1 are as follows. First, the best-fit functions do not depend on the underlying distribution used in the completeness simulation. As seen in Figure \ref{fig:schehter_method1}, the best-fit functions derived from the completeness functions that are based on simulations with underlying Schechter distribution (the dashed lines) are essentially the same as the functions derived from the completeness functions that are based on simulations with underlying flat distribution (the solid lines) at the same flux scatter level. Second, the data with larger flux scatter tend to yield the best-fit functions with larger number densities at the bright end than those derived from the data with less flux scatter. For the best-fit Schechter functions, the change in the bright end starts to appear in the data with flux scatter of $>0.3$ mag, which is shown in the green and red lines in Figure \ref{fig:schehter_method1}). Both features are consistent with what expected based on Section \ref{sec:mock_result}.

Figure\ref{fig:schehter_method2} shows the best-fit functions derived with method 2. Based on Section \ref{sec:mock_result} we expect that, in presence of flux scatter, the derived UVLFs will differ from each other when the completeness function is based on a simulation with an underlying distribution that does not match the true luminosity function. If so, the derived UVLFs that are based on the simulations with an underlying flat distribution (solid lines) should show more variability with flux scatter than those derived with underlying Schechter distribution (dashed lines). We observe this behavior in Figure\ref{fig:schehter_method2}, indeed. The best-fit Schechter functions that were derived from simulations with an underlying Schechter distribution (dashed lines) do not differ from each other by more than 0.1 dex, regardless of the amount of the added flux scatters. The situation is different for those that were derived from simulations with an underlying flat function (solid lines). The number density of galaxies in the brightest bin ($M_\textrm{UV}=-22$) based on the best-fit function derived from the data with flux scatter $\sim0.5$ mag is $>0.2$ dex larger than the number density estimated by the best-fit functions derived from the data with less flux scatter. Again, these results are consistent with those in Section \ref{sec:mock_result}.  


Lastly, we show the best-fit functions derived with method 3 in Figure \ref{fig:schehter_method3}. Based on Section
\ref{sec:mock_result}, we expect this method to derive the same best-fit functions regardless of the underlying distribution used in the completeness simulation and of the amount of flux scatter. We find that all the derived UVLFs in Figure \ref{fig:schehter_method3} do not differ from each other by more than $\sim0.05$ dex in any magnitude bin. This level of variation is smaller than those of the UVLFs derived with other methods, especially at the bright end.

Since the derived UVLFs of different colours in Figure \ref{fig:schehter_method1}-\ref{fig:schehter_method3} are based on the same data sets and completeness simulations but have different levels of flux scatter, the derived UVLFs should not differ from each other if the determination is unbiased with respect to the flux scatter. We have shown that at low levels of flux scatter all methods work well. However, at larger flux scatters ($\gtrsim 0.3$ mag), method 1 and method 2 (only when the completeness function is based on a flat underlying distribution) yield UVLFs with higher number densities at the bright end. The larger the flux scatters, the higher the number densities at the bright end. These results confirm our earlier finding that method 3 is the most robust approach to derive UVLF in the presence of substantial photometric uncertainty (i.e. flux scatter). Nonetheless, we emphasize again that none of our derived UVLFs differ from each other by more than 2 $\sigma$. This is likely because the size of our tested observational data in this section is much smaller than the mock data in our experiment in Section \ref{sec:section3} (half sky), which suggests that small number statistics still dominates the derived high-z UVLFs in current literature. However, as observations are detecting more and more high-z galaxies, we expect that the relative contribution from Poisson noise will become smaller and thus systematic biases among UVLFs derived using different completeness methods may contribute much more significantly to the total uncertainty, as suggested by Section \ref{sec:mock_result}.

\begin{table*}[]
\renewcommand{\arraystretch}{1.7} %
\begin{tabular}{|c|c|c|c|c|c|c|c|c|c|}
\hline
\multirow{2}{*}{$\sigma_\textrm{added}=$} & \multirow{2}{*}{method} & \multicolumn{4}{c|}{Simulations with flat distribution}                                                                  & \multicolumn{4}{c|}{Simulations with Schechter distribution}                                                             \\ \cline{3-10} 
 &  & \multicolumn{1}{c|}{$\alpha$} & \multicolumn{1}{c|}{$\beta$} & \multicolumn{1}{c|}{$M^*$} & \multicolumn{1}{c|}{$\log\phi^*$} &\multicolumn{1}{c|}{$\alpha$} & \multicolumn{1}{c|}{$\beta$} & \multicolumn{1}{c|}{$M^*$} & \multicolumn{1}{c|}{$\log\phi^*$} \\ \hline

\multirow{6}{*}{0.0 mag}& \multirow{2}{*}{method1}& $-1.61 ^{+0.12}_{-0.12}$&--& $-20.84 ^{+0.19}_{-0.21}$& $-2.97 ^{+0.13}_{-0.14}$& \cellcolor{green!25}$-1.61 ^{+0.12}_{-0.12}$&\cellcolor{green!25}--&\cellcolor{green!25}$-20.81 ^{+0.18}_{-0.22}$& \cellcolor{green!25}$-2.94 ^{+0.12}_{-0.14}$\\ \cline{3-10}
 & & $-1.88 ^{+0.11}_{-0.10}$& $-5.17 ^{+0.70}_{-0.85}$& $-21.29 ^{+0.19}_{-0.17}$& $-3.46 ^{+0.14}_{-0.13}$& $-1.87 ^{+0.11}_{-0.10}$& $-5.05 ^{+0.67}_{-0.82}$& $-21.25 ^{+0.20}_{-0.18}$& $-3.43 ^{+0.14}_{-0.13}$\\ \cline{2-10}
& \multirow{2}{*}{method2}& $-1.69 ^{+0.10}_{-0.10}$&--& $-20.80 ^{+0.20}_{-0.21}$& $-2.98 ^{+0.13}_{-0.14}$& \cellcolor{green!25}$-1.61 ^{+0.08}_{-0.09}$&\cellcolor{green!25}--&\cellcolor{green!25}$-20.68 ^{+0.13}_{-0.19}$& \cellcolor{green!25}$-2.82 ^{+0.09}_{-0.12}$\\ \cline{3-10}
 & & $-1.86 ^{+0.11}_{-0.09}$& $-4.37 ^{+0.63}_{-0.87}$& $-21.01 ^{+0.28}_{-0.20}$& $-3.31 ^{+0.19}_{-0.15}$& $-1.74 ^{+0.13}_{-0.10}$& $-4.13 ^{+0.48}_{-0.65}$& $-20.80 ^{+0.28}_{-0.22}$& $-3.08 ^{+0.18}_{-0.15}$\\ \cline{2-10}
& \multirow{2}{*}{method3}& $-1.61 ^{+0.10}_{-0.11}$&--& $-20.64 ^{+0.16}_{-0.21}$& $-2.86 ^{+0.11}_{-0.14}$& \cellcolor{green!25}$-1.58 ^{+0.09}_{-0.10}$&\cellcolor{green!25}--&\cellcolor{green!25}$-20.68 ^{+0.12}_{-0.18}$& \cellcolor{green!25}$-2.81 ^{+0.08}_{-0.12}$\\ \cline{3-10}
 & & $-1.83 ^{+0.12}_{-0.09}$& $-5.18 ^{+0.84}_{-1.06}$& $-21.09 ^{+0.25}_{-0.19}$& $-3.34 ^{+0.18}_{-0.14}$& $-1.77 ^{+0.13}_{-0.10}$& $-5.00 ^{+0.71}_{-0.91}$& $-21.06 ^{+0.24}_{-0.19}$& $-3.24 ^{+0.17}_{-0.14}$\\ \hline
\multirow{6}{*}{0.1 mag}& \multirow{2}{*}{method1}& $-1.62 ^{+0.12}_{-0.12}$&--& $-20.85 ^{+0.20}_{-0.24}$& $-2.98 ^{+0.14}_{-0.17}$& \cellcolor{green!25}$-1.62 ^{+0.12}_{-0.12}$&\cellcolor{green!25}--&\cellcolor{green!25}$-20.85 ^{+0.21}_{-0.24}$& \cellcolor{green!25}$-2.97 ^{+0.15}_{-0.17}$\\ \cline{3-10}
 & & $-1.91 ^{+0.10}_{-0.09}$& $-5.45 ^{+0.80}_{-0.93}$& $-21.39 ^{+0.20}_{-0.17}$& $-3.55 ^{+0.16}_{-0.13}$& $-1.91 ^{+0.11}_{-0.10}$& $-5.34 ^{+0.79}_{-0.95}$& $-21.38 ^{+0.22}_{-0.17}$& $-3.54 ^{+0.17}_{-0.14}$\\ \cline{2-10}
& \multirow{2}{*}{method2}& $-1.66 ^{+0.10}_{-0.10}$&--& $-20.69 ^{+0.18}_{-0.21}$& $-2.91 ^{+0.12}_{-0.14}$& \cellcolor{green!25}$-1.61 ^{+0.06}_{-0.08}$&\cellcolor{green!25}--&\cellcolor{green!25}$-20.61 ^{+0.09}_{-0.16}$& \cellcolor{green!25}$-2.77 ^{+0.06}_{-0.09}$\\ \cline{3-10}
 & & $-1.83 ^{+0.11}_{-0.09}$& $-4.39 ^{+0.62}_{-0.87}$& $-20.92 ^{+0.27}_{-0.20}$& $-3.25 ^{+0.18}_{-0.14}$& $-1.72 ^{+0.13}_{-0.10}$& $-4.25 ^{+0.50}_{-0.65}$& $-20.74 ^{+0.25}_{-0.20}$& $-3.03 ^{+0.16}_{-0.14}$\\ \cline{2-10}
& \multirow{2}{*}{method3}& $-1.61 ^{+0.11}_{-0.11}$&--& $-20.65 ^{+0.16}_{-0.21}$& $-2.86 ^{+0.11}_{-0.14}$& \cellcolor{green!25}$-1.58 ^{+0.09}_{-0.11}$&\cellcolor{green!25}--&\cellcolor{green!25}$-20.68 ^{+0.12}_{-0.19}$& \cellcolor{green!25}$-2.81 ^{+0.08}_{-0.12}$\\ \cline{3-10}
 & & $-1.80 ^{+0.13}_{-0.10}$& $-4.86 ^{+0.73}_{-0.95}$& $-21.00 ^{+0.29}_{-0.21}$& $-3.28 ^{+0.20}_{-0.16}$& $-1.72 ^{+0.14}_{-0.11}$& $-4.69 ^{+0.61}_{-0.78}$& $-20.95 ^{+0.28}_{-0.21}$& $-3.16 ^{+0.19}_{-0.15}$\\ \hline
\multirow{6}{*}{0.3 mag}& \multirow{2}{*}{method1}& $-1.58 ^{+0.12}_{-0.12}$&--& $-20.91 ^{+0.18}_{-0.21}$& $-2.98 ^{+0.12}_{-0.14}$& \cellcolor{green!25}$-1.57 ^{+0.12}_{-0.12}$&\cellcolor{green!25}--&\cellcolor{green!25}$-20.87 ^{+0.18}_{-0.21}$& \cellcolor{green!25}$-2.96 ^{+0.12}_{-0.14}$\\ \cline{3-10}
 & & $-1.89 ^{+0.09}_{-0.08}$& $-6.12 ^{+0.88}_{-0.85}$& $-21.48 ^{+0.13}_{-0.12}$& $-3.57 ^{+0.10}_{-0.09}$& $-1.88 ^{+0.09}_{-0.09}$& $-5.97 ^{+0.88}_{-0.92}$& $-21.45 ^{+0.14}_{-0.13}$& $-3.54 ^{+0.11}_{-0.10}$\\ \cline{2-10}
& \multirow{2}{*}{method2}& $-1.60 ^{+0.10}_{-0.10}$&--& $-20.82 ^{+0.17}_{-0.19}$& $-2.94 ^{+0.12}_{-0.12}$& \cellcolor{green!25}$-1.55 ^{+0.06}_{-0.07}$&\cellcolor{green!25}--&\cellcolor{green!25}$-20.62 ^{+0.08}_{-0.12}$& \cellcolor{green!25}$-2.76 ^{+0.05}_{-0.08}$\\ \cline{3-10}
 & & $-1.84 ^{+0.10}_{-0.08}$& $-4.80 ^{+0.65}_{-0.90}$& $-21.20 ^{+0.22}_{-0.19}$& $-3.38 ^{+0.14}_{-0.13}$& $-1.72 ^{+0.12}_{-0.10}$& $-4.45 ^{+0.46}_{-0.58}$& $-20.87 ^{+0.22}_{-0.18}$& $-3.10 ^{+0.15}_{-0.13}$\\ \cline{2-10}
& \multirow{2}{*}{method3}& $-1.60 ^{+0.10}_{-0.11}$&--& $-20.63 ^{+0.15}_{-0.21}$& $-2.84 ^{+0.10}_{-0.14}$& \cellcolor{green!25}$-1.57 ^{+0.09}_{-0.10}$&\cellcolor{green!25}--&\cellcolor{green!25}$-20.69 ^{+0.12}_{-0.20}$& \cellcolor{green!25}$-2.80 ^{+0.08}_{-0.12}$\\ \cline{3-10}
 & & $-1.84 ^{+0.10}_{-0.08}$& $-6.19 ^{+1.09}_{-0.90}$& $-21.17 ^{+0.17}_{-0.15}$& $-3.39 ^{+0.13}_{-0.11}$& $-1.80 ^{+0.10}_{-0.08}$& $-6.10 ^{+1.04}_{-0.93}$& $-21.19 ^{+0.16}_{-0.14}$& $-3.32 ^{+0.12}_{-0.11}$\\ \hline
\multirow{6}{*}{0.5 mag}& \multirow{2}{*}{method1}& $-1.79 ^{+0.08}_{-0.08}$&--& $-21.38 ^{+0.21}_{-0.23}$& $-3.30 ^{+0.14}_{-0.15}$& \cellcolor{green!25}$-1.77 ^{+0.11}_{-0.10}$&\cellcolor{green!25}--&\cellcolor{green!25}$-21.33 ^{+0.22}_{-0.25}$& \cellcolor{green!25}$-3.26 ^{+0.15}_{-0.17}$\\ \cline{3-10}
 & & $-1.97 ^{+0.07}_{-0.06}$& $-5.33 ^{+0.90}_{-1.04}$& $-21.71 ^{+0.20}_{-0.15}$& $-3.70 ^{+0.13}_{-0.11}$& $-1.98 ^{+0.09}_{-0.08}$& $-5.26 ^{+0.89}_{-1.04}$& $-21.69 ^{+0.21}_{-0.17}$& $-3.70 ^{+0.15}_{-0.12}$\\ \cline{2-10}
& \multirow{2}{*}{method2}& $-1.78 ^{+0.08}_{-0.08}$&--& $-21.15 ^{+0.21}_{-0.21}$& $-3.18 ^{+0.14}_{-0.14}$& \cellcolor{green!25}$-1.70 ^{+0.08}_{-0.07}$&\cellcolor{green!25}--&\cellcolor{green!25}$-20.64 ^{+0.12}_{-0.13}$& \cellcolor{green!25}$-2.84 ^{+0.08}_{-0.09}$\\ \cline{3-10}
 & & $-1.86 ^{+0.15}_{-0.10}$& $-3.87 ^{+0.56}_{-0.82}$& $-21.11 ^{+0.56}_{-0.33}$& $-3.33 ^{+0.34}_{-0.22}$& $-1.90 ^{+0.11}_{-0.08}$& $-4.69 ^{+0.42}_{-0.55}$& $-20.93 ^{+0.25}_{-0.21}$& $-3.23 ^{+0.18}_{-0.16}$\\ \cline{2-10}
& \multirow{2}{*}{method3}& $-1.70 ^{+0.09}_{-0.09}$&--& $-20.67 ^{+0.19}_{-0.25}$& $-2.90 ^{+0.13}_{-0.17}$& \cellcolor{green!25}$-1.68 ^{+0.10}_{-0.10}$&\cellcolor{green!25}--&\cellcolor{green!25}$-20.74 ^{+0.17}_{-0.25}$& \cellcolor{green!25}$-2.87 ^{+0.12}_{-0.17}$\\ \cline{3-10}
 & & $-1.88 ^{+0.12}_{-0.08}$& $-5.35 ^{+1.22}_{-1.34}$& $-21.02 ^{+0.37}_{-0.22}$& $-3.33 ^{+0.25}_{-0.15}$& $-1.87 ^{+0.13}_{-0.09}$& $-5.51 ^{+1.20}_{-1.30}$& $-21.11 ^{+0.31}_{-0.21}$& $-3.31 ^{+0.22}_{-0.15}$\\ \hline

\end{tabular}
\caption{Best-fit parameters for the UVLFs derived with different methods for each combination of flux scatter and underlying distributions in the completeness simulation. For each method, we show the parameters for the best-fit Schechter parameters in the upper row and the best-fit double-power law parameters in the lower row. We highlight in green the preferred model according to the widely applicable Bayesian information criterion \citep{Watanabe2013}.}
\label{table:schechterparameters}
\end{table*}

\subsection{Summary}
\label{sec:observation_summary}
In summary, we use the images from the Hubble Legacy Fields (HLF) \citep{Whitaker2019} and separate them into two smaller regions according to their depths: the XDF and the GOODS region. We select $z\sim5$ LBG galaxies, using both Lyman-break selection and photo-z redshift criteria. We then measured the candidate's $M_\textrm{UV}$ from their best-fit spectra. We then create another three sets of data of added flux scatters in the measured $M_\textrm{UV}$ (0.1, 0.3 and 0.5 mag). 

We run two completeness simulations: one that injects galaxies with an underlying flat distribution, and one with an underlying Schechter distribution. We also add Gaussian noise to the measured $M_\textrm{UV}$ magnitudes in the completeness simulations to match each data set.
Based on the simulations, the final four sets of data have estimated flux scatter ranging from $\sim0.1$ to 0.5 mag in the brightest magnitude bins. For each simulation and added scatter level, we calculate the completeness functions according to the three methods considered in this work. Finally, we fit with two functional forms: a Schechter function and a double power-law function. In total, we have 48 best-fit functions (four flux scatter levels, two underlying distributions in the simulation, three definitions and two functional forms of the fit). 

We find that the Schecther function is preferred over the DPL function in all of the cases. Given the current Poisson noise from $\sim1000$ galaxies ($\lesssim100$ galaxies at the bright end), the fits generally do not differ from each other by more than $2\sigma$ when the flux scatter is $\lesssim$ 0.3 mag. However, they still show the expected behavior found in Section \ref{sec:mock_result}. Method 1 does not depend on the underlying Schechter distribution used in the completeness simulation but yields an excess at the bright end ($\sim0.2$ dex at $M_\textrm{UV}=-22$ mag) when flux scatters are larger than 0.3 mag. Method 2 tolerates flux scatter well if the completeness simulation simulates galaxies with the correct underlying distribution. Otherwise, it can yield an excess at the bright end if the flux scatter is greater than 0.3 mag. Method 3 shows the least change with flux scatter; their variation is within $\sim0.05$ dex across all considered magnitude bins.

\section{Conclusions}
We explore different definitions of the completeness function used in literature and how they affect the derived UV luminosity functions (UVLFs) in presence of flux scatter. We consider three definitions: (method 1) completeness as a function of intrinsic magnitude, (method 2) completeness as a function of recovered magnitude, and (method 3) completeness as a function of both intrinsic and recovered magnitude. In addition, we investigate another subtle difference in the implementation of the completeness simulation across literature, namely the specific shape of the underlying brightness distribution of the injected sources, as UVLF papers in the literature used completeness simulations with artificial galaxies drawn from different underlying distributions.

We first carry out our analysis using mock observations. We assume a mock universe with an intrinsic UVLF that is of the Schechter form. We also assume a mock survey with a limiting magnitude, which `observes' and `recovers' the galaxy luminosities with varying amount of flux scatter. We then create three sets of mock completeness simulations; each injects sources with a different underlying distribution. We found the following:
\begin{enumerate}
    \item The UVLFs derived with method 1 is sensitive to the amount of flux scatter. The derived UVLFs tend to be overestimated at the brightest magnitude bins, especially when flux scatter is $\gtrsim 0.2$ mag. This results in the UVLF whose bright-end slope ($\beta$) is flatter than the intrinsic value. The UVLFs derived with this method do not depend on the underlying distribution of the injected source in the completeness simulation.
    \item Method 2 is, in contrast, sensitive to the underlying distribution used in the completeness simulation when the flux scattering is presence. The derived UVLFs tend to bias toward the shape of the distribution used in the completeness simulation.
    \item Method 3 can recover the intrinsic UVLF well regardless of the flux scatter and the underlying distribution used in the completeness simulation.
\end{enumerate}
We further test the three methods with a direct application to measure the UVLF of $z\sim5$ galaxies from Hubble Legacy Field images with and without added flux scatter. The flux scatter in the derived UV magnitudes of galaxies in the original data set ranges from $\sigma=0.1$ mag in the brightest magnitude bins to $\sigma=0.2$ mag in the fainter bins. We added Gaussian noise to the derived UV magnitudes to create three more data sets with flux scatter of 0.1, 0.3, and 0.5 magnitudes. We found that all derived UVLFs agree with each other within $2\sigma$ uncertainties. However, at flux scatter $\gtrsim 0.3$ mag, the derived UVLFs with method 1 and method 2 demonstrate a trend that agrees well with the results from the mock observation. Method 3 is most robust against flux scatter and the underlying distribution of the injected sources.

Overall, our work indicates that caution is warranted in interpreting the bright-end excess found in high-redshift UVLF by some previous studies \citep[e.g.,][]{Calvi2016,Rojas-Ruiz2020}. Also, while at the moment the systematic bias introduced by a particular choice of completeness simulations is sub-dominant compared to small number statistics, once larger samples of high-redshift galaxies are discovered by the James Webb Space Telescope and the Nancy Grace Roman Space Telescope, we expect that different methods will yield UVLFs that may significantly disagree with each other, especially at the bright end. It is therefore important to keep flux uncertainties in the measured $M_\textrm{UV}$ to be $\lesssim0.2$ mag. This level of flux scatter is typically achievable at $z\lesssim6$ galaxies. However, this critical threshold is on par with typical flux uncertainties for $z>6$ galaxies. This emphasizes the importance of acquiring multi-band images that are sufficiently deep, and also to advance our understanding and modeling of spectra of high-z galaxies to accurately determine the galaxies' intrinsic UV luminosity from broadband photometry. In addition, to a lesser extent and to eliminate this incidental, we recommend the completeness definition and the UV derivation method that fully take both input and output properties into account, i.e. method 3.

To aid the community in reducing as much as possible systematic uncertainty deriving from different implementations of completeness corrections, we publicly release a completeness simulation code, GLACiAR2, that calculates the completeness function as a function of both intrinsic and recovered magnitudes (but it also has the flexibility to be used for methods 1 and 2). Also, the code can be easily modified to suit different sample selection criteria, therefore being an ideal tool to aid with future measurements of the UVLF during the epoch of reionization from upcoming James Webb Space Telescope NIRCAM observations. 

\section*{Acknowledgements}
The authors thank the anonymous referee for helpful suggestions and a careful reading of the paper. We thank Katherine Whitaker for helping us with the SExtractor procedures on the HLFs. We also thank Steven Finkelstein, Sofia Rojas-Ruiz, Rebecca Bowler, Pascal Oesch, and Rychard Bouwens for useful correspondence on UVLF derivation methods. This research was supported by the Australian Research Council Centre of Excellence for All Sky Astrophysics in 3 Dimensions (ASTRO 3D), through project number CE170100013. We acknowledge partial support from NASA through grant JWST-ERS-1342. Lastly, this research made use of Spartan, a High Performance Computing system operated by Research Computing Services at The University of Melbourne \citep{Lafayette2016}.

\section*{Data Availability}
The data underlying this article will be shared on reasonable request to the corresponding author.




\bibliographystyle{mnras}
\bibliography{completeness_arxiv} 

\begin{thebibliography}{}
\makeatletter
\relax
\def\mn@urlcharsother{\let\do\@makeother \do\$\do\&\do\#\do\^\do\_\do\%\do\~}
\def\mn@doi{\begingroup\mn@urlcharsother \@ifnextchar [ {\mn@doi@}
  {\mn@doi@[]}}
\def\mn@doi@[#1]#2{\def\@tempa{#1}\ifx\@tempa\@empty \href
  {http://dx.doi.org/#2} {doi:#2}\else \href {http://dx.doi.org/#2} {#1}\fi
  \endgroup}
\def\mn@eprint#1#2{\mn@eprint@#1:#2::\@nil}
\def\mn@eprint@arXiv#1{\href {http://arxiv.org/abs/#1} {{\tt arXiv:#1}}}
\def\mn@eprint@dblp#1{\href {http://dblp.uni-trier.de/rec/bibtex/#1.xml}
  {dblp:#1}}
\def\mn@eprint@#1:#2:#3:#4\@nil{\def\@tempa {#1}\def\@tempb {#2}\def\@tempc
  {#3}\ifx \@tempc \@empty \let \@tempc \@tempb \let \@tempb \@tempa \fi \ifx
  \@tempb \@empty \def\@tempb {arXiv}\fi \@ifundefined
  {mn@eprint@\@tempb}{\@tempb:\@tempc}{\expandafter \expandafter \csname
  mn@eprint@\@tempb\endcsname \expandafter{\@tempc}}}

\bibitem[\protect\citeauthoryear{{Atek}, {Richard}, {Kneib}  \&
  {Schaerer}}{{Atek} et~al.}{2018}]{Atek2018}
{Atek} H.,  {Richard} J.,  {Kneib} J.-P.,   {Schaerer} D.,  2018, \mn@doi
  [\mnras] {10.1093/mnras/sty1820}, \href
  {https://ui.adsabs.harvard.edu/abs/2018MNRAS.479.5184A} {479, 5184}

\bibitem[\protect\citeauthoryear{{Barone-Nugent}, {Wyithe}, {Trenti}, {Treu},
  {Oesch}, {Bouwens}, {Illingworth}  \& {Schmidt}}{{Barone-Nugent}
  et~al.}{2015}]{Barone-Nugent2015}
{Barone-Nugent} R.~L.,  {Wyithe} J.~S.~B.,  {Trenti} M.,  {Treu} T.,  {Oesch}
  P.,  {Bouwens} R.,  {Illingworth} G.~D.,   {Schmidt} K.~B.,  2015, \mn@doi
  [\mnras] {10.1093/mnras/stv633}, \href
  {https://ui.adsabs.harvard.edu/abs/2015MNRAS.450.1224B} {450, 1224}

\bibitem[\protect\citeauthoryear{Beckwith et~al.,}{Beckwith
  et~al.}{2006}]{Beckwith2006}
Beckwith S. V.~W.,  et~al., 2006, \mn@doi [The Astronomical Journal]
  {10.1086/507302}, 132, 1729

\bibitem[\protect\citeauthoryear{{Behroozi}, {Wechsler}  \&
  {Conroy}}{{Behroozi} et~al.}{2013}]{Behroozi2013}
{Behroozi} P.~S.,  {Wechsler} R.~H.,   {Conroy} C.,  2013, \mn@doi [\apj]
  {10.1088/0004-637X/770/1/57}, \href
  {https://ui.adsabs.harvard.edu/abs/2013ApJ...770...57B} {770, 57}

\bibitem[\protect\citeauthoryear{{Bernard} et~al.,}{{Bernard}
  et~al.}{2016}]{Bernard2016}
{Bernard} S.~R.,  et~al., 2016, \mn@doi [\apj] {10.3847/0004-637X/827/1/76},
  \href {https://ui.adsabs.harvard.edu/abs/2016ApJ...827...76B} {827, 76}

\bibitem[\protect\citeauthoryear{{Bertin} \& {Arnouts}}{{Bertin} \&
  {Arnouts}}{1996}]{BertinArnouts1996}
{Bertin} E.,  {Arnouts} S.,  1996, \mn@doi [\aaps] {10.1051/aas:1996164}, \href
  {https://ui.adsabs.harvard.edu/abs/1996A&AS..117..393B} {117, 393}

\bibitem[\protect\citeauthoryear{{Bouwens}, {Illingworth}, {Blakeslee},
  {Broadhurst}  \& {Franx}}{{Bouwens} et~al.}{2004}]{Bouwens2004}
{Bouwens} R.~J.,  {Illingworth} G.~D.,  {Blakeslee} J.~P.,  {Broadhurst} T.~J.,
    {Franx} M.,  2004, \mn@doi [\apjl] {10.1086/423786}, \href
  {https://ui.adsabs.harvard.edu/abs/2004ApJ...611L...1B} {611, L1}

\bibitem[\protect\citeauthoryear{{Bouwens}, {Illingworth}, {Blakeslee}  \&
  {Franx}}{{Bouwens} et~al.}{2006}]{Bouwens2006}
{Bouwens} R.~J.,  {Illingworth} G.~D.,  {Blakeslee} J.~P.,   {Franx} M.,  2006,
  \mn@doi [\apj] {10.1086/498733}, \href
  {https://ui.adsabs.harvard.edu/abs/2006ApJ...653...53B} {653, 53}

\bibitem[\protect\citeauthoryear{{Bouwens} et~al.,}{{Bouwens}
  et~al.}{2011}]{Bouwens2011}
{Bouwens} R.~J.,  et~al., 2011, \mn@doi [\apj] {10.1088/0004-637X/737/2/90},
  \href {https://ui.adsabs.harvard.edu/abs/2011ApJ...737...90B} {737, 90}

\bibitem[\protect\citeauthoryear{{Bouwens} et~al.,}{{Bouwens}
  et~al.}{2015}]{Bouwens2015}
{Bouwens} R.~J.,  et~al., 2015, \mn@doi [\apj] {10.1088/0004-637X/803/1/34},
  \href {https://ui.adsabs.harvard.edu/abs/2015ApJ...803...34B} {803, 34}

\bibitem[\protect\citeauthoryear{{Bouwens} et~al.,}{{Bouwens}
  et~al.}{2016}]{Bouwens2016}
{Bouwens} R.~J.,  et~al., 2016, \mn@doi [\apj] {10.3847/0004-637X/830/2/67},
  \href {https://ui.adsabs.harvard.edu/abs/2016ApJ...830...67B} {830, 67}

\bibitem[\protect\citeauthoryear{{Bouwens}, {Stefanon}, {Oesch}, {Illingworth},
  {Nanayakkara}, {Roberts-Borsani}, {Labb{\'e}}  \& {Smit}}{{Bouwens}
  et~al.}{2019}]{Bouwens2019}
{Bouwens} R.~J.,  {Stefanon} M.,  {Oesch} P.~A.,  {Illingworth} G.~D.,
  {Nanayakkara} T.,  {Roberts-Borsani} G.,  {Labb{\'e}} I.,   {Smit} R.,  2019,
  \mn@doi [\apj] {10.3847/1538-4357/ab24c5}, \href
  {https://ui.adsabs.harvard.edu/abs/2019ApJ...880...25B} {880, 25}

\bibitem[\protect\citeauthoryear{{Bouwens} et~al.,}{{Bouwens}
  et~al.}{2021}]{Bouwens2021}
{Bouwens} R.~J.,  et~al., 2021, arXiv e-prints, \href
  {https://ui.adsabs.harvard.edu/abs/2021arXiv210207775B} {p. arXiv:2102.07775}

\bibitem[\protect\citeauthoryear{{Bowler} et~al.,}{{Bowler}
  et~al.}{2014}]{Bowler2014}
{Bowler} R.~A.~A.,  et~al., 2014, \mn@doi [\mnras] {10.1093/mnras/stu449},
  \href {https://ui.adsabs.harvard.edu/abs/2014MNRAS.440.2810B} {440, 2810}

\bibitem[\protect\citeauthoryear{{Bowler} et~al.,}{{Bowler}
  et~al.}{2015}]{Bowler2015}
{Bowler} R.~A.~A.,  et~al., 2015, \mn@doi [\mnras] {10.1093/mnras/stv1403},
  \href {https://ui.adsabs.harvard.edu/abs/2015MNRAS.452.1817B} {452, 1817}

\bibitem[\protect\citeauthoryear{{Bowler}, {Jarvis}, {Dunlop}, {McLure},
  {McLeod}, {Adams}, {Milvang-Jensen}  \& {McCracken}}{{Bowler}
  et~al.}{2020}]{Bowler2020}
{Bowler} R.~A.~A.,  {Jarvis} M.~J.,  {Dunlop} J.~S.,  {McLure} R.~J.,  {McLeod}
  D.~J.,  {Adams} N.~J.,  {Milvang-Jensen} B.,   {McCracken} H.~J.,  2020,
  \mn@doi [\mnras] {10.1093/mnras/staa313}, \href
  {https://ui.adsabs.harvard.edu/abs/2020MNRAS.493.2059B} {493, 2059}

\bibitem[\protect\citeauthoryear{{Bradley} et~al.,}{{Bradley}
  et~al.}{2012}]{Bradley2012}
{Bradley} L.~D.,  et~al., 2012, \mn@doi [\apj] {10.1088/0004-637X/760/2/108},
  \href {https://ui.adsabs.harvard.edu/abs/2012ApJ...760..108B} {760, 108}

\bibitem[\protect\citeauthoryear{{Brammer}, {van Dokkum}  \& {Coppi}}{{Brammer}
  et~al.}{2008}]{Brammer2008}
{Brammer} G.~B.,  {van Dokkum} P.~G.,   {Coppi} P.,  2008, \mn@doi [\apj]
  {10.1086/591786}, \href
  {https://ui.adsabs.harvard.edu/abs/2008ApJ...686.1503B} {686, 1503}

\bibitem[\protect\citeauthoryear{{Bridge} et~al.,}{{Bridge}
  et~al.}{2019}]{Bridge2019}
{Bridge} J.~S.,  et~al., 2019, \mn@doi [\apj] {10.3847/1538-4357/ab3213}, \href
  {https://ui.adsabs.harvard.edu/abs/2019ApJ...882...42B} {882, 42}

\bibitem[\protect\citeauthoryear{{Brinchmann} et~al.,}{{Brinchmann}
  et~al.}{2017}]{Brinchmann2017}
{Brinchmann} J.,  et~al., 2017, \mn@doi [\aap] {10.1051/0004-6361/201731351},
  \href {https://ui.adsabs.harvard.edu/abs/2017A&A...608A...3B} {608, A3}

\bibitem[\protect\citeauthoryear{{Bruzual} \& {Charlot}}{{Bruzual} \&
  {Charlot}}{2003}]{BC03}
{Bruzual} G.,  {Charlot} S.,  2003, \mn@doi [\mnras]
  {10.1046/j.1365-8711.2003.06897.x}, \href
  {https://ui.adsabs.harvard.edu/abs/2003MNRAS.344.1000B} {344, 1000}

\bibitem[\protect\citeauthoryear{{Cai}, {Lapi}, {Bressan}, {De Zotti},
  {Negrello}  \& {Danese}}{{Cai} et~al.}{2014}]{Cai2014}
{Cai} Z.-Y.,  {Lapi} A.,  {Bressan} A.,  {De Zotti} G.,  {Negrello} M.,
  {Danese} L.,  2014, \mn@doi [\apj] {10.1088/0004-637X/785/1/65}, \href
  {https://ui.adsabs.harvard.edu/abs/2014ApJ...785...65C} {785, 65}

\bibitem[\protect\citeauthoryear{{Calvi} et~al.,}{{Calvi}
  et~al.}{2016}]{Calvi2016}
{Calvi} V.,  et~al., 2016, \mn@doi [\apj] {10.3847/0004-637X/817/2/120}, \href
  {https://ui.adsabs.harvard.edu/abs/2016ApJ...817..120C} {817, 120}

\bibitem[\protect\citeauthoryear{{Carrasco}, {Trenti}, {Mutch}  \&
  {Oesch}}{{Carrasco} et~al.}{2018}]{Carrasco2018}
{Carrasco} D.,  {Trenti} M.,  {Mutch} S.,   {Oesch} P.~A.,  2018, \mn@doi
  [\pasa] {10.1017/pasa.2018.17}, \href
  {https://ui.adsabs.harvard.edu/abs/2018PASA...35...22C} {35, e022}

\bibitem[\protect\citeauthoryear{{Cullen}, {McLure}, {Khochfar}, {Dunlop}  \&
  {Dalla Vecchia}}{{Cullen} et~al.}{2017}]{Cullen2017}
{Cullen} F.,  {McLure} R.~J.,  {Khochfar} S.,  {Dunlop} J.~S.,   {Dalla
  Vecchia} C.,  2017, \mn@doi [\mnras] {10.1093/mnras/stx1451}, \href
  {https://ui.adsabs.harvard.edu/abs/2017MNRAS.470.3006C} {470, 3006}

\bibitem[\protect\citeauthoryear{{Eddington}}{{Eddington}}{1913}]{Eddington1913}
{Eddington} A.~S.,  1913, \mn@doi [\mnras] {10.1093/mnras/73.5.359}, \href
  {https://ui.adsabs.harvard.edu/abs/1913MNRAS..73..359E} {73, 359}

\bibitem[\protect\citeauthoryear{{Erb}, {Pettini}, {Shapley}, {Steidel}, {Law}
  \& {Reddy}}{{Erb} et~al.}{2010}]{Erb2010}
{Erb} D.~K.,  {Pettini} M.,  {Shapley} A.~E.,  {Steidel} C.~C.,  {Law} D.~R.,
  {Reddy} N.~A.,  2010, \mn@doi [\apj] {10.1088/0004-637X/719/2/1168}, \href
  {https://ui.adsabs.harvard.edu/abs/2010ApJ...719.1168E} {719, 1168}

\bibitem[\protect\citeauthoryear{{Faucher-Gigu{\`e}re}, {Lidz}, {Hernquist}  \&
  {Zaldarriaga}}{{Faucher-Gigu{\`e}re} et~al.}{2008}]{FaucherGiguere2008}
{Faucher-Gigu{\`e}re} C.-A.,  {Lidz} A.,  {Hernquist} L.,   {Zaldarriaga} M.,
  2008, \mn@doi [\apj] {10.1086/592289}, \href
  {https://ui.adsabs.harvard.edu/abs/2008ApJ...688...85F} {688, 85}

\bibitem[\protect\citeauthoryear{{Finkelstein}}{{Finkelstein}}{2016}]{Finkelstein2016}
{Finkelstein} S.~L.,  2016, \mn@doi [\pasa] {10.1017/pasa.2016.26}, \href
  {https://ui.adsabs.harvard.edu/abs/2016PASA...33...37F} {33, e037}

\bibitem[\protect\citeauthoryear{{Finkelstein} et~al.,}{{Finkelstein}
  et~al.}{2015}]{Finkelstein2015}
{Finkelstein} S.~L.,  et~al., 2015, \mn@doi [\apj]
  {10.1088/0004-637X/810/1/71}, \href
  {https://ui.adsabs.harvard.edu/abs/2015ApJ...810...71F} {810, 71}

\bibitem[\protect\citeauthoryear{{Gnedin}}{{Gnedin}}{2016}]{Gnedin2016}
{Gnedin} N.~Y.,  2016, \mn@doi [\apjl] {10.3847/2041-8205/825/2/L17}, \href
  {https://ui.adsabs.harvard.edu/abs/2016ApJ...825L..17G} {825, L17}

\bibitem[\protect\citeauthoryear{{Haardt} \& {Madau}}{{Haardt} \&
  {Madau}}{2012}]{HaardtMadau2012}
{Haardt} F.,  {Madau} P.,  2012, \mn@doi [\apj] {10.1088/0004-637X/746/2/125},
  \href {https://ui.adsabs.harvard.edu/abs/2012ApJ...746..125H} {746, 125}

\bibitem[\protect\citeauthoryear{{Harikane} et~al.,}{{Harikane}
  et~al.}{2018}]{Harikane2018}
{Harikane} Y.,  et~al., 2018, \mn@doi [\pasj] {10.1093/pasj/psx097}, \href
  {https://ui.adsabs.harvard.edu/abs/2018PASJ...70S..11H} {70, S11}

\bibitem[\protect\citeauthoryear{{Hogg} \& {Turner}}{{Hogg} \&
  {Turner}}{1998}]{Hogg1998}
{Hogg} D.~W.,  {Turner} E.~L.,  1998, \mn@doi [\pasp] {10.1086/316173}, \href
  {https://ui.adsabs.harvard.edu/abs/1998PASP..110..727H} {110, 727}

\bibitem[\protect\citeauthoryear{{Hou}, {Aoyama}, {Hirashita}, {Nagamine}  \&
  {Shimizu}}{{Hou} et~al.}{2019}]{Hou2019}
{Hou} K.-C.,  {Aoyama} S.,  {Hirashita} H.,  {Nagamine} K.,   {Shimizu} I.,
  2019, \mn@doi [\mnras] {10.1093/mnras/stz121}, \href
  {https://ui.adsabs.harvard.edu/abs/2019MNRAS.485.1727H} {485, 1727}

\bibitem[\protect\citeauthoryear{{Huang}, {Ferguson}, {Ravindranath}  \&
  {Su}}{{Huang} et~al.}{2013}]{Huang2013}
{Huang} K.-H.,  {Ferguson} H.~C.,  {Ravindranath} S.,   {Su} J.,  2013, \mn@doi
  [\apj] {10.1088/0004-637X/765/1/68}, \href
  {https://ui.adsabs.harvard.edu/abs/2013ApJ...765...68H} {765, 68}

\bibitem[\protect\citeauthoryear{{Ilbert} et~al.,}{{Ilbert}
  et~al.}{2009}]{Ilbert2009}
{Ilbert} O.,  et~al., 2009, \mn@doi [\apj] {10.1088/0004-637X/690/2/1236},
  \href {https://ui.adsabs.harvard.edu/abs/2009ApJ...690.1236I} {690, 1236}

\bibitem[\protect\citeauthoryear{{Ishigaki}, {Kawamata}, {Ouchi}, {Oguri},
  {Shimasaku}  \& {Ono}}{{Ishigaki} et~al.}{2015}]{Ishigaki2015}
{Ishigaki} M.,  {Kawamata} R.,  {Ouchi} M.,  {Oguri} M.,  {Shimasaku} K.,
  {Ono} Y.,  2015, \mn@doi [\apj] {10.1088/0004-637X/799/1/12}, \href
  {https://ui.adsabs.harvard.edu/abs/2015ApJ...799...12I} {799, 12}

\bibitem[\protect\citeauthoryear{{Ishigaki}, {Kawamata}, {Ouchi}, {Oguri},
  {Shimasaku}  \& {Ono}}{{Ishigaki} et~al.}{2018}]{Ishigaki2018}
{Ishigaki} M.,  {Kawamata} R.,  {Ouchi} M.,  {Oguri} M.,  {Shimasaku} K.,
  {Ono} Y.,  2018, \mn@doi [\apj] {10.3847/1538-4357/aaa544}, \href
  {https://ui.adsabs.harvard.edu/abs/2018ApJ...854...73I} {854, 73}

\bibitem[\protect\citeauthoryear{{Khusanova} et~al.,}{{Khusanova}
  et~al.}{2020}]{Khusanova2020}
{Khusanova} Y.,  et~al., 2020, \mn@doi [\aap] {10.1051/0004-6361/201935400},
  \href {https://ui.adsabs.harvard.edu/abs/2020A&A...634A..97K} {634, A97}

\bibitem[\protect\citeauthoryear{{Kimm} \& {Cen}}{{Kimm} \&
  {Cen}}{2013}]{Kimm2013}
{Kimm} T.,  {Cen} R.,  2013, \mn@doi [\apj] {10.1088/0004-637X/776/1/35}, \href
  {https://ui.adsabs.harvard.edu/abs/2013ApJ...776...35K} {776, 35}

\bibitem[\protect\citeauthoryear{{Koekemoer} et~al.,}{{Koekemoer}
  et~al.}{2011}]{Koekemoer2011}
{Koekemoer} A.~M.,  et~al., 2011, \mn@doi [\apjs] {10.1088/0067-0049/197/2/36},
  \href {https://ui.adsabs.harvard.edu/abs/2011ApJS..197...36K} {197, 36}

\bibitem[\protect\citeauthoryear{{Lacey}, {Baugh}, {Frenk}  \&
  {Benson}}{{Lacey} et~al.}{2011}]{Lacey2011}
{Lacey} C.~G.,  {Baugh} C.~M.,  {Frenk} C.~S.,   {Benson} A.~J.,  2011, \mn@doi
  [\mnras] {10.1111/j.1365-2966.2010.18021.x}, \href
  {https://ui.adsabs.harvard.edu/abs/2011MNRAS.412.1828L} {412, 1828}

\bibitem[\protect\citeauthoryear{{Lafayette}, {Sauter}, {Vu}  \&
  {Meade}}{{Lafayette} et~al.}{2016}]{Lafayette2016}
{Lafayette} L.,  {Sauter} G.,  {Vu} L.,   {Meade} B.,  2016, \mn@doi [OpenStack
  Summit] {10.4225/49/58ead90dceaaa}

\bibitem[\protect\citeauthoryear{{Liu}, {Mutch}, {Poole}, {Angel}, {Duffy},
  {Geil}, {Mesinger}  \& {Wyithe}}{{Liu} et~al.}{2017}]{Liu2017}
{Liu} C.,  {Mutch} S.~J.,  {Poole} G.~B.,  {Angel} P.~W.,  {Duffy} A.~R.,
  {Geil} P.~M.,  {Mesinger} A.,   {Wyithe} J. S.~B.,  2017, \mn@doi [\mnras]
  {10.1093/mnras/stw2912}, \href
  {https://ui.adsabs.harvard.edu/abs/2017MNRAS.465.3134L} {465, 3134}

\bibitem[\protect\citeauthoryear{{Livermore}, {Finkelstein}  \&
  {Lotz}}{{Livermore} et~al.}{2017}]{Livermore2017}
{Livermore} R.~C.,  {Finkelstein} S.~L.,   {Lotz} J.~M.,  2017, \mn@doi [\apj]
  {10.3847/1538-4357/835/2/113}, \href
  {https://ui.adsabs.harvard.edu/abs/2017ApJ...835..113L} {835, 113}

\bibitem[\protect\citeauthoryear{{Ma} et~al.,}{{Ma} et~al.}{2018}]{Ma2018}
{Ma} X.,  et~al., 2018, \mn@doi [\mnras] {10.1093/mnras/sty1024}, \href
  {https://ui.adsabs.harvard.edu/abs/2018MNRAS.478.1694M} {478, 1694}

\bibitem[\protect\citeauthoryear{{Ma} et~al.,}{{Ma} et~al.}{2019}]{Ma2019}
{Ma} X.,  et~al., 2019, \mn@doi [\mnras] {10.1093/mnras/stz1324}, \href
  {https://ui.adsabs.harvard.edu/abs/2019MNRAS.487.1844M} {487, 1844}

\bibitem[\protect\citeauthoryear{{Ma}, {Quataert}, {Wetzel}, {Hopkins},
  {Faucher-Gigu{\`e}re}  \& {Kere{\v{s}}}}{{Ma} et~al.}{2020}]{Ma2020}
{Ma} X.,  {Quataert} E.,  {Wetzel} A.,  {Hopkins} P.~F.,  {Faucher-Gigu{\`e}re}
  C.-A.,   {Kere{\v{s}}} D.,  2020, \mn@doi [\mnras] {10.1093/mnras/staa2404},
  \href {https://ui.adsabs.harvard.edu/abs/2020MNRAS.498.2001M} {498, 2001}

\bibitem[\protect\citeauthoryear{{Maraston}}{{Maraston}}{2005}]{Maraston2005}
{Maraston} C.,  2005, \mn@doi [\mnras] {10.1111/j.1365-2966.2005.09270.x},
  \href {https://ui.adsabs.harvard.edu/abs/2005MNRAS.362..799M} {362, 799}

\bibitem[\protect\citeauthoryear{{Mashian}, {Oesch}  \& {Loeb}}{{Mashian}
  et~al.}{2016}]{Mashian2016}
{Mashian} N.,  {Oesch} P.~A.,   {Loeb} A.,  2016, \mn@doi [\mnras]
  {10.1093/mnras/stv2469}, \href
  {https://ui.adsabs.harvard.edu/abs/2016MNRAS.455.2101M} {455, 2101}

\bibitem[\protect\citeauthoryear{{Mason} et~al.,}{{Mason}
  et~al.}{2015a}]{Mason2015}
{Mason} C.~A.,  et~al., 2015a, \mn@doi [\apj] {10.1088/0004-637X/805/1/79},
  \href {https://ui.adsabs.harvard.edu/abs/2015ApJ...805...79M} {805, 79}

\bibitem[\protect\citeauthoryear{{Mason}, {Trenti}  \& {Treu}}{{Mason}
  et~al.}{2015b}]{Mason2015b}
{Mason} C.~A.,  {Trenti} M.,   {Treu} T.,  2015b, \mn@doi [\apj]
  {10.1088/0004-637X/813/1/21}, \href
  {https://ui.adsabs.harvard.edu/abs/2015ApJ...813...21M} {813, 21}

\bibitem[\protect\citeauthoryear{{Mason}, {Naidu}, {Tacchella}  \&
  {Leja}}{{Mason} et~al.}{2019}]{Mason2019}
{Mason} C.~A.,  {Naidu} R.~P.,  {Tacchella} S.,   {Leja} J.,  2019, \mn@doi
  [\mnras] {10.1093/mnras/stz2291}, \href
  {https://ui.adsabs.harvard.edu/abs/2019MNRAS.489.2669M} {489, 2669}

\bibitem[\protect\citeauthoryear{{McLure}, {Cirasuolo}, {Dunlop}, {Foucaud}  \&
  {Almaini}}{{McLure} et~al.}{2009}]{McLure2009}
{McLure} R.~J.,  {Cirasuolo} M.,  {Dunlop} J.~S.,  {Foucaud} S.,   {Almaini}
  O.,  2009, \mn@doi [\mnras] {10.1111/j.1365-2966.2009.14677.x}, \href
  {https://ui.adsabs.harvard.edu/abs/2009MNRAS.395.2196M} {395, 2196}

\bibitem[\protect\citeauthoryear{{McLure} et~al.,}{{McLure}
  et~al.}{2013}]{McLure2013}
{McLure} R.~J.,  et~al., 2013, \mn@doi [\mnras] {10.1093/mnras/stt627}, \href
  {https://ui.adsabs.harvard.edu/abs/2013MNRAS.432.2696M} {432, 2696}

\bibitem[\protect\citeauthoryear{{Morishita}}{{Morishita}}{2021}]{Morishita2021}
{Morishita} T.,  2021, \mn@doi [\apjs] {10.3847/1538-4365/abce67}, \href
  {https://ui.adsabs.harvard.edu/abs/2021ApJS..253....4M} {253, 4}

\bibitem[\protect\citeauthoryear{{Morishita} et~al.,}{{Morishita}
  et~al.}{2018}]{Morishita2018}
{Morishita} T.,  et~al., 2018, \mn@doi [\apj] {10.3847/1538-4357/aae68c}, \href
  {https://ui.adsabs.harvard.edu/abs/2018ApJ...867..150M} {867, 150}

\bibitem[\protect\citeauthoryear{{Moutard} et~al.,}{{Moutard}
  et~al.}{2016}]{Moutard2016}
{Moutard} T.,  et~al., 2016, \mn@doi [\aap] {10.1051/0004-6361/201527294},
  \href {https://ui.adsabs.harvard.edu/abs/2016A&A...590A.103M} {590, A103}

\bibitem[\protect\citeauthoryear{{Oesch} et~al.,}{{Oesch}
  et~al.}{2007}]{Oesch2007}
{Oesch} P.~A.,  et~al., 2007, \mn@doi [\apj] {10.1086/522423}, \href
  {https://ui.adsabs.harvard.edu/abs/2007ApJ...671.1212O} {671, 1212}

\bibitem[\protect\citeauthoryear{{Oesch} et~al.,}{{Oesch}
  et~al.}{2009}]{Oesch2009}
{Oesch} P.~A.,  et~al., 2009, \mn@doi [\apj] {10.1088/0004-637X/690/2/1350},
  \href {https://ui.adsabs.harvard.edu/abs/2009ApJ...690.1350O} {690, 1350}

\bibitem[\protect\citeauthoryear{{Oesch} et~al.,}{{Oesch}
  et~al.}{2010}]{Oesch2010}
{Oesch} P.~A.,  et~al., 2010, \mn@doi [\apjl] {10.1088/2041-8205/709/1/L21},
  \href {https://ui.adsabs.harvard.edu/abs/2010ApJ...709L..21O} {709, L21}

\bibitem[\protect\citeauthoryear{{Oesch} et~al.,}{{Oesch}
  et~al.}{2012}]{Oesch2012}
{Oesch} P.~A.,  et~al., 2012, \mn@doi [\apj] {10.1088/0004-637X/759/2/135},
  \href {https://ui.adsabs.harvard.edu/abs/2012ApJ...759..135O} {759, 135}

\bibitem[\protect\citeauthoryear{{Oesch}, {Bouwens}, {Illingworth}, {Labb{\'e}}
   \& {Stefanon}}{{Oesch} et~al.}{2018}]{Oesch2018}
{Oesch} P.~A.,  {Bouwens} R.~J.,  {Illingworth} G.~D.,  {Labb{\'e}} I.,
  {Stefanon} M.,  2018, \mn@doi [\apj] {10.3847/1538-4357/aab03f}, \href
  {https://ui.adsabs.harvard.edu/abs/2018ApJ...855..105O} {855, 105}

\bibitem[\protect\citeauthoryear{{Ono} et~al.,}{{Ono} et~al.}{2018}]{Ono2018}
{Ono} Y.,  et~al., 2018, \mn@doi [\pasj] {10.1093/pasj/psx103}, \href
  {https://ui.adsabs.harvard.edu/abs/2018PASJ...70S..10O} {70, S10}

\bibitem[\protect\citeauthoryear{{Ren}, {Trenti}  \& {Mason}}{{Ren}
  et~al.}{2019}]{Ren2019}
{Ren} K.,  {Trenti} M.,   {Mason} C.~A.,  2019, \mn@doi [\apj]
  {10.3847/1538-4357/ab2117}, \href
  {https://ui.adsabs.harvard.edu/abs/2019ApJ...878..114R} {878, 114}

\bibitem[\protect\citeauthoryear{{Roberts-Borsani}, {Morishita}, {Treu},
  {Leethochawalit}  \& {Trenti}}{{Roberts-Borsani}
  et~al.}{2021}]{Roberts-Borsani2021}
{Roberts-Borsani} G.,  {Morishita} T.,  {Treu} T.,  {Leethochawalit} N.,
  {Trenti} M.,  2021, arXiv e-prints, \href
  {https://ui.adsabs.harvard.edu/abs/2021arXiv210606544R} {p. arXiv:2106.06544}

\bibitem[\protect\citeauthoryear{{Robertson}, {Ellis}, {Furlanetto}  \&
  {Dunlop}}{{Robertson} et~al.}{2015}]{Robertson2015}
{Robertson} B.~E.,  {Ellis} R.~S.,  {Furlanetto} S.~R.,   {Dunlop} J.~S.,
  2015, \mn@doi [\apjl] {10.1088/2041-8205/802/2/L19}, \href
  {https://ui.adsabs.harvard.edu/abs/2015ApJ...802L..19R} {802, L19}

\bibitem[\protect\citeauthoryear{{Rojas-Ruiz}, {Finkelstein}, {Bagley},
  {Stevans}, {Finkelstein}, {Larson}, {Mechtley}  \& {Diekmann}}{{Rojas-Ruiz}
  et~al.}{2020}]{Rojas-Ruiz2020}
{Rojas-Ruiz} S.,  {Finkelstein} S.~L.,  {Bagley} M.~B.,  {Stevans} M.,
  {Finkelstein} K.~D.,  {Larson} R.,  {Mechtley} M.,   {Diekmann} J.,  2020,
  \mn@doi [\apj] {10.3847/1538-4357/ab7659}, \href
  {https://ui.adsabs.harvard.edu/abs/2020ApJ...891..146R} {891, 146}

\bibitem[\protect\citeauthoryear{{Salmon} et~al.,}{{Salmon}
  et~al.}{2020}]{Salmon2020}
{Salmon} B.,  et~al., 2020, \mn@doi [\apj] {10.3847/1538-4357/ab5a8b}, \href
  {https://ui.adsabs.harvard.edu/abs/2020ApJ...889..189S} {889, 189}

\bibitem[\protect\citeauthoryear{{Schenker} et~al.,}{{Schenker}
  et~al.}{2013}]{Schenker2013}
{Schenker} M.~A.,  et~al., 2013, \mn@doi [\apj] {10.1088/0004-637X/768/2/196},
  \href {https://ui.adsabs.harvard.edu/abs/2013ApJ...768..196S} {768, 196}

\bibitem[\protect\citeauthoryear{{Schlafly} \& {Finkbeiner}}{{Schlafly} \&
  {Finkbeiner}}{2011}]{SchlaflyFinkbeiner2011}
{Schlafly} E.~F.,  {Finkbeiner} D.~P.,  2011, \mn@doi [\apj]
  {10.1088/0004-637X/737/2/103}, \href
  {https://ui.adsabs.harvard.edu/abs/2011ApJ...737..103S} {737, 103}

\bibitem[\protect\citeauthoryear{{Schlegel}, {Finkbeiner}  \&
  {Davis}}{{Schlegel} et~al.}{1998}]{Schlegel1998}
{Schlegel} D.~J.,  {Finkbeiner} D.~P.,   {Davis} M.,  1998, \mn@doi [\apj]
  {10.1086/305772}, \href
  {https://ui.adsabs.harvard.edu/abs/1998ApJ...500..525S} {500, 525}

\bibitem[\protect\citeauthoryear{{Schmidt} et~al.,}{{Schmidt}
  et~al.}{2014}]{Schmidt2014}
{Schmidt} K.~B.,  et~al., 2014, \mn@doi [\apj] {10.1088/0004-637X/786/1/57},
  \href {https://ui.adsabs.harvard.edu/abs/2014ApJ...786...57S} {786, 57}

\bibitem[\protect\citeauthoryear{{Shibuya}, {Ouchi}  \& {Harikane}}{{Shibuya}
  et~al.}{2015}]{Shibuya2015}
{Shibuya} T.,  {Ouchi} M.,   {Harikane} Y.,  2015, \mn@doi [\apjs]
  {10.1088/0067-0049/219/2/15}, \href
  {https://ui.adsabs.harvard.edu/abs/2015ApJS..219...15S} {219, 15}

\bibitem[\protect\citeauthoryear{{Somerville}, {Gilmore}, {Primack}  \&
  {Dom{\'\i}nguez}}{{Somerville} et~al.}{2012}]{Somerville2012}
{Somerville} R.~S.,  {Gilmore} R.~C.,  {Primack} J.~R.,   {Dom{\'\i}nguez} A.,
  2012, \mn@doi [\mnras] {10.1111/j.1365-2966.2012.20490.x}, \href
  {https://ui.adsabs.harvard.edu/abs/2012MNRAS.423.1992S} {423, 1992}

\bibitem[\protect\citeauthoryear{{Stefanon}, {Bouwens}, {Labb{\'e}}, {Muzzin},
  {Marchesini}, {Oesch}  \& {Gonzalez}}{{Stefanon}
  et~al.}{2017}]{Stefanon2017a}
{Stefanon} M.,  {Bouwens} R.~J.,  {Labb{\'e}} I.,  {Muzzin} A.,  {Marchesini}
  D.,  {Oesch} P.,   {Gonzalez} V.,  2017, \mn@doi [\apj]
  {10.3847/1538-4357/aa72d8}, \href
  {https://ui.adsabs.harvard.edu/abs/2017ApJ...843...36S} {843, 36}

\bibitem[\protect\citeauthoryear{{Stefanon} et~al.,}{{Stefanon}
  et~al.}{2019}]{Stefanon2019}
{Stefanon} M.,  et~al., 2019, \mn@doi [\apj] {10.3847/1538-4357/ab3792}, \href
  {https://ui.adsabs.harvard.edu/abs/2019ApJ...883...99S} {883, 99}

\bibitem[\protect\citeauthoryear{{Tacchella}, {Bose}, {Conroy}, {Eisenstein}
  \& {Johnson}}{{Tacchella} et~al.}{2018}]{Tacchella2018}
{Tacchella} S.,  {Bose} S.,  {Conroy} C.,  {Eisenstein} D.~J.,   {Johnson}
  B.~D.,  2018, \mn@doi [\apj] {10.3847/1538-4357/aae8e0}, \href
  {https://ui.adsabs.harvard.edu/abs/2018ApJ...868...92T} {868, 92}

\bibitem[\protect\citeauthoryear{{Trenti} \& {Stiavelli}}{{Trenti} \&
  {Stiavelli}}{2008}]{Trenti2008}
{Trenti} M.,  {Stiavelli} M.,  2008, \mn@doi [\apj] {10.1086/528674}, \href
  {https://ui.adsabs.harvard.edu/abs/2008ApJ...676..767T} {676, 767}

\bibitem[\protect\citeauthoryear{{Trenti}, {Stiavelli}, {Bouwens}, {Oesch},
  {Shull}, {Illingworth}, {Bradley}  \& {Carollo}}{{Trenti}
  et~al.}{2010}]{Trenti2010}
{Trenti} M.,  {Stiavelli} M.,  {Bouwens} R.~J.,  {Oesch} P.,  {Shull} J.~M.,
  {Illingworth} G.~D.,  {Bradley} L.~D.,   {Carollo} C.~M.,  2010, \mn@doi
  [\apjl] {10.1088/2041-8205/714/2/L202}, \href
  {https://ui.adsabs.harvard.edu/abs/2010ApJ...714L.202T} {714, L202}

\bibitem[\protect\citeauthoryear{{Trenti} et~al.,}{{Trenti}
  et~al.}{2011}]{Trenti2011}
{Trenti} M.,  et~al., 2011, \mn@doi [\apjl] {10.1088/2041-8205/727/2/L39},
  \href {https://ui.adsabs.harvard.edu/abs/2011ApJ...727L..39T} {727, L39}

\bibitem[\protect\citeauthoryear{Watanabe}{Watanabe}{2013}]{Watanabe2013}
Watanabe S.,  2013, J. Mach. Learn. Res., 14, 867–897

\bibitem[\protect\citeauthoryear{{Whitaker} et~al.,}{{Whitaker}
  et~al.}{2011}]{Whitaker2011}
{Whitaker} K.~E.,  et~al., 2011, \mn@doi [\apj] {10.1088/0004-637X/735/2/86},
  \href {https://ui.adsabs.harvard.edu/abs/2011ApJ...735...86W} {735, 86}

\bibitem[\protect\citeauthoryear{{Whitaker} et~al.,}{{Whitaker}
  et~al.}{2019}]{Whitaker2019}
{Whitaker} K.~E.,  et~al., 2019, \mn@doi [\apjs] {10.3847/1538-4365/ab3853},
  \href {https://ui.adsabs.harvard.edu/abs/2019ApJS..244...16W} {244, 16}

\bibitem[\protect\citeauthoryear{Wilkins, Feng, Di~Matteo, Croft, Lovell  \&
  Waters}{Wilkins et~al.}{2017}]{Wilkins2017}
Wilkins S.~M.,  Feng Y.,  Di~Matteo T.,  Croft R.,  Lovell C.~C.,   Waters D.,
  2017, \mn@doi [Monthly Notices of the Royal Astronomical Society]
  {10.1093/mnras/stx841}, 469, 2517

\bibitem[\protect\citeauthoryear{{Windhorst} et~al.,}{{Windhorst}
  et~al.}{2011}]{Windhorst2011}
{Windhorst} R.~A.,  et~al., 2011, \mn@doi [\apjs] {10.1088/0067-0049/193/2/27},
  \href {https://ui.adsabs.harvard.edu/abs/2011ApJS..193...27W} {193, 27}

\bibitem[\protect\citeauthoryear{{Wyithe}, {Yan}, {Windhorst}  \&
  {Mao}}{{Wyithe} et~al.}{2011}]{Wyithe2011}
{Wyithe} J. S.~B.,  {Yan} H.,  {Windhorst} R.~A.,   {Mao} S.,  2011, \mn@doi
  [\nat] {10.1038/nature09619}, \href
  {https://ui.adsabs.harvard.edu/abs/2011Natur.469..181W} {469, 181}

\bibitem[\protect\citeauthoryear{{Yan} et~al.,}{{Yan} et~al.}{2012}]{Yan2012}
{Yan} H.,  et~al., 2012, \mn@doi [\apj] {10.1088/0004-637X/761/2/177}, \href
  {https://ui.adsabs.harvard.edu/abs/2012ApJ...761..177Y} {761, 177}

\bibitem[\protect\citeauthoryear{{Yue}, {Ferrara}  \& {Xu}}{{Yue}
  et~al.}{2016}]{Yue2016}
{Yue} B.,  {Ferrara} A.,   {Xu} Y.,  2016, \mn@doi [\mnras]
  {10.1093/mnras/stw2145}, \href
  {https://ui.adsabs.harvard.edu/abs/2016MNRAS.463.1968Y} {463, 1968}

\makeatother
\end{thebibliography}




\appendix
\section{Flux uncertainty calculation}
\label{appendix:flux_uncertainties}
We use the ``empty aperture" method to calculate all flux uncertainties in this work. The procedure is similar to that in \citet{Whitaker2019} (see also \citealt{Trenti2011}). For each filter and region, we create a noise-equalized image (science image divided by its RMS map or a signal-to-noise map). We then place 500 random circular apertures of a fixed diameter in empty region of the field, i.e. those random apertures must not overlap with any SExtracted sources according to the SExtractor's segmentation images. We then measure their median absolute deviation ($\sigma_\textrm{NMAD}$) for that aperture diameter. This process is repeated for a range of aperture diameter: from $0\arcsec.1$ to $1\arcsec$. We then fit this relation with a power-law function:
\begin{equation}
    \sigma_\textrm{NMAD}(N) = \alpha N^\beta
\end{equation}
, where N is defined as the square root of the number of pixels within the aperture \citep{Whitaker2011}. For example, a circular aperture of radius $r$ pixels, N would be equal to $\sqrt{\pi r^2}$. 

\begin{figure}
    \centering
    \includegraphics[width=\textwidth]{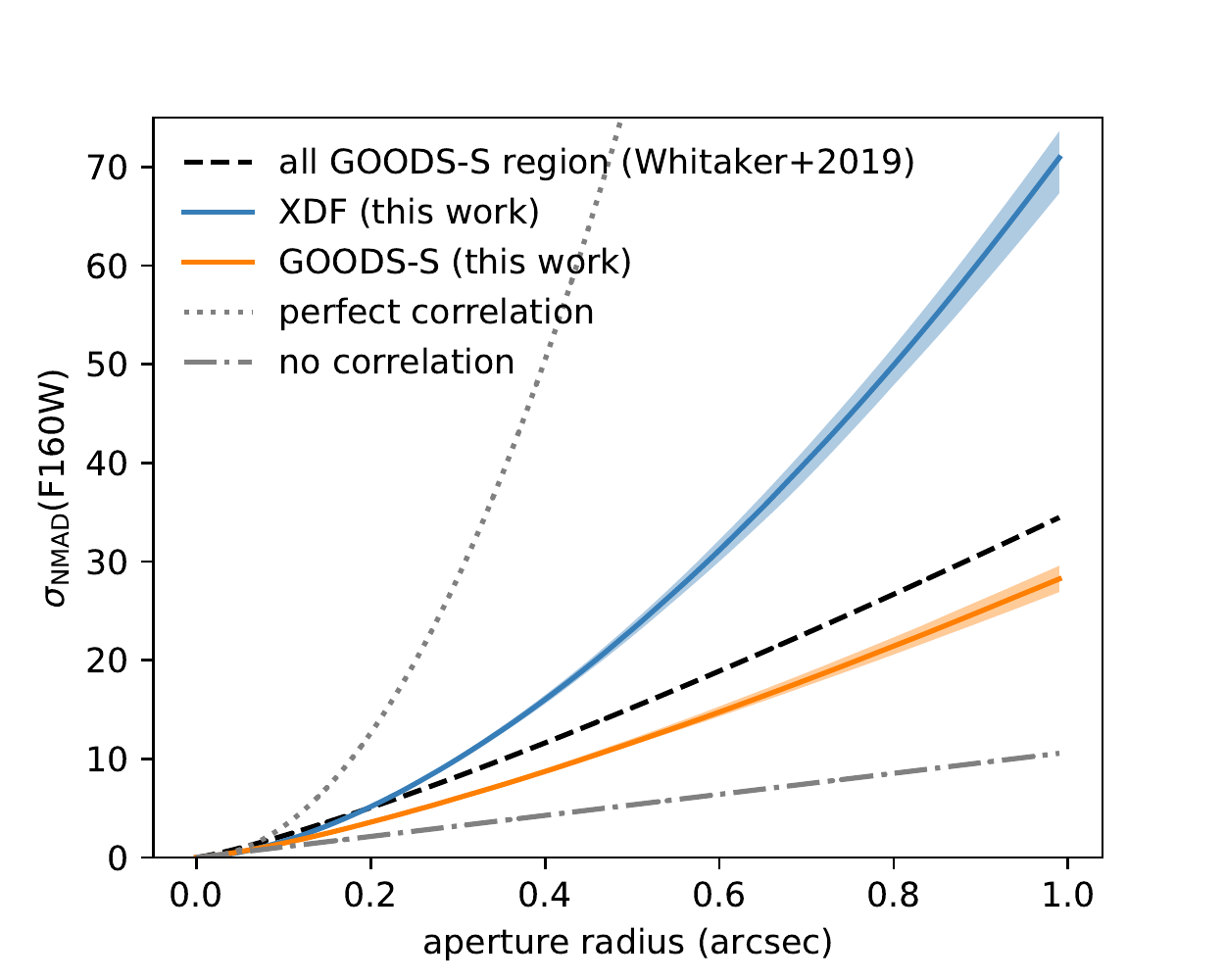}
    \caption{Normalized median absolute deviation, $\sigma_\textrm{NMAD}$, as a function of aperture for the F160W image of each field in the GOODS-S region. The solid line shows the power-law fit derived in this work. The dashed lines shows the fit derived in \citet{Whitaker2019} using the whole GOODS-S field (without splitting into regions of different depths). These $\sigma_\textrm{NMAD}$ relations are used to derived flux uncertainties in this work.}
    \label{fig:nmad}
\end{figure}

The best-fit $\sigma_\textrm{NMAD}$ relations for F160W band of both fields are shown in Figure \ref{fig:nmad}. The interpretation of this figure is that if there is no correlation among the pixels, the relation would be linear i.e. $\beta=1$ (dash-dotted line). If there is a perfect correlations within each aperture, the relation would have $\beta=2$ (dotted line). The $\beta$ value for the XDF region ($\beta=1.63\pm0.04$) is greater than the $\beta$ value for the GOODS region ($\beta=1.28\pm0.03$), suggesting that the background noise in the XDF region is more correlated than that in the GOODS-S field. This is likely because the XDF region is deeper. The larger level of the correlation between adjacent pixels may arise from underlying faint objects that are not picked up by SExtractor, and from a different dither pattern. In Figure \ref{fig:nmad}, we also plot the $\sigma_\textrm{NMAD}$ relation presented in \citet{Whitaker2019} in dashed line, which lies slightly above our derived relation from the GOODS region but smaller than the derived relation from the XDF regions. These characteristics are expected because \citet{Whitaker2019} derived their $\sigma_\textrm{NMAD}$ relation from the combined HLF field that includes both XDF region and GOODS region (which contribute to the majority of the area).

We use the derived $\sigma_\textrm{NMAD}(N)$ relations to calculate the flux uncertainties of each object. For each band and object $i$, we first calculate the flux uncertainties of aperture with radius $r$ pixels by multiplying $\sigma_\textrm{NMAD}(\sqrt{\pi r^2})$ with an average rms at the object's position ($\Bar{\sigma_i}$), which is equal to \texttt{FLUXERR\_APER}$/\sqrt{\pi r^2}$. This is because in SExtractor, when detector gain is large, \texttt{FLUXERR\_APER} is approximately equal to
\begin{equation}
    \texttt{FLUXERR\_APER}_i \approx \sqrt{\sum_{j=1}^{N_i^2}\sigma_j^2} \approx \sqrt{\pi r^2}\Bar{\sigma_i}
\end{equation}
The summation is over all $N_i^2 = \pi r^2$ pixels in the circular aperture at object $i$. These calculations yield the uncertainties of aperture fluxes at each object's position.
To calculate the uncertainty of an object's total flux, we propagate the uncertainty of its $0\arcsec.7$ aperture flux to the uncertainty of the total flux according to the total flux calculation described in Section \ref{sec:flux_measurement}. 

\section{\texttt{GLACiAR2}}
\label{appendix:glaciar}
\texttt{GLACiAR2} is a modification of \texttt{GLACiAR}, an open python tool to estimate the completeness of galaxy surveys \citep{Carrasco2018}. We keep the original core structure that creates and recovers the artificial galaxies. We mainly modified the input parameters and the format of the output catalogues to accommodate various types of completeness definitions and UVLF determination procedures in this work.

The summary of the code structure is as follows. For each field, the code simulates galaxies in redshift and input $M_\textrm{UV}$ bins. For each bin, the code randomly draws $M_\textrm{UV}$ values from the bin according to a specified distribution of the injected galaxies and randomly draws UV spectral slopes to create a set of model spectra. By default, each model spectrum is a piece-wise function consisting of a power-law function and a Lyman break:
\begin{equation}
    F(\lambda) = \begin{cases} 
          0 & \lambda\leq 1216 \\
           a\lambda^\beta & \lambda> 1216.
       \end{cases}
\end{equation}
A number (parameter \texttt{n\_galaxies}) of light profiles are assigned for each model spectrum. The light profiles are of Sersic type that are randomized with different Sersic indexes, inclinations, and eccentricities. They are convolved with user-supplied PSF images, and normalized with the fluxes from the model spectra according to the user's supplied filter throughput. They are finally projected onto the images at randomized positions. The criteria for the projection positions are that they must be on the observed area in the detection image (i.e. non-zero pixels) and are not allowed to overlap with other simulated galaxies. The users can control these input parameters with the file `parameters.yaml'. We provide the description of the input parameters are in Table \ref{table:glaciar_parameters}. 

The code proceeds to recover these injected galaxies and classify their recovery status. It first runs \texttt{SExtractor} on these modified images with user-specified \texttt{SExtractor} parameter and auxiliary files. It then attempt to recover each injected source by searching the segmentation map within a user's specified patrol region around the injected position (a square with the side's length equal to $(2\times\texttt{margin})$). The closest object to the injected position within that margin is saved. If the recovered object has a S/N ratio in the detection band greater than a threshold value (\texttt{min\_sn}), we proceed to check whether it the recovered source is blended with any real objects in the original science image. We do so by checking the pixels in the original segmentation map where the simulated object lies. The blending criteria is slightly modified to the following. We now strictly require that at most 25\% of the pixels of recovered source can overlap with any original source to be considered as detected. If the original source is brighter than the input magnitude of the simulated galaxy, we further require that the flux of the recovered source in the detection band must be within 25\% of the input flux. This is to prevent the code from picking up the original source as a ``recovered'' source when the simulated galaxy is placed on an original source that is bright and extended. The recovering status of all simulated source are recorded in the output files (see Table \ref{table:glaciar_status} for the status description). The user has an option to perform further dropout selections based on these recovered sources by modifying the code dropouts.py. We provide the recovery status, the \texttt{AUTO}, the \texttt{ISO}, and the \texttt{APER} magnitudes, as well as their respective S/N ratios in all photometric bands for the user to use in the dropout criteria. 

The output of the code is the following. First is the detail of each injected source. These files, *RecoveredGalaxies*zx.xx.cat, are specific to each redshift bin. They contain both input and output parameters of each injected source, including: input $M_\textrm{UV}$, input UV spectral slope, the recovery status, the dropout selection status, recovered magnitudes and S/N ratios in all bands (\texttt{AUTO}, \texttt{ISO}, and \texttt{APER} magnitudes), Kron radii, \texttt{SExtractor}'s r90 parameter, and the injected positions. The region files of these galaxies are also provided and colour-coded by their recovery status. The second output file, *RecoveredStats*.cat, contains the summary of the recovery status in each redshift and input magnitude bins.
The third file contains the output matrix $N(M_\textrm{in},m_\textrm{out})$ for the recovered galaxies and those that pass the dropout criteria. These files are *RecoveredStats\_*\_nout.cat, and *RecoveredStats\_*dout.cat respectively. For the former, we treated all galaxies with recovery status $\ge0$ as succesfully recovered. $m_\textrm{out}$ is the \texttt{AUTO} magnitude in the detection band. The latter is outputted only if the user indicates that the injected galaxies should go through the dropout section. Their completeness according to the method 1 (completeness as a function of input magnitudes) are also plotted out. These summary files are provided so that the user to see the overall picture of the result. The user can use the detailed output files to obtain the matrix and/or completeness according to the user's desired definition.

\begin{table}[]
    \centering
    \begin{tabular}{|l|p{0.8\textwidth}|}
    \hline
    Status & Description (The injected galaxy is ...)\\
    \hline
    \ 0    & detected (see \texttt{min\_sn} parameter) and is isolated.\\
    \ 1    & detected but is minimally blended with fainter object ($<25\%$ of the pixel area). \\
    \ 2   & detected but is minimally blended with brighter object (see text).\\
    $-1$  & detected but blended with brighter objects.\\
    $-2$  & detected but significantly blended with fainter object. \\
    $-3$   & detected by \texttt{SExtractor} but the S/N ratio in the detection band is smaller than \texttt{min\_sn}.\\
    $-4$   & not detected by \texttt{SExtractor}.\\
    \hline
    \end{tabular}
    \caption{The description of the recovery status of the injected galaxies in \texttt{GLACiAR2}\label{table:glaciar_status}}
\end{table}

\section{Comparison of candidates in this study to those in Bouwens et al. (2015)}
\label{appendex:compare_candidates}
Different studies may find different numbers of dropout candidates in a given field. This can arise from flux scatter, which can be due to using different image processing methods and or flux measurement method, or from different selection criteria. The difference in principle does not matter as long as the completeness simulation correctly accounts for undetected candidates. We nevertheless compare our $z\sim5$ candidates to those in \citet{Bouwens2015} in this section.

\begin{figure*}
    \centering
    \includegraphics[width=0.7\textwidth]{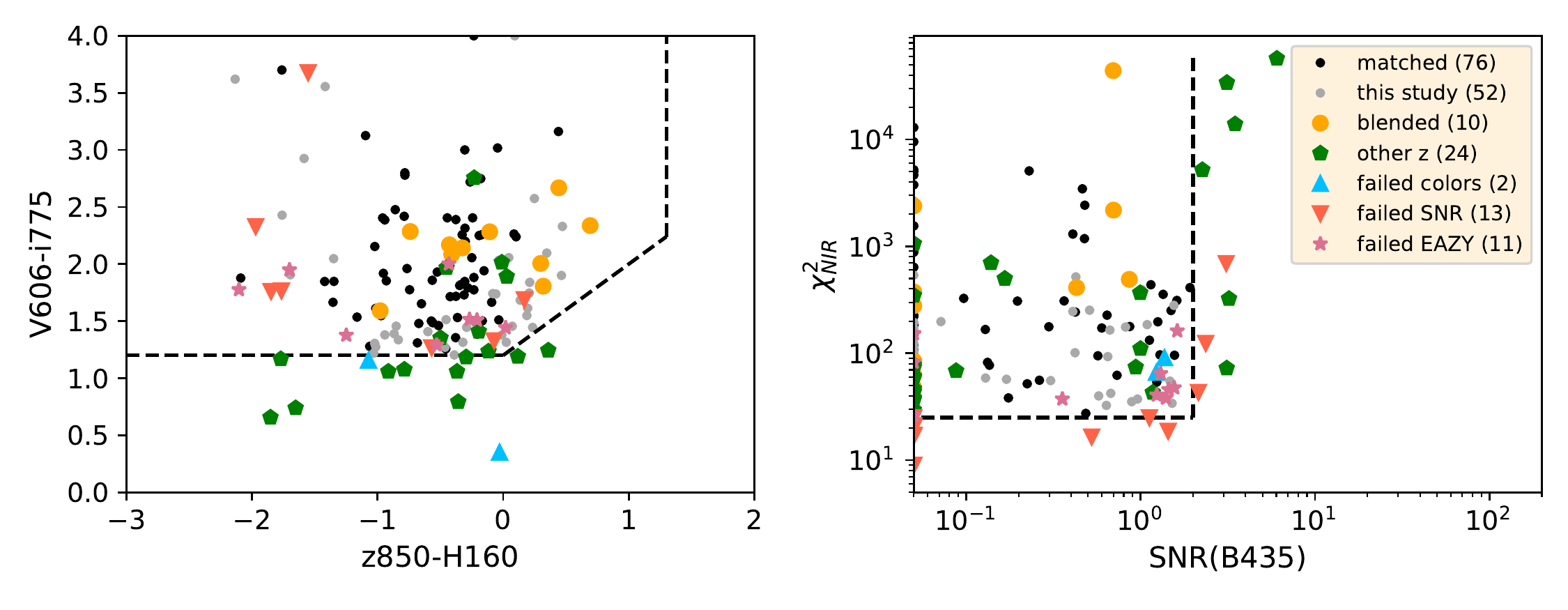}
    \caption{The colour (left) and SNR (right) selection criteria for $z\sim5$ dropouts in the XDF region. The data points show our flux and SNR measurements of our candidates and other galaxies that are SExtracted and matched with the $z\sim5$ candidates in \citet[][B15]{Bouwens2015}. The detailed description of each entry in the legend is as following: matched = candidates that are found in both studies, this study = candidates that are in this study but not in B15, blended = B15 candidates that are classified as blended with bright objects in our study, other z = B15 candidates that satisfy the dropout selections at other redshifts in our study, failed colours = B15 candidates that fail our dropout selections for $z\sim4$ to $z\sim7$ galaxies, failed SNR = B15 candidates that only pass the colour selection for $z\sim5$ dropout in our study but fail the SNR selection, failed EAZY = B15 candidates that failed the photometric redshift criteria in this study. We plot galaxies that fall out of the plotted range at the edges of the figures (i.e. those at the top axis in the left figure and at the left axis in the right figures).
    We note that 11 candidates in B15 are not SExtracted in our study and therefore not shown in the figures.}
    \label{fig:compare_bouwens_XDF}
\end{figure*}

In the XDF region, we found 128 candidates that passed our dropout and photometric redshift selection criteria at $z\sim5$. \citet{Bouwens2015} found 152 candidates at the same redshift within the same area. Among these, 76 candidates are identified as dropouts in both studies. This means that there are 52 objects that are classified as dropouts in our study but not in \citet{Bouwens2015}. On the other hand, there are 76 objects that are classified as dropouts in \citet{Bouwens2015} but are not in out sample. We investigated these 76 objects and found the following. 16 out of the 76 objects are not SExtracted by our study. Most of them are faint and small objects. But there are three objects with $m_\textrm{F160W}<28$ mag, which are blended with other larger galaxies and not SExtracted by our SExtractor parameters. 10 out of the 76 objects are considered as blended with brighter objects according to the blending criteria in Section \ref{sec:selection}. 24 satisfy the dropout selection criteria for other redshifts LBGs (either $z\sim4$ or $z\sim6$). 2 fail the colour selection for any redshifts ($z\sim4$ to $z\sim7$). 13 fail either the SNR(B) upper limit cut or the $\chi^2_\textrm{NIR}$ minimum limit cut. Lastly, there are 11 objects that passed the colour and SNR selection but failed the photometric redshift selection criteria. The summary of these numbers is shown in the colour-colour and SNR plots in Figure \ref{fig:compare_bouwens_XDF}

\begin{figure*}
    \centering
    \includegraphics[width=0.7\textwidth]{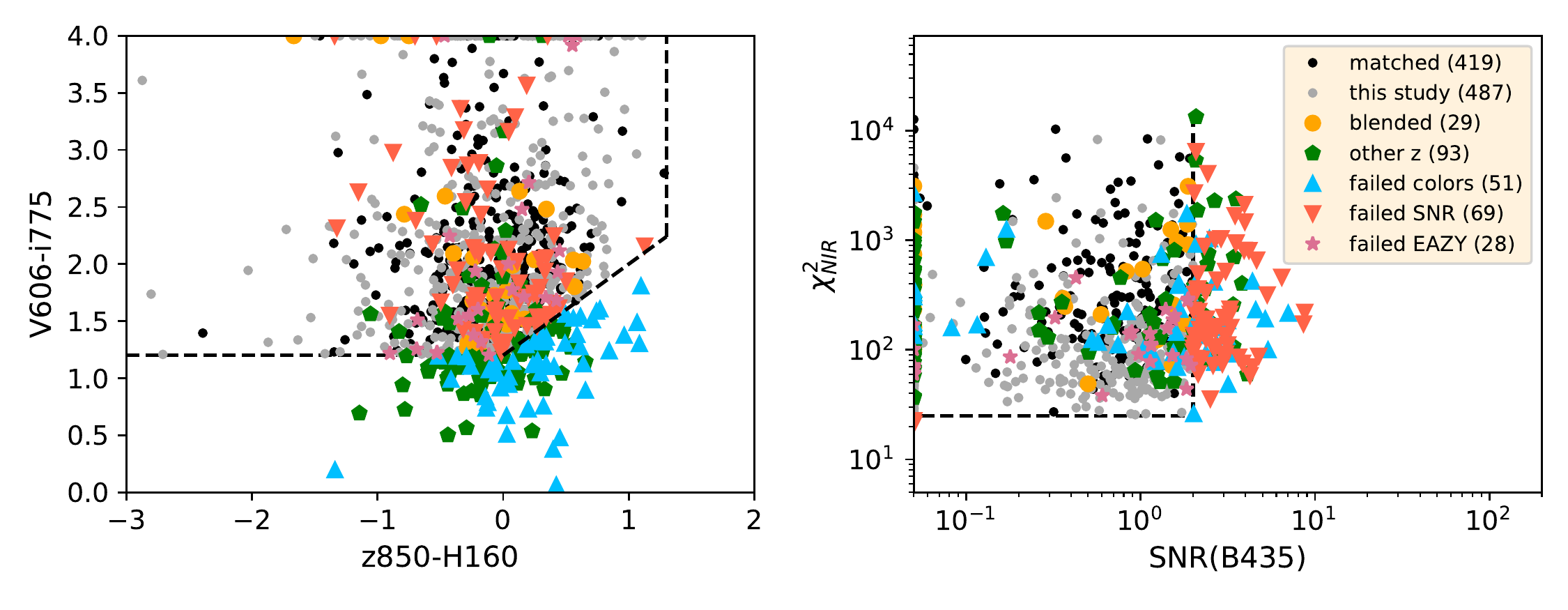}
    \caption{Same as \ref{fig:compare_bouwens_XDF} but for the GOODS region.}
    \label{fig:compare_bouwens_GOODS}
\end{figure*}

We found a similar result for the GOODS region, where about half of the candidates in our study overlaps with those in \citet{Bouwens2015}. However for this region, we found 906 candidates, which is more than the 728 galaxies found in the same area by \citet{Bouwens2015}. The summary is shown in Figure \ref{fig:compare_bouwens_GOODS}.

The overlapping of $\sim50\%$ of the candidates across the two studies is reasonable. Based on the completeness in Figure \ref{fig:completeness_method1}, a galaxy with an intrinsic $M_\textrm{UV}<-19$ generally has a $\sim70\%$ chance of being included in the sample. Assuming that the completeness in \citet{Bouwens2015} is similar, the chance that a galaxy is included in both samples is therefore $0.7^2$ or $\sim50\%$.

\section{Detail of the fitting procedure for each method}\label{appendix:fitting_detail}

Here we describe the fitting procedure we applied on the HLF data for each UVLF derivation method.

\textbf{Method 1:}\label{sec:uvlf_method1}
This method defines completeness as a function of intrinsic magnitude. To determine the UVLF, the number of observed galaxies is compared to the number of predicted galaxies at each intrinsic UV magnitude (i.e. we compare $N^\textrm{exp}(M_\textrm{in},\theta)$ to $N^\textrm{obs}(M_\textrm{in})$). Hence, we need ensure that the derived $M_\textrm{UV}$ of the observed galaxies is as close to the intrinsic values as possible. We follow \citet{Finkelstein2015} to correct the UV magnitudes obtained from SED fitting in Section \ref{sec:selection} with the average difference between the intrinsic and the recovered magnitudes derived from the completeness simulation. For each completeness simulation, we calculate the average difference from all injected galaxies that pass the selection criteria with $M_\textrm{UV,injected}<-18$ for the XDF region and with $M_\textrm{UV,injected}<-19.5$ for the GOODS region. These values are in the range of $\sim0.07$--$0.08$ mag, corresponding to the flux correction of $\sim5$--$10\%$.

We fit the data with the Schechter function:
\begin{multline}
    \phi(M_\textrm{UV}) = \frac{\ln{10}}{2.5}\phi^*\times10^{-0.4(\alpha+1)(M_\textrm{UV}-M_\textrm{UV}^*)}\\\times \exp{[-10^{-0.4(M_\textrm{UV}-M_\textrm{UV}^*)}]}.
    \label{eq:schechter}
\end{multline}
Since a Schechter function implies that the number density of galaxies drops off exponentially at the bright end, we also fit the data with the double power law function to better characterize an alternative (less steep) dropoff at the bright end:
\begin{multline}
    \phi(M_\textrm{UV}) = \frac{\phi^*}{10^{0.4(\alpha+1)(M_\textrm{UV}-M_\textrm{UV}^*)}+10^{0.4(\beta+1)(M_\textrm{UV}-M_\textrm{UV}^*)}}.
    \label{eq:dpl}
\end{multline}
We use a Markov chain Monte Carlo (MCMC) approach to constrain the fitting parameters $\alpha, M_\textrm{UV}^*, \phi^*$ and/or $\beta$. Similar to \citet{Finkelstein2015}, we assume that the probability of detecting a galaxy is Poissonian, whose log-likelihood is:
\begin{multline}
    \log \mathcal{L} = \sum_\textrm{field}\sum_{M_\textrm{in}}\big[N^\textrm{obs}(M_\textrm{in})\ln[N^\textrm{exp}(M_\textrm{in},\theta)]-N^\textrm{exp}(M_\textrm{in},\theta)\\-\ln[N^\textrm{obs}(M_\textrm{in}]\big].
    \label{eq:poisson_likelihood}
\end{multline}
The summation is over both regions (XDF and GOODS) and all $M_\textrm{in}$ bins. We only consider $M_\textrm{in}$ bins with completeness greater than $\sim35\%$, i.e those brighter than or equal to $-19.5$ magnitude for the GOODS region and $-18$ for the XDF region. $N^\textrm{exp}(M_\textrm{in},\theta)$ is calculated by equation \ref{eq:nexp_method1}.

\textbf{Method 2:}\label{sec:uvlf_method2}
We follow the $V_\textrm{max}$ method in \citet{McLure2009} and \citet{Bowler2020}. We first calculate the non-parametric binned LF:
\begin{equation}
    \phi(M_\textrm{bin}) = \sum_{i=1}^N \int_{z_\textrm{min}}^{z_\textrm{max,i}}\frac{C(M_\textrm{bin},z')}{V_\textrm{max,i}(M_i,z')} dz',
\end{equation}
where $C(M_\textrm{bin},z')$ is defined in Section \ref{sec:bowler_method}. The summation is over all galaxies whose recovered $M_\textrm{UV}$ are within the considered $M_\textrm{bin}$. The lower limit of the integration is the redshift below which no simulated galaxy passes the colour selection. We use the minimum redshift simulated in the completeness simulation $z_\textrm{min}=4.1$. The upper limit $z_\textrm{max}$ is specific to each observed galaxy. It is the redshift at which the galaxy's $\chi^2_\textrm{NIR(z)}$ becomes smaller than the S/N requirement of 25, but less than the maximum redshift in the completeness simulation $z_\textrm{max}<6.5$. We then fit these binned LF with the Schechter and the double-power law function. The best-fit parameters are  estimated with the MCMC method as well, using a Gaussian likelihood function. 

\textbf{Method 3:}\label{sec:uvlf_method3}
We  also use the MCMC method to calculate the best-fit parameters for the Schechter and double power-law function. We directly use Equation \ref{eq:nexp_method3} to calculate $N^\textrm{exp}(M_\textrm{recov},\theta)$ and use the Poisson log-likelihood function (Equation \ref{eq:poisson_likelihood}) in the MCMC method. 

\section{An additional test on low signal-to-noise data (direct noise injection in the images)}
\label{appendix:lowsn_data2}
In Section \ref{sec:lowsn_data}, we generated the inferior data sets by adding flux scatter to the measured $M_\textrm{UV}$ to reflect the flux scatters caused by photometric $M_\textrm{UV}$ estimation, while keeping the number of data points the same. In practice, another source of flux scatter can come from having low SNR images due to shallower exposures. In this section, we generate synthetic data sets to investigate this by adding noise to the HLF images directly, which we then use to test the fitting methods considered in this work. 

We generate two additional data sets by adding Gaussian noise to the images of the two region used in this work (XDF and GOODS). For each region, we measure its pixel-scale $1\sigma$ limit, the standard deviation in the values of background pixels. We then create two new sets of images by adding Gaussian noise at the level of $2\sigma$ and $4\sigma$. We also adjust the RMS maps to reflect the noise accordingly. For each new data set, we repeat the procedures in Section \ref{sec:flux_measurement} -- \ref{sec:recovered_MUV} to select the $z\sim5$ candidates, measure their $M_\textrm{UV}$, carry out the completeness simulations, and derive the best-fit UVLFs.

Adding the noise directly to the images mimics observations with shallower depths or observations at higher redshifts. It affects the measurements in two folds: lower number of candidates, and increase flux scatter especially at the faint end. The number of candidates reduces from 906 and 128 (in the GOODS and XDF region respectively) to 371 candidates and 83 candidates in images with $2\sigma$-added noise. The counts drop to 119 and 63 candidates in the images with $4\sigma$-added noise. These numbers of candidates are comparable to the current sample size for candidates at $z\sim7$--$8$ \citep{Bouwens2021}. Based on the results from the completeness simulation, we found that flux scatter at the brightest magnitude bins is only slightly larger than that in the images without added noise. For example, the flux scatter at the brightest magnitude bins in the $4\sigma$-added noise images are $\sim0.12$ mag in the XDF region and $\sim0.18$ mag in the GOODS region, less than 0.05 mag increase as compared to the data without added noise (the blue lines in Figure \ref{fig:mag_uncertainties}). But the transitions from the regime in which the source counts dominate the noise budget to one in which the background dominates happens at roughly 1 magnitude brighter than those in the images without added noise.
At $M_\textrm{UV} = -20$ mag, the flux scatter is 0.5 mag in the GOODS region and 0.25 mag in the XDF region.

We find similar results regarding the best-fit UVLFs by different methods as in the previous section. The best-fit UVLFs by different methods do not differ from each other by more than $1\sigma$ but show the expected behavior. All best-fit Schechter functions from both method 1 and method 2 at the bright end do not differ from each other by more than 0.1 dex. This is because both number of candidates and flux scatters at the bright end in the low-SNR data are similar to those from the images without the added noise. However, there is some noticeable trend at the faint end. For method 1, the best-fit Schechter functions derived from lower SNR data show a steeper faint end, i.e $\alpha=-1.78^{+0.20}_{-0.17}$ from the data with $4\sigma$-added noise and $\alpha=-1.72^{+0.16}_{-0.15}$ from the data with $2\sigma$-added noise. This is to be compared to $\alpha=-1.60^{+0.11}_{-0.12}$ from the data without the added noise. For method 2, at the faint end ($M_\textrm{UV} \gtrsim -20$ mag), the best-fit Schechter functions derived based on the completeness simulations with underlying flat distribution are all $\sim 0.1$ dex lower than those derived based on the completeness simulations with an underlying Schechter distribution. The difference at the faint end is expected since it is where flux scatter in the low SNR data sets is larger than the flux scatter in the original data set. For method 3, the best-fit Schechter functions differ from each other within $0.2$ dex at all magnitude bins without a discernible trend with either SNR of the data sets or the underlying distribution used in the completeness simulations. All these results are consistent with the findings in Section \ref{sec:mock_result} and Section \ref{sec:observational_results}.

\clearpage
\onecolumn
\centering
\begin{longtable}{| p{.15\textwidth} | p{.75\textwidth} |} 
\caption{A simple longtable example from GLACiAR2  \label{table:glaciar_parameters}}\\
\hline
\textbf{Parameter} & \textbf{Description} \\
\hline
\endfirsthead
\multicolumn{2}{c}%
{\tablename\ \thetable\ -- \textit{Continued from previous page}} \\
\hline
\textbf{Parameter} & \textbf{Description} \\
\hline
\endhead
\hline \multicolumn{2}{r}{\textit{Continued on next page}} \\
\endfoot
\hline
\endlastfoot
\texttt{LF\_shape}** & List of the underlying distributions of the injected galaxies. The choices are the following.\newline
`flat' - The distribution of the injected galaxies within each $M_\textrm{in}$ bin is flat. All input magnitude bins have the same number of injected galaxies, \texttt{n\_galaxies} $\times$ \texttt{n\_iterations}.\newline
`schechter\_flat' - The distribution of the injected galaxies within each $M_\textrm{in}$ bin follows a Schechter distribution with the specified parameters in \texttt{LF\_Schechter\_params.txt}. All input magnitude bins have the same number of injected galaxies, \texttt{n\_galaxies} $\times$ \texttt{n\_iterations}.\newline
`schechter' Both distribution of injected galaxies within each $M_\textrm{in}$ bin and the number of galaxies in each magnitude bin follow the Schechter distribution. \newline 
`linear' or `exp' - Both distribution of injected galaxies within each $M_\textrm{in}$ bin and the number of galaxies in each magnitude bin follow a linear (or exponential) relation with a slope \texttt{lin\_slope} (or exponential base \texttt{exp\_base}).\newline
For the latter three options, the number of the injected galaxies in the brightest input magnitude bin is \texttt{n\_galaxies} $\times$ \texttt{n\_iterations}. The nunber at fainter bins will be larger than this and is according to the specified distribution. Use these features with caution because the number can get large (default=`schechter\_flat').
\\ 
\texttt{n\_galaxies}* & Number of injected galaxies per image per iteration when \texttt{LF\_shape} is flat or schechter\_flat. Otherwise, it is the number of galaxies injected per iteration in the brightest magnitude bin -- see above. (default=100) \\ 
\texttt{n\_iterations}* & Number of iterations for each redshift and input magnitude bins. Each iteration will inject \texttt{n\_galaxies} galaxies with the same model spectrum, i.e. same $\beta$ slope and same $M_\textrm{UV}$ drawn from the \texttt{LF\_shape} distribution. But they will have different light profiles, which are randomly assigned according to the specified Sersic index, inclination and ellipticity ranges.  (default = 20).  \\ 
\texttt{n\_inject\_max}** & If the number of injected galaxies per iteration is larger than this number, the code will inject the galaxies in batches of \texttt{n\_inject\_max} galaxies. This is to avoid overcrowding the image with the injected sources. (default = \texttt{n\_galaxies}).  \\ 
\texttt{min\_mag}* & Brightest recovered magnitude bin in the output matrix.  (default = 24.0).  \\ 
\texttt{max\_mag}* & Faintest recovered magnitude bin in the output matrix. (default = 30.0).  \\ 
\texttt{mag\_bins}* & The number of recovered magnitude bins in the output matrix. It is also used to determined the size of the input $M_\textrm{UV}$ magnitude bins. (default = 13)  \\
\texttt{Minput\_min}** & Brightest input $M_\textrm{UV}$ bin.  (Default = the magnitude corresponding to \texttt{min\_mag} and \texttt{max\_z})  \\
\texttt{Minput\_max}** & Faintest input $M_\textrm{UV}$ bin.  (Default = the magnitude corresponding to \texttt{max\_mag} and \texttt{min\_z})  \\
\texttt{min\_z} & Minimum redshift of the simulated galaxies (default = 7.5)  \\
\texttt{max\_z} & Maximum redshift of the simulated galaxies (default = 9.0) \\
\texttt{z\_bins} & The number of desired redshift bins (default = 16) \\
\texttt{ref\_uv\_wl}** & The wavelength in angstrom at which $M_\textrm{UV}$ is determined. (default = 1600)  \\
\texttt{n\_bands} & Number of filters the survey images have been observed in (required)  \\
\texttt{bands} & Name of the bands from \texttt{n\_bands} (required)\\
\texttt{detection\_band}* & Name of the detection band. The recovered magnitude in the output matrix will be in this band. Alternatively put `det' if the detection images are coadded from multiple-band images. (required) \\
\texttt{det\_combination}** & Name of the detection bands used in a coadd. The recovered magnitude in the output matrix will be in the band listed in the first entry of this list. (required if \texttt{detection\_band} is `det') \\
\texttt{coadd\_type}** & The coadd method used to create detection images. The choices are 
1 for a simple coadd or 2 for a coadd of noise-equalized images. This is only used when \texttt{detection\_band} is `det'. (default = 1) \\
\texttt{zeropoints} & Zeropoint values corresponding to each band in \texttt{bands}. They are used to assign pixel values of the simulated galaxies and run \texttt{SExtractor} (Default = 25.0)\\
\texttt{gain\_values} & Gain values for each band in \texttt{bands}. They are used to run \texttt{SExtractor}. (required)\\
\texttt{size\_pix} & Pixel scale for the images in arcsec (default = 0.08) \\
\texttt{margin**} & The size of the square within which the recovery search is performed. It is half the length of the side of a square, centering at the position of each injected galaxy. The unit is arcsecond. (default = 0.3) \\
\texttt{list\_of\_fields} & Text file containing a list with the names of the fields the simulation will run for, which can be one or more. (required) \\
\texttt{path\_to\_images} & Directory where the images are located (required). \\
\texttt{path\_to\_results} & Directory where outputs will be placed. If not specified, it will raise an error. The programme will create a folder inside it with the results (required). \\
\texttt{image\_name} & Name of the images. All science and rms images should have the name as follows: ‘\texttt{image\_name}+\texttt{field}\_\texttt{band}+\texttt{imfits\_end}' and ‘\texttt{image\_name}+\texttt{field}\_\texttt{band}+\texttt{rmsfits\_end}', respectively. (required)\\
\texttt{imfits\_end}** & See above (default = \_drz.fits)\\
\texttt{rmsfits\_end}** & See above (default = \_rms.fits)\\
\texttt{fixed\_psf}** & This is required if the images are psf-matched. It is the file name of the common psf fits image that is put in the Files folder. If the images are not psf-matched, this parameter is not required and the psf in each band is assumed to be Files/psf\_\texttt{band}.fits (optional)\\
\texttt{R\_eff} & Effective radius in kpc for a simulated galaxy at $z = 6$. This value changes with the redshift as $(1+ z)^{-1}$. (default = 1.075 kpc) \\
\texttt{beta\_mean} & Mean value for a Gaussian distribution of the UV spectral slope (default = $-2.2$) \\
\texttt{beta\_sd} & Standard deviation for the for a Gaussian distribution of the UV spectral slope (default = 0.4) \\
\texttt{types\_galaxies} & Number indicating the amount of Sérsic indexes (default = 2) \\
\texttt{sersic\_indices} & Value of the Sérsic index parameter n for the number of \texttt{types\_galaxies} (default = [1, 4]) \\
\texttt{fraction\_type*} & Fraction of galaxies with each specified Sérsic indexes (previous name is \texttt{fraction\_type\_galaxies}, default = [0.5, 0.5]) \\
\texttt{de\_Vaucouleurs} & Boolean indicating whether de Vaucouleurs galaxies (n = 4) will only have circular shape. (default = False)\\
\texttt{ibins} & Number of inclination angle bins. The inclinations vary from $0^{\circ}$ to $90^{\circ}$, e.g. if 9 bins are chosen, the inclinations of each galaxy will be randomly assigned as $0^{\circ}$, $10^{\circ}$,..., or $80^{\circ}$.  (default = 9)\\
\texttt{ebins} & Number of eccentricity bins. The values vary from 0 to 1, e.g.. if 5 bins are chosen, eccentricity of each galaxy will be randomly assigned as 0, 0.2,..., or 0.8. One bin indicates only circular shapes. (default = 5) \\
\texttt{min\_sn} & Minimum S/N ratio in the selected band for an artificial object to be considered detected. If an object is SExtracted to have S/N ratio below this threshold, it will not go through the blending and/or dropout classification. Note that the user can also specify a stricter S/N criteria in the dropout selection. (default = 3)  \\
\texttt{dropouts} & Boolean that indicates whether the user desires to run a dropout selection (default = False)\\
\hline
\multicolumn{2}{|l|}{* The parameter is modified from the original \texttt{GLACiAR} code \citep{Carrasco2018}, ** new parameters in \texttt{GLACiAR2}} \\
\end{longtable}
\clearpage
\twocolumn

\bsp	
\label{lastpage}
\end{document}